\documentclass[12pt, letterpaper]{JHEP3}
\usepackage{amsmath,epsf,amssymb,latexsym,cite,graphics,bbm}
\usepackage[matrix,arrow,frame,import,curve,color]{xy}
\DeclareFontFamily{U}{rsf}{}
\DeclareFontShape{U}{rsf}{m}{n}{
  <5> <6> rsfs5 <7> <8> <9> rsfs7 <10-> rsfs10}{}
\DeclareMathAlphabet\Scr{U}{rsf}{m}{n}

\def\C{{\mathbb C}}

\def\P{{\mathbb P}}
\def\R{{\mathbb R}}
\def\Z{{\mathbb Z}}

\def\Aut{\operatorname{Aut}}
\def\End{\operatorname{End}}

\def\Res{\operatorname{Res}}

\def\diag{\operatorname{diag}}

\def\im{\operatorname{im}}

\def\smdet{\operatorname{det}}  %different from the math \det, which makes \det_{a} put the "a" under the "det."

\def\SL{\operatorname{SL}}
\def\GL{\operatorname{GL}}

\def\GU{\operatorname{U{}}}

\def\GE{\operatorname{E}}

\def\p{\partial}

\def\la{\langle}
\def\ra{\rangle}
\def\lad{\langle\!\langle}
\def\rad{\rangle\!\rangle}

\def\ff#1#2{{\textstyle\frac{#1}{#2}}}
\def\half{\frac{1}{2}}

\def\cD{{\cal D}}
\def\cE{{\cal E}}
\def\cF{{\cal F}}

\def\cJ{{\cal J}}
\def\cK{{\cal K}}
\def\cL{{\cal L}}
\def\cM{{\cal M}}
\def\cN{{\cal N}}
\def\cO{{\cal O}}

\def\cQ{{\cal Q}}

\def\ep{{\epsilon}}

\def\Up{{\Upsilon}}

\newcommand\gammab{\overline{\gamma}}

\newcommand\epb{\overline{\ep}}

\newcommand\thetab{\overline{\theta}}

\newcommand\lambdab{\overline{\lambda}}

\newcommand\sigmab{\overline{\sigma}}
\newcommand\taub{\overline{\tau}}

\newcommand\phib{\overline{\phi}}
\newcommand\chib{\overline{\chi}}
\newcommand\psib{\overline{\psi}}

\newcommand\etat{\widetilde{\eta}}

\newcommand\xit{\widetilde{\xi}}

\newcommand\Gammab{\overline{\Gamma}}

\newcommand\Sigmab{\overline{\Sigma}}
\newcommand\Upsilonb{\overline{\Upsilon}}
\newcommand\Phib{\overline{\Phi}}

\newcommand\sbr{\overline{s}}  %\sb already defined.  no idea what it does.
\newcommand\tb{\overline{t}}

\newcommand\zb{\overline{z}}

\newcommand\qt{\tilde{q}}

\newcommand\Eb{\overline{E}}

\newcommand\Gb{\overline{G}}

\newcommand\Jb{\overline{J}}
\newcommand\Kb{\overline{K}}

\newcommand\Pb{\overline{P}}
\newcommand\Qb{\overline{Q}}

\newcommand\Ht{\widetilde{H}}

\newcommand\Jt{\widetilde{J}}

\newcommand\Wt{\widetilde{W}}

\def\cDb{\overline{\cD}}
\def\cQb{\overline{\cQ}}
\def\cJb{\overline{\cJ}}
\def\cOb{\overline{\cO}}

\def\Autt{\widetilde{\Aut}}

\newcommand{\bea}{\begin{eqnarray}}
\newcommand{\eea}{\end{eqnarray}}
\newcommand{\be}{\begin{equation}}
\newcommand{\ee}{\end{equation}}
\newcommand{\bma}{\begin{pmatrix}}
\newcommand{\ema}{\end{pmatrix}}

\newcommand{\del}{{\partial}}

\title{Summing the Instantons in Half-Twisted Linear Sigma Models}
\author {Jock~McOrist\\
\normalsize Enrico Fermi Institute \\
\normalsize University of Chicago \\
\normalsize Chicago, IL 60637, USA
}
\author{Ilarion V.~Melnikov\\
\normalsize Max-Planck-Institut f\"ur Gravitationphysik (Albert-Einstein-Institut),\\
\normalsize Am M\"uhlenberg 1, D-14476 Golm, Germany
}
\abstract{We study half-twisted linear sigma models relevant to (0,2) compactifications of the heterotic string.  Focusing on theories with a (2,2)
locus, we examine the linear model parameter space and the dependence of genus zero half-twisted correlators on these parameters.  We show that in
a class of theories the correlators and parameters separate into A and B types, present techniques to compute the dependence, and apply these to
some examples.  These results should bear on the mathematics of (0,2) mirror symmetry and the physics of the moduli space and Yukawa couplings in heterotic compactifications.}

\preprint{AEI-2008-079\\EFI-08-26}

\begin{document}
\section{Introduction}
Quantum corrections to classical geometric notions play a key role in the study of string vacua.  Quantum
effects are known to resolve classical singularities, to connect seemingly disparate moduli spaces, to provide quantitative tests
of string dualities, and even to destabilize classical string vacua.  The degree to which these quantum effects are understood
is closely related to the number of space-time and world-sheet supersymmetries preserved by the background.

In this work we will be concerned with quantum corrections in $N=1$, $d=4$ compactifications of the perturbative heterotic
string.  This is probably the simplest string compactification that leads to ``almost familiar'' models of $N=1$
SUSY particle physics coupled to gravity.  The apparent simplicity of these backgrounds is due to the rather direct relation
between space-time physics and the (0,2) superconformal theory on the string world-sheet.  As long as the theory is
at weak string coupling, the study of these $N=1$ compactifications is reduced to two-dimensional physics.  When the
world-sheet SCFT is based on a large radius geometry, the two-dimensional physics reduces to the study of
geometry of holomorphic vector bundles over certain complex manifolds.

Despite such a well-understood conceptual framework, even theories with a weakly coupled large radius limit remain mysterious.
What are the quantum corrections to the classical moduli space?  Where does the world-sheet theory become singular, thereby requiring some non-perturbative string phenomena to resolve the singularity?  Can we compute the moduli dependence of some simple quantities
such as Yukawa couplings of charged matter fields?  Answers to these questions are crucial to the study of non-perturbative
effects in the heterotic string, moduli stabilization in these backgrounds, and quantitative applications to phenomenology.
Yet, they remain relatively unexplored even in the heterotic string on the Calabi-Yau quintic hypersurface in $\P^4$!

Motivated by these questions, we concentrate on a tame set of (0,2) theories:  those with a (2,2) locus and a geometric
interpretation as a sigma model for a Calabi-Yau target-space equipped with a rank $3$ holomorphic vector bundle.  The (2,2) locus corresponds to setting
the holomorphic bundle to be the tangent bundle of the Calabi-Yau manifold, and the (0,2) deformations correspond to deformations of the
tangent bundle.   Even within this class of examples, it is possible for quantum effects
to lift classical moduli~\cite{Braun:2008sf}.  Early on, it was shown that generically world-sheet instantons contribute to a potential for deformations
that break (2,2) supersymmetry~\cite{Dine:1986zy}.  However, it was subsequently persuasively argued that in a large class of
models these instanton effects are either entirely absent~\cite{Distler:1986wm}, or cancel among
themselves~\cite{Silverstein:1995re,Basu:2003bq,Beasley:2003fx}. This class includes the sigma models with target-space a
Calabi-Yau hypersurface in a toric variety, and it is these stable theories that we study.

These theories provide a fertile ground for exploring (0,2) deformations.  On the one hand, the (2,2) locus is
well-understood: mirror symmetry elegantly answers the basic questions raised above, and the computational aspects of mirror
symmetry are well developed through the use of simple field-theoretic tools such as topological field theories and
linear sigma models.  On the other hand, they have (0,2) deformations that, while appearing drastic from the world-sheet
perspective, seem entirely benign from the space-time point of view: the low energy theory is still a supersymmetric
$\GE_6\times\GE_8$ chiral gauge theory coupled to $\cN=1$ supergravity.  Accordingly,
the effect of small (0,2) deformations should just be to slightly shift various K\"ahler potentials and Yukawa couplings.

Could the world-sheet theory also be affected less drastically than first thought?  Is there a sensible extension of mirror symmetry
that would allow computations of quantum corrections in the presence of (0,2) deformations?  Are the tools developed to study the
(2,2) models useful off the (2,2) locus?  Over the years, a number of results have suggested this is the case.

First,  as in the (2,2) case, there are exactly soluble (0,2) SCFTs where a mirror isomorphism may be explicitly constructed~\cite{Blumenhagen:1996vu}.  Second, the familiar A and B chiral rings continue to make sense off the (2,2) locus~\cite{Adams:2003zy,Adams:2005tc}.  That is, the (0,2) theories on a genus zero world-sheet have two finite topological rings, each computed by an appropriate half-twisted
theory.  We refer to these as the A/2 and B/2 twists.  Finally, studies of half-twisted massive (0,2) linear sigma models and
Landau-Ginzburg theories have shown that these rings are eminently computable and provide a  non-trivial generalization of quantum cohomology~\cite{Adams:2003zy,Katz:2004nn,Guffin:2007mp}. These findings suggest that there may be a well-defined mirror map, exchanging A/2-twisted and B/2-twisted theories.

In this work, we add to these results an analysis of half-twisted linear sigma models for a Calabi-Yau hypersurface in a toric variety. The (0,2)-theories we consider are given by small deformations away from the (2,2) locus. Our aim is to elucidate the role of non-perturbative quantum corrections and bundle parameters in physical observables.  Let us now summarize the results we obtain.

\subsection{A Summary of the Results}
Our first set of results relates to (0,2) linear sigma models for projective toric varieties.
Following~\cite{Morrison:1994fr}, we refer to such a theory as a {\em V-model}.  This theory admits the A/2 twist, and the natural parameters in the
A/2-twisted V-model Lagrangian are divided into two classes:  complexified K\"ahler parameters, collectively denoted by $q$, which
preserve (2,2) supersymmetry; and the {\em E-parameters} describing the (0,2) deformations.

For technical reasons, we separate the E-parameters into two classes:  the linear and the non-linear.  As one might guess from the
terminology, the dependence of the twisted correlators on the first class is easy to compute~\cite{McOrist:2007kp}, while the second class remains a challenge.  Some computations in examples suggest that there are circumstances where the half-twisted
correlators do not depend on these non-linear parameters, but we do not have a proof that this is so.

By relating the parameters in the Lagrangian to the geometry of $V$, it is easy to see that the E-parameters should roughly be thought of as deformations of the tangent bundle of the variety $V$.  We say ``roughly,'' because to match the deformations of the bundle, this space must be modded out by a certain group related to the group of automorphisms of $V$.  Although this quotient is difficult to define globally, it does give us some idea of
the space of deformations in a small neighborhood about a suitably generic point.  We refer to these deformations as the {\em E-deformations}.
We expect that the A/2-twisted V-model should only depend on the E-deformations, and not a particular choice of the E-parameters, which means
there must be field redefinitions in the theory that act on the E-parameters but do not affect properly normalized amplitudes.

Our first result, obtained in section~\ref{s:edefs}, is to describe the relevant field redefinitions and use these to count the
E-deformations. This corrects a formula in our earlier work~\cite{McOrist:2007kp}, where only linear E-deformations
were considered.

There are two sets of techniques available to compute correlators in the A/2-twisted V-model: the approach of~\cite{Guffin:2007mp}, which uses algebraic techniques to compute sheaf cohomology on the instanton moduli space; and an approach that computes the entire instanton series by extending (2,2) Coulomb branch techniques~\cite{Melnikov:2006kb} to include linear E-parameters~\cite{McOrist:2007kp}.  The first method is powerful---for instance, it should be able to determine any dependence on the non-linear E-deformations---but requires a bit of commutative algebra machinery and work at the level of {\v C}ech co-chains. The second method, though currently restricted to linear deformations, is computationally simpler to use and provides a
quick route to quantum cohomology.  In section~\ref{s:toricint} we propose a third method that avoids some of the complications
of~\cite{Guffin:2007mp} and closely resembles the familiar toric intersection theory on instanton moduli space available on the (2,2) locus.

Next, we turn to the {\em M-model}, the linear sigma model for a Calabi-Yau hypersurface $M\subset V$.  Our first task, as in the
V-model, is to count the parameters in the M-model Lagrangian modulo field redefinitions.  The parameters are divided into
the complexified K\"ahler and E-parameters already familiar from the V-model and the new {\em J-parameters}
describing the choice of Calabi-Yau hypersurface in $V$, as well as the restriction of the E-deformed bundle to it.  The E- and J-parameters
are restricted by (0,2) supersymmetry to satisfy a number of bilinear constraints, collectively denoted by $E\cdot J = 0$.

As in the V-model, we expect that a number of these parameters may be absorbed by field redefinitions into
irrelevant D-terms.  In section~\ref{s:mdefs}, we describe what we believe to be the complete set of such redefinitions modulo certain
genericity assumptions.  Combining the count of parameters in the Lagrangian modulo the $E\cdot J$ constraint and the field redefinitions,
we obtain a count of linear model deformations.  These do not completely describe the full space of marginal deformations of the
SCFT; however, we hope that they will play an analogous role to the {\em toric} and {\em polynomial} deformations of (2,2) models.

Next, we turn to a study of the A/2-twisted M-model.   We use localization properties of the half-twisted path integral to show that the
genus zero A/2-twisted correlators are independent of the J-parameters and reduce to computations in the associated V-model.  This
(0,2) extension of the quantum restriction formula of~\cite{Morrison:1994fr} is derived in section~\ref{ss:a2qrestrict}.  Combining this with our
results on the V-model, we obtain the complete dependence of the A/2-twisted M-model correlators on the $q$ and the linear E-parameters.
In addition, we compute the (0,2) analogue of the discriminant locus in the model, and show that the correlators obtained by quantum
restriction do show the expected divergences.  We apply our results to some interesting models, including the bi-cubic hypersurface in
$\P^2\times\P^2$.

Having obtained a reasonable understanding of the A/2-twist, we turn to the B/2-twist of the M-model, where our results are not as
complete.  We again rely on localization of the B/2-twisted path integral, and by analysing the zero mode sector, we derive
in section~\ref{ss:vanishing} sufficient conditions for the genus zero B/2-twisted correlators to be independent of the $q$ parameters, and, therefore, to
reduce to classical geometric computations on $M$.  The conditions are satisfied in a number of models, such as the bi-cubic hypersurface
in $\P^2\times\P^2$ and the two-K\"ahler parameter hypersurfaces in weighted $\P^4$.

A priori, this analysis does not guarantee the B/2-twisted correlators to also be independent of the E-parameters; however, we show in
section~\ref{ss:lg} that B/2-twisted theories that are independent of K\"ahler parameters and have a Landau-Ginzburg phase are automatically
independent of E-deformations.

\subsection{A Brief Glimpse of Applications}
Our results show that the dependence on bundle moduli of certain {\em un-normalized} Yukawa couplings is readily computable. There are many new hints of various non-renormalization results, such as those obtained in the B/2 theories we study, as well as explicit computations of how quantum effects modify expectations from classical geometry.  For instance, the expression we derive for the discriminant locus of the A/2-twisted M-model implies that the K\"ahler moduli and the E-parameters enter on the same footing, with the former resolving classical bundle singularities, and the latter smoothing singularities in (2,2) SCFTs.

On the (2,2) locus the linear model parameters, often termed {\em algebraic} coordinates, turn out to be particularly suited to the study of
mirror symmetry.   The existence of these coordinates and the corresponding {\em global monomial-divisor mirror map} was explored
in~\cite{Morrison:1994fr,Morrison:1995yh} based on earlier work of~\cite{Witten:1991zz,MR1269718,Hosono:1993qy,Aspinwall:1993rj}.
It was shown that in terms of these coordinates mirror symmetry becomes a comparison of rational functions, and the choice of
canonical coordinates for the SCFT (i.e. special coordinates in case of (2,2) supersymmetry) becomes a question that can be studied in the
classical B-model.

A similar structure may exist at least in a neighborhood of the (2,2) locus. In general, the untwisted (0,2) M-model depends
on the K\"ahler parameters $q$, as well as the deformations contained in the E- and J-parameters.  The E- and J-parameters are
difficult to disentangle because of the supersymmetry constraint $E\cdot J=0$, as well as ambiguities introduced by the field redefinitions.

Our analysis suggests that locally in moduli space the deformations may be decomposed into the K\"ahler and E-deformations
and the J-deformations.  In terms of these, we have shown that the A/2-twisted correlators are independent of the J-deformations, and we have
presented evidence that the B/2-twisted correlators are independent of the K\"ahler and E-deformations.  It is then natural to guess that the
action of (0,2) mirror symmetry should exchange these sets of deformations.

The computational techniques we have developed for counting parameters and computing the dependence of correlators
on K\"ahler, E- and J-parameters should be of use to check the purported mirror pair, and we may be able to formulate
a (0,2) mirror map in terms of linear model parameters.  No doubt, the
details are bound to be more involved, but the effort promises high returns. If successful, it may help to determine the  K\"ahler
potential in these theories,  lead to a quantitative understanding of the moduli space in the neighborhood of
the (2,2) locus, and allow us to compute normalized Yukawa couplings in this class of models.  Our results and techniques could
also  shed light on aspects of the moduli space far from the (2,2) locus, such as the transitions between disparate
linear sigma model descriptions and resolutions of singularities studied in~\cite{Distler:1996tj,Chiang:1997kt}.

It should be noted that phenomenologically interesting heterotic compactifications (e.g.~\cite{Bouchard:2005ag}) do not possess a (2,2) locus, and
our results are not directly applicable to those theories.  Nevertheless, we believe the techniques we have developed should generalize to
those examples, at least for compactifications without torsion and an extra $\GU(1)$ left-moving current algebra.  The half-twisted
correlators should still be amenable to solution via localization, and phenomenologically interesting examples should merely require a
more involved notation and book-keeping.  It is less clear how to apply our ideas to heterotic compactifications with torsion and
non-K\"ahler target-space, but a careful study of the half-twisted theories based on the linear model constructed
in~\cite{Adams:2006kb} should be a useful first step.

\subsection{Organization of the Paper}
The rest of the paper is organized as follows. To keep our work reasonably self-contained, we begin with a review of (2,2) linear sigma
models and some details of relevant toric geometry.  In the next three sections, we tackle the A/2-twisted V-model (this is also mostly review),
followed by the A/2 and B/2 twists of the M-model.  We conclude with
a discussion of outstanding issues and what we feel to be the next obvious questions to pursue.  We have included an appendix with our
conventions for (0,2) supersymmetry and the half-twists that are used throughout the paper. Note that some mistakes were made in the counting of parameters below; the corrected counting appears in \cite{Kreuzer:2010ph}.

\section{The Linear Model on the (2,2) Locus} \label{s:22gen}

The material in this section is largely a review of the results obtained in~\cite{Witten:1993yc,Morrison:1994fr}.
The reader is referred to those references for a further discussion of the linear sigma models we study.

The gauged linear sigma model (GLSM)~\cite{Witten:1993yc} has proven to be a versatile tool in exploring the moduli space of non-trivial
superconformal theories.  The utility of the GLSM often amounts to relating questions about quantum geometry to classical geometric notions.
For example, it provides a physical realization for the construction of~\cite{MR1269718} of mirror pairs of Calabi-Yau hypersurfaces in Fano toric
varieties, and reduces many computations in these models to a study of toric geometry.
Before we discuss the details of the gauge theory, we will remind the reader of some aspects of toric geometry relevant to
the physics of linear models.  A more detailed and precise discussion of these properties is given in~\cite{Cox:2000vi}.

\subsection{Toric Geometry Basics} \label{ss:toricbase}
The toric varieties that we will encounter in this paper will be smooth and projective.
However, many of the tools we use apply to the larger class of Fano toric varieties
with certain restrictions on the allowed singularities.  It is this larger class that is relevant for the constructions of~\cite{MR1269718}.

A toric variety $V$ of dimension $d$ has a quotient presentation
\begin{equation}
\label{eq:holoquotient}
V\simeq  \frac{\C^n -F}{[\C^\ast]^{r}},
\end{equation}
where $d=n-r$, and the $\C^\ast$ action on $\C^n$ is given by
\begin{equation}
z^i \to \prod_{a=1}^r (t_a)^{Q^a_i} z^i,~~~t_a\in [\C^\ast]^r,
\end{equation}
where $Q^a_i$ is a matrix of integral charges.  The  {\em exceptional set} $F$ is a union of intersections of hyperplanes in $\C^n$.

This data is encoded by the toric fan $\Sigma_V$.  Recall that a fan in $\R^d$ is a collection of strongly convex rational polyhedral cones such
that: (a) the face of any cone is also in the collection, and (b) the intersection of any two cones is a face of each.  We say that $V$ is
{\em simplicial} if every full-dimensional cone in the fan has $d$ generators.  In this case, the quotient construction above is a standard
geometric quotient, and the $z^i$ are profitably thought of as homogeneous coordinates on $V$.  Smooth toric varieties are always
simplicial.

Let the one-dimensional cones of $\Sigma_V$ be denoted by $\rho^i \in \R^d$, $i=1,\ldots,n$.  The $\rho^i$ are linearly dependent,
and an integral basis for the relations yields a basis for the $(\C^\ast)^{r}$ action on the $z^i$.
The exceptional set is also determined by $\Sigma_V$:  for each collection $\{\rho^i\}_{i\in I}$ that does not belong to a full-dimensional
cone, $F$ contains the intersection of hyperplanes $\cap_{i\in I} \{ z^i = 0\} $.

The $\rho^i$ are in one-to-one correspondence with the torus-invariant divisors on $V$:  $\rho^i \to D_i$, with $D_i$ the image under the
quotient of the hyperplane $\{z^i = 0\}$ to $V$.  These divisors are dual to $\xi_i \in H^{1,1}(V)$, which satisfy a number of properties:
\begin{enumerate}
\item $\xi_i$ generate $H^{k,k}(V)$ under the wedge product, subject to the Stanley-Reisner relations: for each irreducible set $I$ in $F$, we
          have $\wedge_{i\in I} \xi_i = 0.$
\item The top exterior powers have a canonical normalization.  Denoting $\int_V \xi_{i_1} \cdots \xi_{i_d}$ by $\#(\xi_{i_1}\cdots \xi_{i_d})$, we
         find that for non-zero wedge products $\xi_{i_1} \cdots \xi_{i_d} \in H^{d,d}(V)$ we have
         \begin{equation}
         \label{eq:22norm}
         \#(\xi_{i_1}\cdots \xi_{i_d} ) = |\det(\rho^{i_1}, \ldots,\rho^{i_d})|^{-1}.
         \end{equation}
\item The $\xi_i$ are linearly dependent: $\xi_i = \sum_a Q_i^a \eta_a$, where $\{\eta_1,\ldots,\eta_r\}$  is an
         integral basis for $H^2(V)$.
\item This is a complete description of the de Rham cohomology of $V$.
\end{enumerate}
For later use, we note that the normalization condition could be equivalently written in terms of the $Q_i^a$, since
\begin{equation}
\det(\rho^{i_1}, \ldots,\rho^{i_d}) = \pm \smdet_p Q,
\end{equation}
where
\begin{equation}
\label{eq:detQ}
\smdet_p Q = \ep^{i_1\cdots i_d i_{d+1} \cdots i_n} Q^1_{i_{d+1}} \cdots Q^{r}_{i_n},
\end{equation}
and $\ep^{i_1\cdots i_n}$ is the usual fully antisymmetric tensor.

In addition to these aspects of toric intersection theory, we will also have use for some properties of $\Aut(V)$, the group of automorphisms
of a complete, simplicial toric variety $V$ with homogeneous coordinate ring $S = \C[z^1,\ldots,z^n]$~\cite{Cox:1992bob}. These properties are:
\begin{enumerate}
\item $\Aut (V)$ fits into an exact sequence
\begin{equation}
\xymatrix{ 1\ar[r] &[\C^\ast]^{r} \ar[r] & \Autt(V)  \ar[r] & \Aut(V)   \ar[r]  &1.}
\end{equation}
\item $\Autt (V)$ is an affine algebraic group of complex dimension
\begin{equation}
\dim \Autt(V) = \sum_{i=1}^n  |S_{i}|,
\end{equation}
where $S_{i}$ is the set of all monomials in $S$ that have the same charges as $z^i$.
\item The connected component of $\Autt(V)$ is naturally isomorphic to the group of graded $\C$-algebra automorphisms of $S$, meaning
that $\Autt(V)$ has a natural action on $\C^n-F$.
\end{enumerate}
With these tools in hand, we are ready to explore the linear sigma models.

\subsection{The V-Model}\label{ss:22vmodel}
The {\em V-model } is a (2,2) supersymmetric abelian gauged linear sigma model that, for suitably chosen parameters,
flows to a non-linear sigma model with target-space a $d$-dimensional toric variety $V$.  It is easiest to present its
action in terms of (2,2) superspace.
The field content is $n$ chiral superfields $\Phi^i$ and $r$ real vector multiplets $V_a$.
It is also useful to consider the
gauge field-strength superfields $\Sigma_a$, which are {\em twisted chiral} multiplets.  In terms of these, the Lagrangian takes
the form
\begin{equation}
\label{eq:22Vaction}
\cL = \int d^4\theta K + \left\{ \int {d\theta^+d\thetab^-} ~\Wt(\Sigma) +\text{h.c.}\right\},
\end{equation}
where $K$ is the K\"ahler potential and $\Wt$ is the twisted superpotential.   These are given by
\begin{equation}
 K = \sum_{i=1}^n \Phib^i \exp\left[2 \sum_{a=1}^{r} Q_i^a V_a\right] \Phi^i - e_0^{-2} \sum_{a=1}^{r} \Sigmab_a \Sigma_a,~~~
\Wt = -\ff{i}{2\sqrt{2}} \sum_{a=1}^{r}\tau^a \Sigma_a.
\end{equation}
Here $e_0$ is the dimensionful coupling of the gauge theory, the $Q_i^a$ are the gauge charges, and the $\tau^a$ are the
complexified Fayet-Ilioupoulos parameters: $\tau^a = i\rho^a + \theta^a/2\pi.$

For a suitable choice of $\rho^a$ the low  energy well approximated by a non-linear sigma model (NLSM) with target-space the
classical moduli space of the gauge theory,
\begin{equation}
\cM_{0} (r) = \left\{ D_a = {\textstyle\sum_i} Q_i^a |\phi^i|^2 -\rho^a = 0 \right\}/ [\GU(1)^r],
\end{equation}
and complexified K\"ahler class $B+iJ$ linear in $\tau^a$.

A useful  notion for the study of the V-model is the cone $\cK_c \subset \R^{r}$ defined as the set of $\rho^a\in\R^{r}$ for which
the $D$-terms have a solution.  The assumption that $V$ is projective (or more generally Fano) ensures that $\cK_c$ is a pointed polyhedral cone in $\R^{r}$.
$\cK_c$ is subdivided into sub-cones by hyperplanes where a gauge group becomes un-Higgsed.  For each of these sub-cones, the
target-space is a toric variety birational to $V$.  Each of these may be given a holomorphic quotient description as in eqn.~(\ref{eq:holoquotient}), with the various quotients differing only in the exceptional set $F$. In keeping with standard physics terminology, we refer to
the subcones of $\cK_c$ as {\em phases.}  By definition, in the V-model there exists a subcone of $\cK_c$ where $\cM_{0}(r)$ is the
variety $V$.  The region outside of $\cK_c$ is also quite interesting, and we will return to it later.

\subsection{The A-twisted V-model}
The V-model admits the A-twist, a shift of the Lorentz generator by the vectorial R-symmetry\cite{Witten:1993yc}.  Since the supercharges
$\cQ_\pm,\cQb_\pm$ carry R-charge, their spins are also modified, and as a result, the twisted field theory possesses a nilpotent BRST
operator $Q_T = \cQb_+ + \cQ_-,$  whose cohomology isolates the chiral ring of the V-model.  Writing the action of the theory
as a sum of $Q_T$-closed and $Q_T$-exact terms, we discover that the twisted theory is a topological field theory (TFT).
An examination of the action of $Q_T$ reveals that, at least as far as local, gauge-invariant  operators are concerned, this cohomology is
spanned by the $\sigma_a$ fields---the lowest components of the $\Sigma_a$ multiplets.   To determine the ring structure, we must,
therefore, compute the correlators $\la\sigma_{a_1}(x_1)\cdots\sigma_{a_k}(x_k)\ra$ in the TFT.

Even without any detailed computations, it is easy to see that the correlators must be holomorphic functions of the $\tau^a$, since  the $\taub^a$ only appear in the action via  $Q_T$-exact terms that decouple from correlators
of $Q_T$-closed observables.  In addition, they must be independent of the $x_i$, since the energy-momentum tensor of the theory is $Q_T$-exact.

The ring structure is eminently computable by localization of the path-integral.  This localization is a consequence of the
fermionic world-sheet scalar symmetry~\cite{Witten:1991zz}.  The basic point is that a non-trivial orbit of such a fermionic symmetry cannot contribute
to the path-integral for a correlator of $Q_T$-invariant operators, and non-zero contributions come entirely from an arbitrarily small neighborhood of
the fixed-point set.  Thus, the path-integral reduces to an integration over the fixed points of $Q_T$ with a measure that may be determined by
expanding the action around the invariant configurations.  Supersymmetry ensures that the contributions from the non-zero modes in
the expansion will cancel in pairs, thereby reducing the correlator to a finite-dimensional integral.  Provided that the fixed-point set is smooth and
compact, the correlator is easy to compute without any additional input.

In the case at hand, an examination of the action of $Q_T$ identifies the $Q_T$ fixed points to be the configurations satisfying
\begin{equation}
\label{eq:Vinstanton}
d\sigma_a = 0,~~~{\textstyle \sum_a Q_i^a \sigma_a \phi^i = 0}~~(\text{no sum on}~i),~~~\nabla_{\zb} \phi^i = 0,~~~D_a + f_a = 0,
\end{equation}
where $f_a$ is the gauge field strength of the $a$-th gauge field.  The solutions to these equations depend on the choice of phase of the
V-model.  Choosing a phase with subcone $\cK \subset \cK_c$, we find that the first two equations require $\sigma=0$, and the last
two are solved by gauge instanton configurations, whose topological class is labelled by instanton numbers
\begin{equation}
n_a = -\ff{1}{2\pi} \int f_a,
\end{equation}
which are restricted to lie in the dual cone $\cK^\vee$.\footnote{Recall that given a cone $\cK \subset \R^r$, the
dual cone $\cK^\vee \subset (\R^r)^\vee$ is the set of all dual vectors with a non-negative pairing with all generators of $\cK$.}

A  consequence of the R-symmetry of the untwisted theory is that
non-zero contributions to $\la \sigma_{a_1}\cdots \sigma_{a_k}\ra$ only come from instanton sectors obeying $k = d + \sum_i Q_i^a n_a$.

\subsubsection{Gauge Instanton Moduli Space} \label{ss:V2Mn}

Instanton configurations with instanton number $n_a\in \cK^\vee$ have a remarkably simple moduli space:  it is a compact toric
variety $\cM_n$, with combinatorics determined by the fan $\Sigma_V$ and the instanton numbers $n_a$.  More precisely, let
$d_i = Q_i^a n_a$ and consider the following replacements in the holomorphic quotient description of $V$:
\begin{enumerate}
\item $\C^n \to Y = \oplus_i H^0(\cO (d_i)) \simeq \oplus_{i|d_i\ge 0} \C^{d_i+1}$, with coordinates
         $$ z^i \to \left\{ \begin{array}{cl} z^{ij}, & j=0,\ldots d_i,~~\text{for}~~d_i \ge 0, \\ 0 & \text{for}~~ d_i <0. \end{array} \right.$$
\item $F \to F_n$, where for each intersection $\cap_{i\in I} \{z^i = 0\} \subset F,$
$F_n \subset Y$ contains the intersection $\cap_{i\in I_+} \cap_j \{z^{ij} = 0\},$  where $I_+\subseteq I$ is the set of $i\in I$ with $d_i \ge 0$.
\item $(\C^\ast)^{r} \to (\C^\ast)^{r}$, with action $z^{ij} \to \prod_a t_a^{Q_i^a} z^{ij}$ for all $j$.
\end{enumerate}
The instanton moduli space is
\begin{equation}
\label{eq:22Mn}
 \cM_{n} = \frac{Y- F_n}{[\C^\ast]^{r}},
\end{equation}
a toric variety of dimension $d+ \sum_{i| d_i \ge 0} (1+d_i) -n$.  Its toric divisors $\{\xi_{i0},\xi_{i1},\ldots\xi_{i d_i}\}$ are linearly dependent:
\begin{equation}
\xi_{i0} = \xi_{i1} = \cdots \xi_{i d_i} \equiv \xi_{i} = \sum_i Q_i^a \eta_a,
\end{equation}
where the $\eta_a$ furnish an integral basis for $H^2(\cM_n, \Z)$.

The intersection theory on $\cM_{n}$ is now easy to compute by the same combinatoric methods that yield the intersection theory on
$V$.   It is convenient to extend the definition of $\#(\cdots)_{\cM_n}$ from that of $\#(\cdots)_V$:  we set
\begin{equation}
\#(\eta_{a_1}\cdots \eta_{a_k} )_{\cM_n} = 0~~~\text{unless}~~~ k = \dim \cM_n.
\end{equation}
If $k = \dim\cM_n$, then the intersection is given by the toric formulas described in section~\ref{ss:toricbase}.

\subsubsection{Correlators in the GLSM and NLSM}

This description of the moduli space and intersection theory on it leads to a formula for the correlators:
\begin{equation}
\label{eq:V22corr}
\la \sigma_{a_1}\cdots\sigma_{a_k} \ra = \sum_{n\in \cK^\vee}\#(\eta_{a_1}\cdots\eta_{a_k} \chi_{n} )_{\cM_n}\prod_{a=1}^{r} q_a^{n_a},
\end{equation}
where
\begin{equation}
q_a = e^{2\pi i \tau^a},
\end{equation}
$\sigma_a \to \eta_a$ is a canonical identification of the operator $\sigma_a$ with $\eta_a \in H^2(\cM_n,\Z)$, and $\chi_n$ is the Euler class of a certain obstruction bundle, explicitly given by
\begin{equation}
\chi_n = \prod_{i| d_i < 0} \xi_i^{-1-d_i}.
\end{equation}
A moment's thought shows that the expression is consistent with the selection rule that follows from the anomalous ghost number symmetry.
This completely determines the A-twisted correlators of the V-model.

We mentioned that under RG flow the untwisted GLSM flows to the non-linear sigma model with target-space $V$.  That theory also has an $A$-twist,
and the resulting path-integral localizes onto the usual world-sheet instantons of the non-linear model.  The world-sheet instantons have
non-compact moduli spaces, making explicit computations difficult.  The V-model gauge-instantons provide a toric compactification of
that non-compact moduli space, with the two differing only in positive co-dimension.  Thus, it is not surprising that the $\tau_a$ are
the K\"ahler coordinates on the moduli space of the non-linear sigma model, and the  instanton sums are directly related to generating
functions for Gromov-Witten invariants of the variety $V$.

\subsection{The M-Model}
The {\em M-model} is a linear sigma model for $M$---a Calabi-Yau hypersurface in the Fano toric variety $V$.  A clue how to construct such
a theory is provided by the R-symmetry of the V-model.  The classical $\GU(1)_L\times \GU(1)_R$ R-symmetry  is violated by gauge instantons, with anomaly  proportional to $\sum_i Q_i^a n_a$ in a background with instanton number $n_a$.

A way to fix this problem is to add an additional matter superfield $\Phi^0$ with charges $Q_0^a = -\sum_i Q_i^a$.  The
resulting theory, dubbed the $\text{V}^+$-model in~\cite{Morrison:1994fr}, is a linear sigma model for a toric Calabi-Yau manifold of dimension
$d+1$---namely the total space of the anticanonical bundle over $V$.  Since $V^+$ is non-compact, we have the possibility of introducing a
non-trivial superpotential coupling for the matter fields.  We take
\begin{equation}
W = \Phi^0 P(\Phi^1,\ldots,\Phi^n),
\end{equation}
where $P$ is a polynomial
of multi-degree $\sum_i Q_i^a$.  This, finally, is the M-model.  Note that the R-symmetry preserved by the M-model is not the naive R-symmetry
of the $\text{V}^+$-model but rather assigns to the  $\Phi^0$ multiplet charges $(1,1)$ under $\GU(1)_L\times\GU(1)_R.$

The classical moduli space of the M-model consists of the D-term constraints of the $V^+$ theory, as well as new F-term constraints:
\begin{equation}
\phi^0 P_{,i} = 0~~\text{for}~i>0,~~~\text{and}~~~P(\phi) = 0.
\end{equation}
For generic choice of coefficients in $P$, $P=0$  is a smooth hypersurface in $V$, so that the only solution to the first set of conditions is
to set $\phi^0=0$.  This reduces the D-term constraints to those of the V-model, and the remaining F-term constraint $P=0$ leads to the desired
result:  the resulting moduli space is the Calabi-Yau hypersurface $M \subset V$.  This theory is believed to flow to a non-trivial IR fixed point
that is in part characterized by the structure of the familiar (a,c) and (c,c) rings.

An important property of the M-model is that $\cK_c$ is no longer pointed, but rather covers all of $\R^r$. The K\"ahler
moduli space is still conveniently divided into phases, and the interpretation of the low energy theory varies significantly
from phase to phase. We will make use of this in our study of the B/2-twisted M-model.  In what follows, we will refer to any
phase containing, possibly as a multiple of a generator, the vector $\sum_i Q_i^a$ as a {\em geometric phase}.   We will
also apply this terminology to the phases of the V-model.

\subsection{A-Twist of the M-Model: Quantum Restriction}
Like the V-model, the M-Model admits the A-twist.  The only subtlety in performing the twist and localization is due to the non-trivial R-charge of $\phi^0$. Working in a geometric phase, we find that under the twist $\phi^0$ becomes a holomorphic one-form on the world-sheet, denoted by
$\phi^0_z$, whose kinetic term has no zero modes on a genus zero world-sheet. Details of this aspect of the twist have been worked out
recently in~\cite{Guffin:2008kt}.

The $Q_T$-cohomology of local gauge-invariant observables is spanned by the $\sigma_a$.  These observables correspond to a subset of the (a,c) ring of the SCFT:  the $\sigma_a$ correspond
to elements of $H^{1,1}(M)$ that are pull-backs of elements of $H^{1,1}(V)$.  The corresponding deformations of the M-Model are known as
{\em toric} K\"ahler deformations~\cite{Aspinwall:1993nu,Cox:2000vi}.  The ring structure of these toric deformations is captured by the genus
zero correlators in the A-twisted M-model, which we denote by $\lad \sigma_{a_1} \cdots \sigma_{a_k} \rad$ to distinguish them from those of
the V-model.

The selection rule of the V-model is modified, since the ghost number symmetry is no longer anomalous, and there is an extra
multiplet $\Phi^0$ with R-charge $1$.  The upshot is that the M-model correlators vanish unless $k = {d-1}$, and we expect
all instantons to make contributions to the non-zero correlators.

Given the close relationship between the observables of the M- and V-models, it is perhaps not surprising that correlators of the  former are related
to those of the latter.  This relationship is elucidated by considering in further detail the twist and localization of the M-model.  The localization conditions are those of the V-model  (eqn.~(\ref{eq:Vinstanton})), supplemented by
\begin{equation}
\phi^0_z = 0,~~\text{and}~~ P(\phi) = 0,
\end{equation}
so that the path-integral localizes onto subsets $\cM_{n;P}$ of the compact toric moduli spaces $\cM_n$.

Although still compact, the subsets $\cM_{n;P} \subset \cM_n$ are difficult to describe, and an explicit computation of the A-twisted
M-model correlators remains to be carried out.  In contrast to the V-model, the generic gauge instanton in the M-model
does {\em not} correspond to a world-sheet instanton of the corresponding non-linear sigma model.  This suggests that the correlators
computed in the M-model are not simply related to the chiral ring of the SCFT.  This disappointing observation is tempered by powerful
(2,2) non-renormalization theorems which leave just one loop-hole for the disagreement:  the correlators may differ by some non-trivial map relating the
complexified K\"ahler parameters $t^a$ of the SCFT and the $\tau^a$ of the linear model~\cite{Witten:1993yc}.  Presumably, this renormalization could
be derived by integrating out the point-like instantons, but this has not been explicitly demonstrated.

While it may be difficult to find the map $\tau(t)$ directly in the M-model, we may still ask how to compute the correlators in terms
of the $\tau^a$.  Here, we have an important simplification: the chiral superpotential couplings are $Q_T$-exact, and so the correlators
$\lad \sigma_{a_1} \cdots \sigma_{a_d} \rad$ must be independent of the details of the hypersurface.  This is the familiar
statement that the A-model is independent of the complex structure moduli.  In~\cite{Morrison:1994fr} this, combined with degree
considerations implied by the ghost number symmetry and an analysis of the singular locus of the theory, was used to relate the M-model
correlators to those of the V-model:
\begin{equation}
\label{eq:22restrict}
\lad \sigma_{a_1} \cdots \sigma_{a_d} \rad = \la  \sigma_{a_1} \cdots \sigma_{a_d}  \frac{-K}{1-K} \ra,
\end{equation}
where
\begin{equation}
-K = \sum_{i=1}^n Q_i^a \sigma_a
\end{equation}
corresponds to the anti-canonical divisor on $V$.  This is the (2,2) quantum restriction formula.

\subsection{B-Twist of the M-Model}
The M-Model also admits the B-twist, where the axial R-symmetry is used to define new Lorentz transformations
of the fields.  Under this twist the topological BRST charge is $Q_T = \cQb_+ + \cQb_-$, and its cohomology captures a subset of the (c,c) ring of
the SCFT.  The local, gauge-invariant operators are the monomials $O_\alpha = \phi^0 f_\alpha(\phi^i)$ found in the superpotential
\begin{equation}
W = \phi^0 P(\phi^i ) = \sum_\alpha O_\alpha.
\end{equation}
The natural correlators to consider are $\lad O_{\alpha_1} \cdots O_{\alpha_{d-1}} \rad$.   These B-twisted correlators
are independent of the $\tau_a$ (this time it is the {\em twisted} chiral superpotential that is $Q_T$-exact), and an analysis of the $\Qb_\pm$-fixed
points shows that the path-integral localizes onto constant maps from the world-sheet to $M$.  This is just what one expects for the (c,c) ring
of the SCFT based on the Calabi-Yau manifold $M$.

\subsection{Parameters in the (2,2) M-model}
Naively, the M-model action contains the $r$ complexified K\"ahler parameters already familiar from the V-model, as well as the
coefficients of monomials in the superpotential.  It is well-known that these explicit parameters of the linear theory may not
capture all the deformations of the M-model.  The hypersurface $M$ may have K\"ahler classes that are not obtained as restrictions
of classes from $V$, and it may have complex structure deformations that cannot be described as deformations of the defining
polynomial\cite{Green:1987rw,Aspinwall:1993nu,Hosono:1993qy}.  Deformations by these ``non-toric'' and ``non-polynomial'' parameters
are difficult to study in the linear theory.  Nevertheless, the restriction to polynomial and toric deformations is a sensible one.  For instance,
under mirror symmetry the toric deformations are mapped to polynomial deformations of the mirror.

A naive count of the (2,2) M-model complex structure parameters is given by the number of monomials in the superpotential.
This obviously produces a gross over-counting:  in the example of the quintic, there are $126$ monomials
in $P$, but we know very well that $25$ of these are redundant.  To understand how this redundancy manifests itself in the linear model,
consider the set of field redefinitions of the corresponding M-model allowed by gauge-invariance and R-symmetry:
\begin{equation}
\Phi^0 \to u \Phi^0,~~~ \Phi^i \to U^i_j \Phi^j,~~~u\neq 0,~~~U \in \GL(5,\C).
\end{equation}
By using these transformations, we may absorb parameters from the superpotential into the (presumably irrelevant) D-terms.  How many parameters may be eliminated in this fashion?  We must remember that these field redefinitions contain the complexified gauge symmetry, which leaves the superpotential invariant.  Moreover, expanding about a generic superpotential, this is the only non-R symmetry of the superpotential.   Thus, we expect that of the $\GL(5,\C)$ transformations precisely one cannot be used to eliminate parameters in $P$.  Denoting the number of monomials in $P$ by $\#(P)$, we conclude that there are
\begin{equation}
N^{2,2}_{\text{c-x}} (\text{quintic}) = \#P - (1+\dim \GL(5,\C) - 1) = 101
\end{equation}
complex structure deformations of the quintic.

The example of the quintic generalizes to an arbitrary M-model: $\GL(5,\C)$ is replaced by $\Autt(V)$, and the
gauge group has rank $r$, leading to
\begin{equation}
N^{2,2} (M) = r +N^{2,2}_{\text{c-x}} = r+ \#(P) - \dim \Autt(V) + (r-1)\label{eq:nm_22}
\end{equation}
toric and polynomial deformations of the M-model.  This correctly reproduces the count of toric and polynomial
deformations for $M \subset V$ obtained by Batyrev in~\cite{MR1269718}.  Table~\ref{table:parameter} lists
additional hands-on examples.

While this simple counting gives an indication of the dimension of the moduli space near a generic point, there are important and, in
general, not well-understood subtleties in making sense of the quotient of the naive parameter space by $\Autt(V)$~\cite{Aspinwall:1993rj}.
To avoid these issues, we will always assume the theory to be near a suitably generic point in the moduli space, where these difficulties
should not arise.

\subsection{The Virtues of Localization}
Many of the results discussed in this section, and in particular those to do with explicit computation of correlators in the twisted theories, rely
on the localization argument. Localization is also at the heart of why many results obtained for twisted (2,2) theories generalize to (0,2)
half-twisted theories.  Not only do both the twisted and half-twisted path integrals localize, but in fact they localize onto intimately related
sets.  An example of this is already familiar from the half-twisted Landau-Ginzburg theories studied in~\cite{Melnikov:2007xi}.  In what follows,
we will see that similar results hold in the GLSM:  the A and A/2 twisted path-integrals localize onto the same configurations;  the fixed-point
set of the B-twisted theory is in general a subset of the fixed-point set of the B/2 twisted theory, but in many examples we can show the two
to be identical.

In short, it is the localization of the path-integral that makes our computations possible. This feature is expected to persist for arbitrary (0,2)
deformations, as well as (0,2) theories without a (2,2) locus.  This makes us confident that many of our results will generalize to the more phenomenologically interesting theories.

\section{A/2 Twist and Projective Toric Varieties} \label{s:vmodel}
Although our main interest lies in the A/2 twist and (0,2) deformations of the M-model, experience with the (2,2) theories suggests that it
behooves us to first examine the A/2 twisted V-model.  In this section we review several approaches to solving the (0,2) deformed
A/2 twisted GLSM with target-space a smooth Fano toric variety $V$.
\subsection{(0,2) Superspace}
To discuss the (0,2) deformations, we first describe the (2,2) locus in terms of (0,2) superspace, with coordinates
$x^\pm,\theta^+,\thetab^+$, superspace covariant derivatives $\cD_+$, $\cDb_+$, and supercharges $\cQ_+,\cQb_+$.\footnote{The reader will find additional details in appendix~\ref{app:conv}.  We mostly follow the conventions in ~\cite{Witten:1993yc}.} Under this decomposition, the matter superfields $\Phi^i_{(2,2)}$ appearing in eqn.~(\ref{eq:22Vaction}) decompose as
\begin{equation}
\Phi^i_{(2,2)} \to \Phi^i,~~\Gamma^i,
\end{equation}
where $\Phi^i$ is a (0,2) chiral superfield, and $\Gamma^i$ is a Fermi superfield.  The vector multiplet $V_a^{(2,2)}$ decomposes into
a (0,2) vector multiplet and a chiral superfield, and the twisted chiral field-strength multiplets split up as
\begin{equation}
\Sigma_a^{(2,2)} \to \Sigma_a, \Up_a,
\end{equation}
where $\Sigma_a$ is a (0,2) chiral superfield, and $\Up_a$ is another Fermi multiplet.
Let us describe these multiplets in a little more detail.

Working in Wess-Zumino gauge, we find that the vector field and its field-strength have the superspace expansion
\begin{eqnarray}
V_{a,-} & = & v_{a,-} - 2i \theta^+ \lambdab_{a,-} - 2i \thetab^+ \lambda_{a,-} + 2 \theta^+ \thetab^+ D_a, \nonumber\\
\Upsilon_a & = & i \cDb_+ V_{a,-} + \theta^+ \p_- v_{a,+} \nonumber\\
~&=& -2 (\lambda_{a,-} - i \theta^+(D_a -i f_{a,01}) -i \theta^+\thetab^+ \p_+ \lambda_{-,a} ).
\end{eqnarray}
The bosonic multiplets have an expansion involving gauge-covariant derivatives $\nabla$:
\begin{eqnarray}
\Phi^i &=& \phi^i + \sqrt{2} \theta^+ \psi_+^i -i \theta^+\thetab^+ \nabla_+ \phi^i, \nonumber\\
\Sigma_a & = & \sigma_a +\sqrt{2} \theta^+\lambda_{a,+} - i \theta^+\thetab^+ \p_+\sigma_a.
\end{eqnarray}
These fields obey a chirality constraint
\begin{equation}
\cDb_+ \Phi^i = \cDb_+ \Sigma_a = 0.
\end{equation}
The fermionic matter multiplets $\Gamma^i$ are the most interesting new structures to emerge from
the (2,2)$\to$(0,2) reduction.  These fields are not chiral, but rather satisfy
\begin{equation}
\label{eq:GE}
\cDb_+\Gamma^i = \sqrt{2} E^i(\Phi,\Sigma),
\end{equation}
where on the (2,2) locus the $E^i$ are given by
\begin{equation}
\label{eq:22E}
E^i = i \sqrt{2} \sum_a Q_i^a \Phi^i \Sigma_a.
\end{equation}
The explicit superspace expansion is given by
\begin{eqnarray}
\Gamma^i &=& \gamma_-^i - \sqrt{2} \theta^+ G^i - i \theta^+\thetab^+ \nabla_+\gamma_-^i - \sqrt{2}\thetab^+ E^i(\Phi,\Sigma) \nonumber\\
~&= & \gamma_-^i -\sqrt{2} \theta^+ G^i -\sqrt{2}\thetab^+ E^i(\phi,\sigma)\nonumber\\
~&~&~ - i\theta^+\thetab^+\left[\nabla_+\gamma_-^i + 2i E^i_{~,j} \psi_+^j + 2 i E^i_{~,a} \lambda_{a,+} \right].
\end{eqnarray}

The action is a sum of a kinetic term, written as an integral over the whole superspace, and a (0,2) superpotential term:
\begin{equation}
\label{eq:FI}
\cL_{\text{F-I}} = \ff{1}{4}\int d\theta^+ \sum_{a=1}^{r}\tau^a \Upsilon_a + \text{h.c.}.
\end{equation}
The action has an important classical symmetry, $\GU(1)_L \times \GU(1)_R$, with charges display in table~\ref{table:U1vmodel}.
\begin{table}[t]
\begin{center}
\begin{tabular}{|c|c|c|c|c|c|}
\hline
$~			$&$\theta^+ 	$&$\Phi^i		$&$\Gamma^i	$&$\Sigma_a	 $&$\Upsilon_a 	 $\\ \hline
$\GU(1)_R	$&$1			$&$0			$&$0			$&$1			 $&$1			 $\\ \hline
$\GU(1)_L		$&$0			$&$0			$&$-1		$&$-1		 $&$0			 $\\ \hline
\end{tabular}
\end{center}
\caption{The $\GU(1)_L\times\GU(1)_R$ symmetry charges for the V-model.}
\label{table:U1vmodel}
\end{table}
On the (2,2) locus these are just the classical left-moving and right-moving R-symmetries, and the vectorial subgroup may be used to define the
(half-)twist.

\subsection{The E-Parameters and E-Deformations}\label{s:edefs}
Having described the (2,2) locus, we are now ready to contemplate (0,2) deformations.   With the matter content as above, there
is not much choice in how to deform the theory while preserving the global symmetries: we must deform the chirality constraints
of the $\Gamma^i$ multiplets to the most general polynomials in the chiral fields allowed by gauge invariance and the classical
$\GU(1)_L\times\GU(1)_R$ symmetry.  The result is the set of E-parameters.

Recall from section~\ref{ss:toricbase} that for each $i$ we introduced the finite set $S_i$ containing the monomials
$\prod_j (\Phi^j)^{n_j}$ with charges $Q^a_i$.  Each of these monomials is allowed to appear in $E^i$ by gauge invariance
and global symmetries.  A look at the symmetry charges shows that the $E^i$ must remain linear in the $\Sigma_a$ to maintain
the classical $\GU(1)_L\times\GU(1)_R$ symmetry.  Thus, the most general form of E-parameters takes the form
\begin{equation}
\label{eq:02Efull}
E^i = i\sqrt{2} \sum_{a=1}^r \Sigma_a E^{ai} (\Phi)  = i \sqrt{2} \sum_{a=1}^r \sum_{\mu \in S_i}  E^{ai}_{~\mu} ~\mu~ \Sigma_a,
\end{equation}
where the $E^{ai}_{~\mu}$ are complex parameters.

Since the monomials in the $S_i$ correspond to generators of the component of $\Autt(V)$ connected to the
identity,  there is a direct relation between the E-parameters and the elements of the group $\Autt(V)$ discussed in~section~\ref{ss:toricbase}.
Evidently, the $E^i$ introduce $r \times \dim \Autt(V)$ continuous parameters into the action.

It is important to recall that the V-model is believed to be a massive theory.  As such, it might seem strange to discuss
``parameters'' of this model.  However,  the massive theories we consider do have topological
rings that are accessed by the half-twisted theory~\cite{Adams:2005tc}; it is in these half-twisted theories that we
count parameters.  The expected geometric interpretation suggests that the half-twisted theory
should depend on the $r$ K\"ahler parameters any deformation parameters of the tangent bundle $T_V$.  In favorable circumstances,
the latter are counted by $\dim H^1(V,\End T_V)$, but in general there may be elements of $H^1(V,\End T_V)$ that cannot be integrated
to finite deformations.  As we will see shortly, the E-parameters describe {\em unobstructed} deformations of $T_V$, so we should
expect
$$ \#(\text{E-deformations}) \le \dim H^1(V,\End T_V).$$
A look at a few simple examples (e.g. $V\simeq \P^1\times\P^1$) shows that the number of E-{\em parameters} is greater than
$\dim H^1 (V,\End T_V)$, making it clear that not all E-parameters correspond to bundle deformations.

The resolution to this over-count is similar to the one we already encountered in counting (2,2) deformations.  A correct
count is obtained if the following field redefinitions are used to absorb parameters:
\bea
\label{eq:22redefs}
\Phi^i  &\to& \sum_{\mu \in S_i} U^i_{\mu} \mu, \nonumber\\
\Gamma^i & \to & \sum_{\mu \in S_i} U^i_{\mu} \frac{\p \mu}{\p \Phi^k} \Gamma^k,\nonumber\\
\Sigma_a  & \to & G_a^b \Sigma_b,
\eea
where $U^i_{\mu}$ label $\Autt(V)$ parameters of the redefinition and $G_a^b\in \GL(r,\C)$.   As in our counting of the deformations
of the (2,2) M-model, we must remember that $r$ of these redefinitions are global gauge symmetries, which do not
act on the E-parameters.  Thus we find that the V-model should have
\begin{equation}
N(V) = 2r+(r-1) \dim\Autt(V) -r^2,\label{eq:ne_1}
\end{equation}
deformations.

A simple test of this formula is obtained by taking $V$  to be
\begin{equation}
V = \underset{m~\text{times}} { \underbrace{\P^1\times\cdots\times\P^1}},
\end{equation}
a product of $m$ factors of $\P^1$.  In this case, $r = m$, and $\dim\Autt(V) = 4 m$, leading to $N(V) = m+3m(m-1)$.  This matches $H^1(V, \End T_V)$, which in this case is computable from elementary facts about line bundles on $\P^1$.
We will now give a geometric argument for the origin of this formula.

\subsubsection{A Geometric Interpretation}

The geometric import of the E-deformations is simple to see in terms of the low-energy NLSM.  While
the right-handed fermions continue to couple to the tangent bundle of the toric variety $T_V$, the left-handed
fermions couple to a deformation of $T_V$, a bundle $\cE \to V$.  Just as $T_V$ may be built as the quotient
\begin{equation}
\xymatrix{ 0\ar[r] &\cO^{r} \ar[r]^-{Q_i^a z^i} & \oplus_i\cO(D_i)  \ar[r] & T_{V}   \ar[r]  &0, }
\end{equation}
we may define $\cE$ via the exact sequence
\begin{equation}
\xymatrix{ 0\ar[r] &\cO^{r} \ar[r]^-{E} & \oplus_i\cO(D_i)  \ar[r] & \cE  \ar[r]  &0. }
\end{equation}

Let us describe these quotients in a more hands-on way that should be familiar to any devoted reader of~\cite{GSW2}.  Consider
the space of vector fields on $\C^n-F$, $v = v^i \p/\p z^i$.  To obtain vector fields on $V$, i.e. sections of $T_V$,
we must impose the equivalence relations
\begin{equation}
v \sim v + \sum_{a,i} \lambda_a Q^a_i z^i \frac{\p}{\p z^i},~~~\lambda_a \in \C^r.
\end{equation}
Note that the Euler vector fields $e^a = Q^a_i z^i \p/\p z^i$ make sense under the $\C^\ast$ action on the coordinates $z^i$.
In order for the quotient to produce a smooth bundle, the Euler vector fields
must span an $r$-dimensional subspace of the tangent space at every point $p \in \C^n-F$.  When $V$ is smooth, this is
guaranteed to be the case.

To construct the bundle $\cE$ instead of $T_V$, we merely modify the vector fields as follows:
\begin{equation}
\sum_i Q^a_i z^i \frac{\p}{\p z^i} \to \sum_{i}E^{ai}(z) \frac{\p}{\p z^i}.
\end{equation}
The modified fields are still well-defined with respect to the toric action, and furthermore, for small deformations the rank
condition remains preserved for every $p\in \C^n-F$.  Thus, we expect to get a smooth bundle $\cE$.

This explicit description makes it clear that two sets of vectors $e^a$ and $e'^a$ define equivalent holomorphic bundles
when exist $f\in \Autt(V)$, and  $g \in\GL(r,\C)$ such that
\begin{equation}
e^a = g^a_b df (e'^a),
\end{equation}
where $df$ denotes the push-forward map associated to $f$.  Recalling that the $f$ include the $[\C^\ast]^r$ action which
leaves the $e^a$ invariant, we see that the E-deformations of the GLSM are just the deformations of $\cE$ obtained by
deforming the defining exact sequence.

\subsubsection{Linear and Non-Linear Deformations}
The non-linear E-deformations, i.e. those that involve monomials $\mu \neq \Phi^j$ for some $j$, turn out to be
more difficult to study than the linear ones. We suggested in~\cite{McOrist:2007kp} that these non-linear deformations should not
affect the $A/2$-twisted V-model, but this is probably too naive.  It may be that such an independence holds when the
linear parameters are sufficiently close the (2,2) locus, but we have not shown this to be the case.

To organize the linear E-parameters, it is convenient to assemble the matter content into sets of fields with
the same gauge charges.  Labeling these sets by index $\alpha$ and the corresponding charges $Q^a_{(\alpha)}$,
we then recast
\begin{equation}
\cDb_+ \Gamma^i = \sum_{a,j} E^{ai}_{~j} \Phi^j \Sigma_a
\end{equation}
as
\begin{equation}
\label{eq:Edef}
\cDb_+ \Gamma_{(\alpha)} = 2 i M_{(\alpha)} \Phi_{(\alpha)},~~~M_{(\alpha)} = \sum_{a=1}^{r} \Sigma_a E^a_{(\alpha)},
\end{equation}
where $M_{(\alpha)}$ is a $k_\alpha \times k_\alpha$ matrix.  Clearly, $\sum_\alpha k_\alpha = n$.

\subsection{The A/2-twisted V-model in the Geometric Phase}
The A/2 Twist of the (0,2) NLSM with toric target-space was considered in~\cite{Katz:2004nn}.  The point of view advocated in~\cite{Katz:2004nn}
was to combine the familiar structure of (2,2) worldsheet instantons with the notion that in (0,2) theories
the basic A/2 twisted observables (the $\sigma_a$ in our case) should correspond to classes in $H^1(V,\cE^\vee)$.  Classically (i.e. for constant
maps), the computation of a correlator is reasonably clear: $\la \sigma_{a_1}\cdots \sigma_{a_d} \ra$ should correspond to the
intersection pairing
\begin{equation}
H^1(V,\cE^\vee) \times H^1(V,\cE^\vee) \times \cdots \times H^1(V,\cE^\vee) \to H^d(V, \wedge^d \cE^\vee) \simeq H^{d,d}(V)\simeq \C.
\end{equation}
The second-to-last isomorphism automatically holds in theories with a (2,2) locus~\cite{Adams:2005tc}.

By using the universal instanton construction, the authors of~\cite{Katz:2004nn} described how to pull back the bundle (more generally, sheaf)
$\cE$ to a sheaf on $\cM_n$, how to construct the obstruction sheaf (the source of the $\chi_n$ insertion on the (2,2)
locus), as well as how to in principle compute the induced intersection pairing on the instanton moduli space.   As usual in
NLSM computations, these results required some choice of compactification of the instanton moduli space.  In the case
when $V$ is a toric variety, the GLSM naturally provides such a compactification.  The ideas in~\cite{Katz:2004nn} were
refined and developed in~\cite{Guffin:2007mp}, culminating in a general method for computing the A/2 correlators in the V-model.
The result should be thought of as a quantum deformation of the intersection ring on $H^\ast (V,\wedge^k \cE^\vee)$.

While the method of~\cite{Katz:2004nn,Guffin:2007mp} is well-motivated and leads to sensible results, a number of questions naturally
arise.  First, can we be sure that the linear model path-integral is compputed  by this sheaf cohomology on the instanton moduli space?
Second, to derive the quantum
cohomology relations, one must first compute correlators and then extract relations they satisfy.  Can these relations be
obtained in a more straight-forward fashion?  Finally, we know that on the (2,2) locus toric geometry techniques reduce the intersection
theory on $\cM_n$ to simple combinatorics.  Is there a formulation of the (0,2) sheaf cohomology reminiscent of the toric geometry
structures? We will now argue that these questions are answered affirmatively.

\subsection{The Half-Twist in the Coulomb Phase}
We mentioned in our discussion of the (2,2) V-model that when $V$ is Fano, the cone $\cK_c$, where the D-terms have a solution is pointed.
Thus, there exists a region in the $\rho^a$ parameter space where SUSY appears to be broken.  This turns out to be an artifact of the
classical analysis.  In this {\em non-geometric phase} the SUSY vacua are there are discrete Coulomb vacua, where the $\sigma$ fields obtain large VeVs, the $\Phi^i$ matter multiplets get massive, and the dynamics of the $\Sigma_a$ multiplets are determined by an effective twisted
superpotential $\Wt_{\text{eff}}(\Sigma)$~\cite{Witten:1993yc,Morrison:1994fr}.  This effective superpotential encodes the quantum cohomology relations
of the A-twisted V-model, and localization techniques applied in the non-geometric phase yield the correlators in the V-model\cite{Melnikov:2006kb}.

In~\cite{McOrist:2007kp}  we argued that a similar situation holds in the A/2-twisted V-model.  By working in the non-geometric phase and
assuming linear E-parameters, we were able to integrate out the $\Phi^i,\Gamma^i$ multiplets and obtain an effective description of the
remaining light degrees of freedom in terms of a massive Landau-Ginzburg theory with an effective (0,2) potential
\begin{equation}
\label{eq:Coulomb}
\cL_{\text{eff}} =   \int d \theta^+ \sum_{a=1}^{r}\Upsilon_a \Jt_a(\Sigma)|_{\thetab^+=0} + \text{h.c.},
\end{equation}
with
\begin{equation}
\label{eq:J}
\Jt_a  =  \log \left[ q_a^{-1} \prod_\alpha \det M_{(\alpha)}^{Q_{(\alpha)}^a} \right],
\end{equation}
where the $M_{(\alpha)}$ are described in~\ref{eq:Edef}.

In the case of linear E-deformations, the effective potential immediately yields the quantum cohomology relations
\begin{equation}
\label{eq:02QC}
\la \sigma_{a_1} \cdots \sigma_{a_k} \prod_{\alpha|Q^a_{(\alpha)}>0} \det M_{(\alpha)}^{Q^a_{(\alpha)}} \ra =
q_a \la \sigma_{a_1} \cdots \sigma_{a_k} \prod_{\alpha|Q^a_{(\alpha)}<0} \det M_{(\alpha)}^{-Q^a_{(\alpha)}} \ra~~\text{for all}~a.
\end{equation}

As on the (2,2) locus, it is easy to extend this description to an explicit formula for the genus zero A/2-twisted correlators.  A simple generalization
of the localization formulae in half-twisted Landau-Ginzburg models yields the correlators as a sum over the common zeroes of the $\Jt_a(\sigma)$:
\begin{equation}
\label{eq:Coulcorr}
\la \sigma_{a_1} \cdots \sigma_{a_k} \ra = \sum_{\sigma| \Jt = 0}\sigma_{a_1}\cdots\sigma_{a_k} \left[ \det_{a,b} \Jt_{a,b} \prod_\alpha \det M_{(\alpha)}
\right]^{-1}.
\end{equation}
As expected, the correlators are position-independent, given by meromorphic functions of the $q_a$ and the
E-deformations, and satisfy the quantum cohomology relations.  When applied to the example
of $V\simeq \P^1\times \P^1$, the results are in agreement with the computations of~\cite{Guffin:2007mp}.

\subsection{``Toric'' (0,2) Intersection Theory}
\label{s:toricint}
In this section, we return to the geometric phase and  obtain the instanton contributions in an alternative way
that closely resembles the familiar (2,2) computations.  We restrict attention to linear E-parameters and assume
$V$ to be a smooth projective toric variety.

At first sight, it is not clear why the (0,2)-deformed V-model should have any toric-like structure, since the E-deformations break the toric
symmetries.  On the (2,2) locus, the toric symmetries are easy to see:  the Lagrangian is invariant under
\begin{equation}
(\Phi^i,\Gamma^i) \to (e^{i\alpha_i} \Phi^i, e^{i\alpha_i}\Gamma^i).
\end{equation}
While a rank $r$ subgroup of this action is gauged, the remaining $d$ symmetries correspond to the $\GU(1)^{d}$ torus action on the toric
variety $V$.  Generic E-deformations break this symmetry completely.  Essentially, this is the statement that the bundle $\cE\to V$
is not toric, and it is not clear that any of the familiar features of toric intersection theory should apply to the sheaf cohomology
$H^k(V,\wedge^k\cE^\vee)$.

A closer look at the localization conditions of the half-twisted theory suggests a more optimistic perspective.  Examining the action of
$\cQ_T = \cQb_+$ given in section~\ref{ss:Aloc}, we find these are given by
\begin{equation}
\label{eq:02Vloc}
\p_{\zb} \sigma_a = 0,~~~{ E^{ai} \sigma_a = 0}~~(\text{no sum on}~i),~~~\nabla_{\zb} \phi^i = 0,~~~D_a + f_a = 0,
\end{equation}
Comparing these conditions to those of the topological theory at the (2,2) locus (eqn.~(\ref{eq:Vinstanton})), we see that as long as
$E^{ai}(\phi) $ has rank $r$ for all $\phi$ outside the exceptional set (this will be true for small E-deformations), the only solution to the
first two conditions is $\sigma_a=0$, and the resulting moduli space of solutions is again the collection of gauge-instanton moduli
spaces---the familiar compact toric $\cM_n$!

In the (2,2) case, once we knew how to do intersection theory on $V$ and the form of $\chi_n$, we had the tools
for determine the instanton contributions to the correlators as well.  It is reasonable to suspect that the same holds in the (0,2) theories:
once we have the tools to study the ring structure on $H^\ast(V,\cE^\vee)$, we should be able to extend the results to $\cM_n$
without too much trouble.  Let us emphasize that this extension is precisely what has been described in~\cite{Katz:2004nn}.  Our
goal here is simply to obtain their results in a more ``toric'' fashion.  Taking our inspiration from the A/2 twisted action evaluated at
instanton number zero, as well as the familiar form of the (2,2) results, we
have developed a conjectured procedure to determine the (0,2) intersection ring on $V$.  We will now describe our conjecture and the
tests it satisfies.

We begin by introducing a set of anti-commuting objects $\pi_i$, and a set of commuting objects $\etat_a$.  The former keep track
of ``bundle'' indices, while the latter should be thought of as a basis for $H^1(V,\cE^\vee)$.  Given these, we define anti-commuting
objects $\xit_i$ by
\begin{equation}
\xit_i = \pi_j \etat_a E^{aj}_{~i}.
\end{equation}
As the notation suggests, the $\xit_i$ are to play a role similar to the $\xi_i$ in the toric intersection theory.

The next step is to construct the analogue of the Stanley-Reisner relations.  There is an obvious guess:  for each
irreducible component of the exceptional set $F$, say labelled by a set $I$, we set
\begin{equation}
\label{eq:02SR}
\prod_{i\in I} \xit_i = 0.
\end{equation}
Let us see that this leads to sensible results on the (2,2) locus, where $E^{aj}_{~i} = Q^a_i \delta^j_i$.  Plugging this in,
we find
\begin{equation}
\prod_{i\in I} \xit_i =\pi_{i_1} \pi_{i_2} \cdots \pi_{i_{|I|}}~ \etat_{a_1} Q^{a_1}_{i_1}  \etat_{a_2} Q^{a_2}_{i_2} \cdots   \etat_{a_{|I|}} Q^{a_{|I|}}_{i_{|I|}} ~
=0,~\text{ ( no sum on the $i$ indices)}.
\end{equation}
Without additional assumptions on the $\pi_i$, the only way for this to hold is if
\begin{equation}
 \prod_{i\in I} \etat_{a} Q^{a}_{i}=0,
\end{equation}
which is the usual Stanley-Reisner relation, provided we identify $\eta_a = \etat_a$.

These relations take on an elegant form when we re-cast the $E^{aj}_{~i}$ in terms of the $E^a_{(\alpha)}$ defined in eqn.~(\ref{eq:Edef}).
Recall that the exceptional set $F$ is the set of $[\C^\ast]^{r}$ orbits in $\C^n$ for which the
D-terms have no solution.  This immediately implies that given two fields $\phi^{i_1},\phi^{i_2}$ with identical gauge
charges and some irreducible component of $F$ labelled by the set $I$, $i_1\in I$ if and only if $i_2 \in I$.  Thus,
we may replace $I$ with a set $A(I) =\{\alpha_1,\cdots,\alpha_k\}$. A little thought
then shows that eqn.~(\ref{eq:02SR}) may be re-written as
\begin{equation}
\prod_{\alpha\in A(I)} \det \left[\etat_a E^a_{(\alpha)}\right] = 0.
\end{equation}

To complete the story, we must find a way to normalize the top cup product of the $\etat_a$.  On the (2,2) locus this
was easy to do in terms of non-zero intersections of $d$ $T$-invariant divisors $D_{i_1},\ldots, D_{i_d}$.  These have
a non-trivial intersection if and only if the corresponding one-dimensional cones $\rho^{i_1}, \ldots, \rho^{i_d}$ belong to a full-dimensional
cone $\sigma_p\in \Sigma_V$, in which case the divisors intersect at the $T$-invariant point $p\in V$.  This is
the origin of the normalization described in eqns.~(\ref{eq:22norm},\ref{eq:detQ}).

Taking our cue from this (2,2) result, we conjecture that for every $T$-invariant point $p$, there are normalization conditions
\begin{equation}
\label{eq:inter}
\#(\xit_{i_1}\cdots \xit_{i_d} )  =  \#(\etat_{a_1}\cdots \etat_{a_d} )~ \#(\pi_{j_1}\cdots \pi_{j_d})|_{p} ~E^{a_1 j_1}_{~i_1}\cdots E^{a_d j_d}_{~i_d},
\end{equation}
where $\#(\eta_{a_1}\cdots \eta_{a_d})$ is the symmetric product to be determined, and
\begin{eqnarray}
\label{eq:inter2}
\#(\xit_{i_1}\cdots \xit_{i_d} ) & = & \smdet_p Q, \nonumber\\
 \#(\pi_{j_1}\cdots \pi_{j_d})|_p & = & \left|\smdet_p Q\right|~ \ep_{j_1\cdots j_d j_{d+1} \cdots j_n}
                                                           ~\left[ \ep_{i_1\cdots i_d i_{d+1} \cdots i_n} \right]^2~E^{1,j_{d+1}}_{~ i_{d+1}} \cdots E^{r, j_n}_{~i_n}.\nonumber\\
                                                        ~&~&~
\end{eqnarray}

Besides passing obvious checks such as anti-symmetry of $\#(\xit_{i_1}\cdots\xit_{i_d})$ and the symmetry of $\#(\etat_{a_1}\cdots\etat_{a_d})$,
eqns.~(\ref{eq:inter}, \ref{eq:inter2}) pass a number of non-trivial checks.  First, on the (2,2) locus we immediately recover the familiar normalization
conditions.  Second, the $\#(\etat_{a_1}\cdots\etat_{a_k})$ so obtained match the $q_a\to 0$ limit of correlators studied by Coulomb branch
techniques in~\cite{McOrist:2007kp}.  Given the intricate structure of those amplitudes (see eqns.~(6.10,6.20)), this amounts to an important test
of the formula.  Finally, when the conjecture is extended to higher instanton numbers, it continues to match the explicit
computations in all the cases we checked.

The details of the extension from $\#(\cdots)_V$ to $\#(\cdots)_{\cM_n}$ are easily guessed by examining the zero mode structure of the
A/2-twisted V-model. The result is that in addition to the modifications discussed in section~\ref{ss:V2Mn}, for each $\alpha$ with non-negative
degree  $d_\alpha = \sum_a n_a Q^a_{(\alpha)}$, we replace
\begin{equation}
E^a_{(\alpha)}  \to E^a_{(\alpha)}\otimes {\mathbbm{1}}_{(d_\alpha+1)^2},
\end{equation}
in the normalization formulas.  Extra fermion zero modes for each $d_\alpha <0$ lead to an extra factor of
\begin{equation}
\chi_n = \prod_{\alpha|d_\alpha <0} \det (\etat_a E^a_{(\alpha)})^{-1-d_\alpha}
\end{equation}
inserted in $\#(\cdots)_{\cM_n}$.  As in the A-twisted theory, we set $\#(\etat_{a_1}\cdots\etat_{a_k})_{\cM_n} = 0$ unless
$k = \dim\cM_n$.

Putting all of this together, we have the conjecture that the A/2-twisted V-model correlators are given by
\begin{equation}
\la \sigma_{a_1}\cdots \sigma_{a_k}\ra = \sum_{n\in \cK^\vee} \#(\etat_{a_1}\cdots\etat_{a_k} \chi_n)_{\cM_n}\prod_{a=1}^{r} q_a^{n_a}.
\end{equation}

We stress that this conjecture does not compute correlators we could not have computed before, however, if true, it has
some intrinsic mathematical interest as a simple generalization of the usual toric intersection theory, and, practically speaking,
it allows a computation of individual instanton contributions with minimal geometric input.

\subsection{V-Model Examples}
We now give a few examples of computations in the A/2 twisted V-models.

\subsubsection{$V\simeq \P^{n}$.}
 The tangent bundle of $\P^n$ is rigid, and these models have no E-deformations.  The GLSM has $n+1$ matter fields coupled to one
 gauge field with charge $1$.  The D-term constraint is  $\sum_i |\phi^i|^2 = \rho$, so that $\cK_c$ consists of the ray $\rho\ge 0$.
 To solve the theory, we will use the Coulomb branch techniques.  The effective superpotential is
\begin{equation}
\Jt = \log\left[q^{-1} \sigma^{n+1}\right],
\end{equation}
which leads to the quantum cohomology relation $\sigma^{n+1} = q$.  Using eqn.~(\ref{eq:Coulcorr}), we obtain an expression for
the non-zero correlators:
\begin{equation}
\label{eq:Pn}
\la\sigma^{n+k(n+1)}\ra = q^k.
\end{equation}

\subsubsection{Example 2: $V\simeq \P^1\times\P^1$.}

The simplest V-model with (0,2) deformations is the $\P^1\times\P^1$ example studied
in~\cite{Adams:2003zy,Katz:2004nn,Guffin:2007mp,McOrist:2007kp}.  This model has four matter fields, two K\"ahler parameters
and charges
\begin{equation}
Q = \left(\begin{array}{cccc}1 & 1 & 0 & 0 \\ 0 & 0 & 1 & 1\end{array}\right).
\end{equation}
The (0,2) deformations are labelled by six parameters $\ep_{1,2,3},\gamma_{1,2,3}$ in the $M_{(\alpha)}$.
For example, we may take
\begin{equation}
M_{(1)} = \left(\begin{array}{cc} \sigma_1 + \ep_1 \sigma_2	& ~\ep_2\sigma_2	\\
				                 \ep_3 \sigma_2				& ~\sigma_1
		      \end{array}
	       \right),~~
M_{(2)} = \left(\begin{array}{cc} \gamma_1\sigma_1 + \sigma_2	&~ \gamma_2\sigma_1	\\
				                 \gamma_3 \sigma_1			&~ \sigma_2
		      \end{array}
	       \right).
\end{equation}
Plugging these into the $\Jt_a$ yields the quantum cohomology relations
\begin{eqnarray}
\label{eq:P1P1qc}
\sigma_1^2 + \ep_{1} \sigma_1 \sigma_2 - \ep_2 \ep_3 \sigma_2^2 & = & q_1, \nonumber\\
\sigma_2^2 + \gamma_{1} \sigma_1 \sigma_2 -\gamma_2\gamma_3 \sigma_1^2 & = & q_2.
\end{eqnarray}

The computation of correlators is not much harder than in the previous example.  First, we note
that $\la\sigma_1^a\sigma_2^b \ra = 0$ unless $a+b$ is even.  This implies that the non-zero correlators
may be put in the form $\la \sigma_1^{2a} \sigma_2^{2b} (\sigma_1\sigma_2)^k\ra$.  The quantum cohomology
relations determine insertions of $\sigma_1^{2a}$ and $\sigma_2^{2b}$ in terms of $(\sigma_1\sigma_2)^k$:
\begin{equation}
\left(\begin{array}{c} \sigma_1^2 \\ \sigma_2^2 \end{array} \right) =
\frac{1}{R_3} \left( \begin{array}{c} A_1 - R_1 \sigma_1\sigma_2 \\ A_2 -R_2 \sigma_1\sigma_2 \end{array} \right),
\end{equation}
where
\begin{equation}
A_1 = q_1+\ep_2\ep_3 q_2,~~A_2 = q_2 + \gamma_2\gamma_3 q_1,
\end{equation}
and
\begin{equation}
R_1 = \ep_1 +\ep_2\ep_3\gamma_1,~~
R_2 = \gamma_1 + \ep_1 \gamma_2\gamma_3, ~~R_3 = 1 - \ep_2\ep_3\gamma_2\gamma_3.
\end{equation}
Thus, the only non-trivial correlators are $\la y^k \ra$, where $y = \sigma_1\sigma_2$.  To compute these, we use
the quantum cohomology relations to recast eqn.~(\ref{eq:Coulcorr}) in terms of $y$.  We find
\begin{equation}
\la (\sigma_1\sigma_2)^k \ra = 2 \times \sum_{y=y_\pm} {y^k} \left[\ff{2}{R_3} (B +2 Dy ) \right]^{-1},
\end{equation}
where
\begin{equation}
y_\pm = \alpha\pm\beta = -\frac{B}{2D} \pm \frac{\sqrt{B^2+4 A_1 A_2 D}}{2 D},
\end{equation}
are the solutions to
\begin{equation}
y^2 = \sigma_1^2\sigma_2^2 = R_3^{-2} (A_1 - R_1 y) (A_2 -R_2 y),
\end{equation}
with
\begin{equation}
B = A_1 R_2 + A_2 R_1,~~~ D = R_3^2 - R_1 R_2.
\end{equation}
The over-all factor of $2$ is due to the two solutions to $\Jt_a (\sigma) = 0$ for each $y$.
Simplifying this a little further, we find
\begin{equation}
\la (\sigma_1\sigma_2)^k \ra = \frac{R_3}{D}\times \frac{ (\alpha+\beta)^k - (\alpha-\beta)^k}{2 \beta}.
\end{equation}
For example, we find
\begin{equation}
\la \sigma_1^2 \ra =     -\frac{R_1}{D},~~
\la \sigma_1\sigma_2 \ra  =  \frac{R_3}{D},~~
\la \sigma_2^2 \ra  =  -\frac{R_2}{D}.
\end{equation}
%\begin{equation}
%D = (1-\ep_2\ep_3\gamma_2\gamma_3)^2- (\ep_1 + \ep_2 \ep_3 \gamma_1 ) (\gamma_1 +\ep_1 \gamma_2\gamma_3).
%\end{equation}
%and as easily we obtain
%\begin{equation}
%\la (\sigma_1\sigma_2)^{10} \ra= -\frac{R_3 B (5 A_1^2 A_2^2 D^2 + B^4 +5 B^2 A_1 A_2 D)
% 								  (B^4+3B^2 A_1 A_2 D + A_1^2 A_2^2 D^2)}{D^{10}}.
%\end{equation}

\subsubsection{Resolved $\P^4_{1,1,2,2,2}.$}\label{sss:P411222}

This model will be relevant for the (0,2) quantum restriction formulas discussed below.  The (2,2) GLSM was studied in~\cite{Morrison:1994fr}, and
its (0,2) deformations were considered in~\cite{McOrist:2007kp}.  In this case the V-model has six matter fields and two
gauge fields, with charges
\begin{equation}
Q = \left(\begin{array}{cccccc} 0 & 0 & 1 & 1 &1 & 1 \\ 1 & 1 & 0 & 0 & 0& -2\end{array}\right).
\end{equation}
We consider a (0,2) deformation with
\begin{eqnarray}
\label{eq:Edefex2}
&&M_{(1)} = \left(\begin{array}{cc} \sigma_2 + \ep_1 \sigma_1	& ~\ep_2\sigma_1	\\
				                 \ep_3 \sigma_1				& ~\sigma_2
		      \end{array}
	       \right),~~
M_{(2)} = \diag(\sigma_1,\sigma_1,\sigma_1),\cr
&&M_{(3)} = \sigma_1 - 2\sigma_2.
\end{eqnarray}
Eqns.~(\ref{eq:02QC}) yield the quantum cohomology relations
\begin{eqnarray}
\sigma_1^3(\sigma_1 - 2\sigma_2) &=& q_1,\nonumber\\
{\sigma_2^2 + \ep_1\sigma_1\sigma_2 -\ep_2\ep_3\sigma_1^2} &=& q_2{(\sigma_1-2\sigma_2)^2},
\end{eqnarray}
and the non-zero correlators:
\begin{eqnarray}
\label{eq:VP411222}
\langle \sigma_1^a \sigma^{4m-a}_2\rangle &=& 4q_1^{m-1}\sum_{z|P(z)=0}\frac{z^{4m-a}}{(1-2z)^{m-1} H(z)},
\end{eqnarray}
where
\begin{eqnarray}
P(z) &=& z^2 + \ep_1 z - \ep_2\ep_3 - q_2 (1-2z)^2, \nonumber\\
H(z) & = & 4 (\ep_1 -4\ep_2\ep_3 + 2(1+\ep_1) z ).
\end{eqnarray}

This example is a simple setting where non-linear E-parameters are allowed.  These yield eighteen additional parameters
and take the form
\begin{equation}
\begin{pmatrix}  \Delta E^3 \\ \Delta E^4 \\ \Delta E^5 \end{pmatrix} = \Sigma_a \left[ K^a \right]
\begin{pmatrix} (\Phi^1)^2 \\ \Phi^1 \Phi^2 \\ (\Phi^2)^2 \end{pmatrix}\Phi^6,
\end{equation}
where the $K^1$ and $K^2$ are $3\times 3$ matrices of parameters.

\section{A/2 Twist and Hypersurfaces: Quantum Restriction} \label{s:restrictA}
Having discussed the A/2 twist of the V-model, we now come to our real interest:  the (0,2) deformations of the
A/2 twist of the M-model.  As in our V-model discussion, we begin with the (0,2) supersymmetric action.
\subsection{Parameters in the M-model}
\label{s:mdefs}
Recall that the field content of the M-model is that of the V-model, plus an additional (2,2) multiplet $\Phi^0_{(2,2)}$ with gauge charges
$Q_0^a = -\sum_{i=1}^n Q_i^a$.  Like the other matter fields, $\Phi^0_{(2,2)}$ decomposes into a (0,2) chiral multiplet $\Phi^0$, and
a Fermi multiplet $\Gamma^0$ with chiral constraint $\cDb_+ \Gamma^0 = 2i Q_0^a\Sigma_a \Phi^0.$  The chiral superpotential
couplings are written in terms of a (0,2) superpotential:
\begin{equation}
\label{eq:02Pot}
\cL_{\cJ} = \int d\theta^+ \left[ \Gamma^0 P(\Phi^1,\cdots,\Phi^n) + \sum_{i=1}^n \Gamma^i \Phi^0 P_{,i}\right] + \text{h.c.},
\end{equation}
where $P_{,i} = \p P/\p\Phi^i.$
$\cL_{\cJ}$ will be gauge-invariant if the polynomial $P$ has charges $-Q_0^a$.  Since the $\Gamma^i$ are
not chiral, it is not obvious that $\cL_{\cJ}$ preserves  (0,2) supersymmetry.   An explicit computation shows that the general superpotential
$\cL_{\cJ} = \int d \theta^+ \Gamma^I \cJ_I$ will be (0,2) supersymmetric provided that the $\cJ_I$ and the chiral constraints $E^I$ in
$\cDb_+\Gamma^I = \sqrt{2} E^I$ are chosen to satisfy
\begin{equation}
\label{eq:02constraint}
\sum_I E^I \cJ_I = 0.
\end{equation}
On the (2,2) locus, the constraint reduces to
\begin{equation}
\Phi^0 \Sigma_a \left[Q_0^a P + \sum_i Q_i^a P_{,i} \right] = 0,
\end{equation}
where the equality follows from the quasi-homogeneity properties of $P$ implied by gauge invariance.
Clearly, this is not the only way to satisfy the constraint.  Replacing the $P_{,i}$ with polynomials $J_i$ of
same charge, and choosing more general $E^i$ as we did in the V-model, we will find a theory with
(0,2) supersymmetry if
\begin{equation}
E^0 P + \Phi^0 E^i J_i = 0.
\end{equation}

We see that the M-model has two types of (0,2) parameters:  the E-parameters familiar from the V-model, and the J-parameters.
The two sets are not independent but must satisfy the (0,2) SUSY constraint.  We find it convenient to label the linear E-parameters
of the M-model in terms of the $M_{(\alpha)}$ of the V-model given in eqn.~(\ref{eq:Edef}), as well as $M_{(0)}$ in
\begin{equation}
\cDb_+ \Gamma^0 = 2i M_{(0)} \Phi^0.
\end{equation}
On the (2,2) locus $M_{(0)} = -\sum_i Q_i^a \sigma_a$.

The geometric structure encoded by the $E$ and $J$ is a choice of bundle $\cF$ on the Calabi-Yau hypersurface $M \subset V$.
$\cF$ is a deformation of $T_M$, whose sections are described as the cohomology of the sequence
\begin{equation}
\xymatrix{ 0\ar[r] &\cO^{r}|_M \ar[r]^-{E} &\oplus_i\cO(D_i)|_{M}  \ar[r]^-{J} & \cO(\sum_i D_i)|_{M}\ar[r] & 0 },
\end{equation}
$\cF = \ker J / \im E$.
Physically, this sequence arises in the geometric phase of the GLSM as a description of the fermions in the
low energy NLSM~\cite{Witten:1993yc,Distler:1995bc}.

\subsection{Counting (0,2) Deformations of the M-model}
\label{ss:mdefs}

A naive count of the (0,2) M-model parameters is given by summing the parameters of the theory modulo the (0,2) SUSY constraint: there are $r$ K\"ahler parameters; $r  (1+D) $  E-parameters\footnote{The extra $1$ comes from $E^0$.}, where $D = \dim\Autt(V)$; there are $\#(J)$ monomials in
the $J_i$, and $\#(P)$ monomials in $P$.  The $E\cdot J = 0$ constraint imposes $r \#(P)$ conditions.
Thus,
\begin{equation}
N_{\text{naive}}(M) = r (2 + D ) + \#(J) -(r-1)\#(P).
\end{equation}

Clearly, this is a vast over-parametrization, and as on the (2,2) locus, we expect that field redefinitions will help to cut down on the number
of parameters of the low energy theory.   Let us consider the field redefinitions
\begin{eqnarray}
\Phi^0 &\to& u \Phi^0, \nonumber\\
\Gamma^0 &\to& v \Gamma^0, \nonumber\\
\Phi^i  &\to& \sum_{\mu \in S_i} U^i_{\mu} \mu, \nonumber\\
\Gamma^i & \to & \sum_{\mu \in S_i} V^i_{\mu} \frac{\p \mu}{\p \Phi^k} \Gamma^k,
\end{eqnarray}
depending on (2D+2) parameters $u,v,U^i_\mu,V^i_\mu$.  We recall that $S_i$ is the set of monomials $\mu$ of charge $Q_i^a$.
Setting $U=V$ and $u=v$, we find the familiar (2,2) redefinitions from eqn.~(\ref{eq:22redefs}).
Since we only demand (0,2) SUSY, there is no longer any reason to have the same set of $U^i_\mu$ for the $\Phi^i$ and the $\Gamma^i$.\footnote{
The reader should note that the same argument would naively apply to the V-model as well, but that would not match the answer expected on
geometric grounds.  A clue to the difference in the sets of redefinitions is provided by the anomalous $\GU(1)_R$ symmetry of the V-model,
but we do not have a fully satisfactory argument for the difference.}

In addition, since the $\Sigma_a$ are no longer tied by supersymmetry to the $\Upsilon_a$ (the normalization of these is fixed by
the periodicity of the $\theta$-angles), we may also perform a $\GL(r,\C)$ rotation on the $\Sigma_a$. Thus, there are
$2(1+D)+r^2$ field redefinitions that may be used to absorb parameters.

As before, not all of these terms modify the holomorphic couplings and the chirality constraint, since the transformations include
global gauge symmetries, as well as $\GU(1)_L$ rotations.  Thus, $r+1$ transformations leave the holomorphic couplings and chiral
constraints invariant. Taking this into account,  we obtain a count of parameters in the (0,2) M-model:
\begin{equation}
N(M) =  r + (r-2) D + \#(J) - (r-1)\#(P) -(r-1)^2.\label{eq:nm_1}
\end{equation}
Of these $N^{(2,2)}(M)$ parametrize motions along the (2,2) locus, while
\begin{equation}
N^{(0,2)} = N(M) - N^{(2,2)}(M)
\end{equation}
counts the E- and J-deformations.

Let us check the validity of the expressions in some simple examples.  First, we consider the quintic, where $r=1$, $D=25$, and $\#(J) = 350$,
which yields
\begin{equation}
N(\text{quintic}) = 326 = 1 + 101 + 224.
\end{equation}
Another simple example is the bi-cubic in $\P^2\times\P^2$ already investigated in section~\ref{sss:P2P2example}.  Here we have
$r=2$, $D=18$, $\#(J) = 360$, and $\#(P) = 100$.  Putting all this together, we find $N(\text{bi-cubic}) = 261$.
A geometric analysis counting the parameters in this theory was performed in~\cite{Distler:1988jj,Eastwood:1989sd} (see~\cite{Hubsch:1992nu} for  a pedagogical discussion), where it was found that there are: $h^{1,1}(M) = 2$ K\"ahler parameters; $h^{1,2}(M) = 83$ complex structure
deformations, all of which are known to be polynomial; and finally, $\dim H^1(M,\End T_M) = 176$ bundle deformations.  These indeed add up to $261$.
We list in table \ref{table:parameter} the counts for a few other models.

The formulas that count the deformations become less cluttered if we use $\dim\Autt(V) = \dim\Aut(V) +r$.\footnote{We thank B.~Nill for
suggesting this.}  Making the substitution, we find that eqns. (\ref{eq:ne_1}, \ref{eq:nm_22},\ref{eq:nm_1}) become
\bea
\label{eq:countphys}
N(V) &=& r+(r-1) \dim\Aut(V),\cr
N^{(2,2)} (M) &=& r+ \#(P) - \dim \Aut(V) -1,\cr
N(M) &=&  r + (r-2) \dim\Aut(V) + \#(J) - (r-1)\#(P) -1.
\eea

\TABLE[t]{
%\begin{center}
\begin{tabular}{|c|c|c|c|c|c|}
\hline
 $V$ 			     	& $N(V)$	& $h^{1,1}_{\text{toric}}(M)$ & $h^{2,1}_{\text{poly}}(M)$  & $N(M)$	 & $N^{(0,2)}(M)$ \\ \hline
$\P^4$ 		     	&1		& 1 					   & 101 				        & 326 	& 224			   \\ \hline
$\P^2\times \P^2$ 	& 18		& 2 					   & 83 				        & 261	& 176			   \\ \hline
$\P^4_{1,1,2,2,2}$ 	& 23		& 2 					   & 83 				        & 276	& 191			   \\ \hline
$\P^4_{1,1,2,2,6}$ 	& 46		& 2 					   &126 				        & 494	& 366			   \\ \hline
$\P^4_{1,2,2,3,4}$ 	& 20		& 2 					   & 70 				        & 232	& 160			   \\ \hline
$\P^4_{1,2,2,2,7}$ 	& 35		& 2 					   & 107 				        & 372	& 263			   \\ \hline
$\P^4_{1,1,1,6,9}$ 	& 105	& 2 					   & 272 				        & 1158	& 884			   \\ \hline
\end{tabular}
%\end{center}
\caption{Linear deformations of the M-model, listed by the associated V-model.  The last column is a counting of
``polynomial'' bundle deformations.  In favorable circumstances this should be $\dim H^1(M,\End T_M)$. More generally,
 it yields a subset of the unobstructed deformations of the tangent bundle.}
\label{table:parameter}
%\end{table}
}

\subsection{The A/2 Twist}
By construction, the E- and J-deformations preserve not only the (0,2) supersymmetry, but also the $\GU(1)_R$ R-symmetry and a global
symmetry $\GU(1)_L$.  The latter becomes the left-moving R-symmetry on the (2,2) locus.  On the (2,2) locus these symmetries
are believed to be the R-symmetries of the IR fixed point~\cite{Witten:1993yc,Silverstein:1994ih,Distler:1995bc}.  Even off the
(2,2) locus the $\GU(1)_L$ plays a distinguished role in the heterotic compactification:  it provides a non-linearly realized component
of the space-time gauge symmetry, and it includes a $\Z_2$ symmetry that may be used to construct a chiral GSO
projection~\cite{Distler:1995bc}.  In our conventions these symmetries act on the field-content with charges given in table~\ref{table:SimpleR_2}.
%\TABLE[t]{
\begin{table}[t]
\begin{center}
\begin{tabular}{|c|c|c|c|c|c|c|c|}
\hline
$~			$&$\theta^+ 	$&$\Phi^i		$&$\Gamma^i	$&$\Phi^0		$&$\Gamma^0	$&$\Sigma_a	 $&$\Upsilon_a 	 $\\ \hline
$\GU(1)_R	$&$1			$&$0			$&$0			$&$1			$&$1			$&$1			 $&$1			 $\\ \hline
$\GU(1)_L		$&$0			$&$0			$&$-1		$&$1			$&$0			$&$-1		 $&$0			 $\\ \hline
\end{tabular}
\end{center}
\caption{The $\GU(1)_R$ and $\GU(1)_L$ symmetry charges for the M-model in the geometric phase.}
\label{table:SimpleR_2}
\end{table}
%\label{table:SimpleR_2}
%\caption{The $\GU(1)_R$ and $\GU(1)_L$ symmetry charges.}
%}
These $\GU(1)$ symmetries lead to the existence of two distinct
half-twists of the theory.  To explain this structure, we must perform an analytic continuation to Euclidean space, and carefully
consider the charges of the fields under the Lorentz group and under the $\GU(1)_R\times\GU(1)_L$ symmetry.  We label
the generator of the former as $J_T$, and we consider the linear combinations
\begin{equation}
J_A  = \ff{1}{2} ( J_R + J_L), ~~~ J_B = \ff{1}{2} (J_R - J_L)
\end{equation}
of the $\GU(1)_R\times\GU(1)_L$ generators. To twist, we redefine the Lorentz charge by
\begin{equation}
J_{T'} = J_{T} - J_A~ (\text{A/2-twist}),~~~ J_{T''} = J_T - J_B~ (\text{B/2-twist}).\label{eq:twist}
\end{equation}
The details of the half-twists are given in appendix~\ref{app:conv}.  We find that under the A/2-twist the
spins of the fields are shifted as follows:  the spins of the fields in the $\Sigma_a,\Upsilon_a$ multiplets are now taken to be
\begin{equation}
\begin{array}{cc}
 \begin{array}{llll}
   \sigma_a 		&\to& \sigma_a 			& \left[\in \Gamma(\cO)\right], \\
   \lambda_{a,+} 	&\to& \lambda_{a,\zb} 		& \left[\in \Gamma (\Kb)\right],\\
   \lambdab_{a,+} 	&\to& \lambdab_a			& \left[ \in \Gamma (\overline{\cO})\right],
 \end{array}
&
 \begin{array}{llll}
  \sigmab_a		&\to& \sigmab_a 			& \left[\in \Gamma( \overline{\cO} )\right], \\
   \lambda_{a,-} 	&\to& \lambda_{a,z}	 		& \left[\in \Gamma(K)\right],\\
   \lambdab_{a,-} 	&\to& \chib_{a}				& \left[ \in \Gamma(\cO)\right],
 \end{array}
\end{array}
\end{equation}
where $K$ is the canonical bundle of the world-sheet (we always work on $\P^1)$.  The zero modes of the kinetic
operator for each of the fields are in one to one correspondence with (anti)holomorphic sections of the corresponding bundles.

In the background of a gauge field with instanton number $n_a$, the twisted matter fields and their zero modes are as
follows:
\begin{equation}
\begin{array}{cc}
 \begin{array}{llll}
   \phi^i	 		&\to& \phi^i	 			& \left[\in \Gamma(\cO(d_i))\right], \\
   \psi_+^i 			&\to& \psi^i_{\zb}	 		& \left[\in \Gamma (\Kb \otimes \cOb({-d_i}))\right],\\
  \psib_+^i 		&\to&  \psib^i				& \left[ \in \Gamma (\cOb({d_i}))\right],\\
   \phi^0	 		&\to& \phi^0_{z}	 			& \left[\in \Gamma(K\otimes\cO(d_0))\right], \\
   \psi_+^0			&\to& \psi^0	 				& \left[\in \Gamma (\cOb({-d_0})\right],\\
  \psib_+^0 		&\to& \psib^0_{\zb}				& \left[ \in \Gamma (\Kb\otimes\cOb({d_0}))\right],
 \end{array}
&
 \begin{array}{llll}
   \phib^i			&\to& \phib^i	 			& \left[\in \Gamma( \cOb({d_i}) )\right], \\
   \gamma_-^i	 	&\to& \gamma^i	 		& \left[\in\Gamma(\cO(d_i))\right],\\
  \gammab_-^i	 	&\to& \gammab^i_{z}			& \left[ \in \Gamma(K\otimes \cO(-d_i))\right],\\
    \phib^0			&\to& \phib^0_{\zb}	 			& \left[\in \Gamma(\Kb\otimes \cOb({d_0}) )\right], \\
   \gamma_-^0	 	&\to& \gamma^0_z	 			& \left[\in\Gamma(K\otimes\cO(d_0))\right],\\
  \gammab_-^0	 	&\to& \gammab^0				& \left[ \in \Gamma(\cO(-d_0))\right].
 \end{array}
\end{array}
\end{equation}
Recall that the degrees are given by
\begin{equation}
d_i = Q_i^a n_a,~~~i=0,\ldots,n.
\end{equation}
As in~\cite{Katz:2004nn,Guffin:2008kt} a Hermitian metric on $\cO(d_i) \to \P^1$ has been used in some of the definitions, so that
the sections of the bundles in braces count zero modes of the kinetic operator in a fixed instanton background.

Under this twist the supercharge $\cQb_+$ becomes the nilpotent world-sheet scalar operator $Q_T$.  The action of $Q_T$, as well
as the decomposition of the action for the theory into $Q_T$-closed and $Q_T$-exact components are described in appendix~\ref{app:conv}.
The details most important to our analysis are that:
\begin{enumerate}
\item $\sigma_a$ represent non-trivial elements in the $Q_T$ cohomology;
\item the anti-holomorphic couplings $\taub_a$ and the couplings in $\Pb, \Eb^i$ and $\Jb_i$ only appear in $Q_T$-exact terms.
\end{enumerate}
By arguments familiar from BRST gauge-fixing or cohomological topological field theories, it follows that correlators of the $\sigma_a$
depend holomorphically on the parameters of the theory. To determine this dependence, we now turn to localization.

\subsection{Localization and Quantum Restriction} \label{ss:a2qrestrict}
The A/2-twisted Lagrangian and action of $Q_T$ are described in appendix~\ref{ss:Aloc}. An examination of the action of $Q_T$ in a geometric
phase of the $M$-model reveals that the A/2-twisted path-integral localizes onto the same configurations as the A-twisted
M-model on the (2,2) locus. That is, the fixed-point set is described by $\cM_{n;P} \subset \cM_n$, where $\cM_n$ is the familiar compact and
toric moduli space of instantons in the V-model.

Thus, to compute $\lad \sigma_{a_1}\cdots \sigma_{a_{d-1}} \rad$, we need to expand the
action around the configurations in $\cM_{n;P}$ and use this as a measure on $\cM_{n;P}$.  This is not simple, and at first one might think
it will be much harder than in the A-twisted theory.  There, the couplings split up into chiral and twisted chiral superpotentials, and the former
couplings were easily shown to be $Q_T$-exact.  In the A/2-twisted M-model there is only one sort of superpotential, and it is not at all
obvious why the correlators of the $\sigma_a$ should be independent of the parameters in $P(\phi)$ or $J(\phi)$.

Nevertheless, we will now show that a few conservative assumptions about the half-twisted theory lead to a quantum restriction formula
off the (2,2) locus.  We assume that
\begin{enumerate}
\item The path-integral reduces to a finite integral over $\cM_{n}$.
\item The semi-classical expansion about points in $\cM_{n}$ is exact.
\item The non-zero modes of the kinetic operators cancel in the one-loop determinants, so we can
         restrict attention to the zero modes.
\item The correlators do not depend on anti-holomorphic couplings.
\end{enumerate}

With those assumptions in mind, we work in a geometric phase of the M-model with a
K\"ahler cone $\cK$.  Recall that throughout this work we assume that V is Fano, which implies
that $\sum_i Q_i^a \in \cK$. Fixing an instanton number $n_a \in \cK^\vee$, and corresponding
degrees $d_0 = -\sum_{i,a} Q_i^a n_a,$ $d_i = \sum_a Q_i^a n_a$, it is evident that in the geometric
phase we have $d_0 \le 0$.

Let us write the action for the M-model as $S_M = S_V + S_0$, where $S_0$ contains all the terms involving fields from the
$\Phi^0,\Gamma^0$ multiplets, while $S_V$ contains the rest.  A quick look at the action reveals that
$S_V$ is just the action for the V-model.  The next step is to localize onto the $Q_T$-invariant configurations.
Instead of localizing directly to $\cM_{n;P}$, we do a partial localization to the larger moduli space $\cM_{n}$,
leaving $P$ and $\Pb$ in the action.

The zero modes of the fields in the $\Phi^i,\Gamma^i$ multiplets are the same as they are in the V-model, while among the fields in
the $\Phi^0,\Gamma^0$ multiplets only the fermions $\psi^0$ and $\gammab^0$ have zero modes when $d_0 \le 0$.  In fact, each of them has
$1-d_0$ zero modes, so that the zero mode integral in the $n_a$-instanton sector may be written as
\begin{equation}
\lad \sigma_{a_1} \cdots \sigma_{a_{d-1}} \rad_n = \int D[\text{fields}]_{V;\cM_n} e^{-S_V} \sigma_{a_1} \cdots \sigma_{a_d}~\int D[\psi^0\gammab^0] e^{-S_0'},
\end{equation}
where
\begin{equation}
S_0' = [ P \Pb]_0 - \sum_{\alpha=1}^{1-d_0} \left( i\sqrt{2} \gammab^0_\alpha (-M_{(0)}) \psi^0_\alpha - \gammab^0_\alpha [\psib^i \Pb_{,i}]_\alpha
+ [\gamma^i J_i]_\alpha \psi^0_\alpha \right),
\end{equation}
and $[\cdots]_\alpha$ denotes projection onto the $\alpha$-th zero mode.  Let us suppose the measure may be chosen so that
\begin{equation}
D[\psi^0 \gammab^0] = - \prod_{\alpha=1}^{1-d_0} \frac{d\gammab^0_\alpha d\psi^0_\alpha}{i\sqrt{2}}.
\end{equation}
In that case, we have essentially reduced the correlator in the M-model to a computation in the V-model, since we have
\begin{equation}
\label{eq:pre_restrict}
\lad\sigma_{a_1}\cdots \sigma_{a_{d-1}} \rad_n = - \int D[\text{fields}]_{V;\cM_n} e^{-S_V}e^{-[P\Pb]_0}(  M_{(0)}^{1-d_0}+ g(M_{(0)},J,\Pb) )
\sigma_{a_1}\cdots\sigma_{a_{d-1}},
\end{equation}
where $g(M_{(0)},J,\Pb)$ is a polynomial where each monomial contains at least one power of $\Pb$.

We now come to our last assumption, namely that the correlators of the A/2-twisted M-model do not depend on anti-holomorphic parameters.
Thus, formally, the limit $\Pb \to 0$ should not change the correlators. Such limits of $Q_T$-exact parameters should generally be considered
with caution.  If the limit changes the large field asymptotics and, for example, leads to non-compact directions in the moduli space, then the
correlators may well jump in the limit.  The crucial point is that in the situation at hand there are no signs of such difficulties.  The integral
in eqn.~(\ref{eq:pre_restrict}) remains perfectly well-behaved, and we expect the correlators to be invariant as $\Pb$ is sent to zero.
Thus, we obtain
\begin{equation}
\lad\sigma_{a_1}\cdots \sigma_{a_{d-1}} \rad_n = - \la \sigma_{a_1} \cdots \sigma_{a_{d-1}} M_{(0)}^{1-d_0}\ra_n.
\end{equation}

In fact, we can do a little bit better by recalling the selection rule in the A-twisted V-model:
\begin{equation}
\la \sigma_{a_1} \cdots \sigma_{a_k} \ra_n = 0~~~\text{unless}~~~ k = d + \sum_i Q_i^a n_a.
\end{equation}
This rule remains unmodified by any (0,2) deformations, since these do not break the classical ghost number symmetry.  Thus,
\begin{equation}
\lad\sigma_{a_1}\cdots \sigma_{a_{d-1}} \rad_n = \la\sigma_{a_1}\cdots \sigma_{a_{d-1}} (-M_{(0)}) \sum_{m=1}^\infty M_{(0)}^m \ra _n,
\end{equation}
since only one term in the sum (namely, $m= -d_0$) contributes.  Finally, exchanging the sums on $n$ and $m$, we arrive at our (0,2)
quantum restriction formula:
\begin{equation}
\label{eq:Arestrict}
\lad\sigma_{a_1}\cdots \sigma_{a_{d-1}} \rad = \la \sigma_{a_1}\cdots \sigma_{a_{d-1}} \frac{ - M_{(0)}}{1 - M_{(0)}} \ra.
\end{equation}

On the (2,2) locus $M_{(0)} = K = -\sum_{i,a} Q_i^a \sigma_a$, and our result reduces to that obtained in~\cite{Morrison:1994fr} and
given in eqn.~(\ref{eq:22restrict}).  This agreement provides a basic justification of our assumption on the choice of measure.
Off the (2,2) locus, we find that the A/2-twisted correlators are independent of $P$ and $J_i$ and depend holomorphically on the
K\"ahler parameters $q_a$ and the E-parameters contained in $M_{(0)}$ and $M_{(\alpha)}$.  In fact, by using some of the field
redefinitions, we may always set $M_{(0)} = K$.  We always make this choice in what follows.

We emphasize that, just as on the (2,2) locus, we expect a non-trivial map between the (0,2) parameters of the NLSM
and these linear model coordinates.  It may well be that this map actually depends on $q$,E and also J-parameters.  Even so,
it is likely that the linear coordinates will still be useful in unravelling the structure of (0,2) theories.

\subsection{The Singular Locus of the A/2 twisted M-model} \label{ss:a2discrim}
The quantum restriction formula gives a simple way to compute the A/2-twisted M-model correlators.  From these
we may extract the quantum cohomology relations and determine the locus in the $q_a, M_{(\alpha)}$ parameter space
where the correlators have poles.  As on the (2,2) locus, these singularities should signal a singularity in the (0,2)
SCFT.  As in type II theories, we expect that here world-sheet perturbation theory breaks down and non-perturbative effects
are necessary to resolve the SCFT singularity.  These effects are not well understood in the heterotic string, and a
parametrization of the singular locus in parameter space is an important step in studying this phenomenon.

In (2,2) theories it is well-known that the singular locus of the GLSM may be determined without computing a single correlator.
The basic tool used is the effective potential governing the $\Sigma_a$ multiplets at large $\sigma_a$ VeVs.  This potential
is easily obtained by integrating out the $\Phi^i$ multiplets at one loop.  We have already discussed how a similar potential
may be computed off the (2,2) locus, and we will now use it to study the singular locus of the theory.  As a by-product, we will
obtain another check on the quantum restriction formula.

We begin by discussing the effective potential in the V-model.  It is convenient to work in a special basis for the gauge
charges, where $\sum_i Q_i^a = 0$ for $a >1 $, and $\sum_i Q_i^1 = \Delta >0$.   A moment's thought will convince the reader
that any $Q_i^a$ may be brought to this form by an ${\SL}(r,\Z)$ transformation.  So, let us return to the potential in eqn.~(\ref{eq:Coulomb}),
and study the solutions to $\Jt_a(\sigma) = 0$.  Working in the special basis, it is useful to define the ratios $z_a = \sigma_a/\sigma_1$, and
write the $\Jt_a = 0$ equations in terms of $k_\alpha\times k_\alpha$ matrices $M_{(\alpha)}(z)$ defined via
\begin{equation}
M_{(\alpha)} (\sigma_1,\ldots,\sigma_{r}) = \sigma_1^{k_\alpha} M_{(\alpha)} (1,z_2,\ldots,z_{r}).
\end{equation}
The result is
\begin{eqnarray}
\label{eq:VJ}
\sigma_1^\Delta \prod_\alpha \det M_{(\alpha)}(z)^{Q^1_{\alpha} }& = & q_1, \nonumber\\
\prod_\alpha \det M_{(\alpha)}(z)^{Q^a_{\alpha}} & = & q_a~~~\text{for}~ a>1.
\end{eqnarray}
These are $r$ equations for $r$ variables, so that for generic $q_a$ there is a zero-dimensional solution set.  While these isolated
$\sigma$ vacua are important in computations of correlators in non-geometric phases, they do not give rise to non-compact directions
in field space, and therefore do not lead to a singularity in the theory.

For certain special values of the $q_a$, $a >1$ and the E-parameters singularities may arise when some of the vacua run off to
infinity, or when the mass matrix for the matter fields becomes degenerate.  These components of the singular locus are
easy to identify in particular models~\cite{McOrist:2007kp}, and on the (2,2) locus there is an algorithmic procedure for finding
them~\cite{Morrison:1994fr}.  These singularities are independent of $q_1$, and, aside from the trivial singularity at
$q_1 = \infty$, the singular locus of the V-model is $q_1$-independent.

Now let us consider the singular locus of the corresponding M-model.  Adding in the $\Phi^0,\Gamma^0$ multiplets with charges
$Q_0^1 = \Delta$, $Q_0^a = 0$ for $a>1$, and integrating out the matter fields, we find that the $\Jt_a$, $a>1$ are identical to
those of the V-model, while $\Jt_1$ is modified.  Thus, eqn.~(\ref{eq:VJ}) is modified to
\begin{eqnarray}
\label{eq:MJ}
\prod_\alpha \det M_{(\alpha)}(z)^{Q^1_{\alpha} }& = & (\Delta)^{\Delta} q_1, \nonumber\\
\prod_\alpha \det M_{(\alpha)}(z)^{Q^a_{\alpha}} & = & q_a~~~\text{for}~ a>1.
\end{eqnarray}
These are now $r$ equations for $r-1$ variables $z_a$, $a>1$.  On the one hand the equations do not have solutions for generic parameters,
but on the other hand, a common solution to these algebraic equations signals a singularity in the theory, since the $\sigma_a$ are only
fixed up to an over-all scale.  The corresponding non-compact direction in field space will lead to a divergence in the $\sigma$ correlators.

On the (2,2) locus the (complex) co-dimension one subvariety in the $q_a$ parameter space where eqns.~(\ref{eq:MJ}) have a solution is
known as the {\em principal component} of the singular locus.   We will use this same terminology off the (2,2) locus as well.  The principal
component of the singular locus is then given by a multi-variate resultant of a polynomial system in $r-1$ variables.  Combining this component
with the singularities from the V-model, we obtain the complete singular locus of the A/2-twisted M-model.

\subsection{The Singular Locus and Quantum Restriction}
Since the $\Jt_a$ for $a>1$ are identical in the M- and V-models, there is a simple relation between the discrete $\sigma$ vacua of the
V-model and the principal component of the singular locus of the M-model.  Let us fix to some generic values of the $q_a$, $a>1$ and the
$M_{(\alpha)}$, $\alpha > 0$.  In this case the V-model is non-singular, while the M-model is on the principal component
of the singular locus if and only if the discrete $\sigma$ vacua of the V-model satisfy $ K^\Delta = 1$.

This should be compared with the quantum restriction formula of eqn.~(\ref{eq:Arestrict}).  Using our special basis of gauge charges,
the restriction formula is equivalently written as
\begin{equation}
\lad\sigma_{a_1}\cdots \sigma_{a_{d-1}} \rad = \la \sigma_{a_1}\cdots \sigma_{a_{d-1}} \frac{ - K}{1 - K^\Delta} \ra.
\end{equation}
For generic values of $q_2,\ldots,q_r$ and the E-parameters, the V-model correlators on the right-hand side are non-singular, and the only
way for the left-hand side to develop a singularity is if the sum over the V-model correlators diverges.  This only takes place
when $K^\Delta = 1$, which implies that the M-model is on the principal component of the discriminant locus.  Thus, the restriction
formula correctly predicts the $q_1$-dependent singularities of the M-model.  The observation that this holds in the example of a
hypersurface in $\P^4_{1,1,2,2,2}$ was an important motivation in our search for the (0,2) quantum restriction formula.

\subsection{Quantum Restriction for Complete Intersections}
The (0,2) quantum restriction formula may be easily generalized to linear sigma models for Calabi-Yau complete intersections in
toric varieties.  On the (2,2) locus quantum restriction for complete intersections was considered in~\cite{Noyvert:1997}.  We will
give a compact expression valid off the (2,2) locus.

The models we now study are a simple generalization of the M-model.  We again begin with a V-model for some Fano toric variety,
but instead of adding a single $\Phi^0,\Gamma^0$ multiplet, we add multiplets $\Phi^I,\Gamma^I$, $I=0,\ldots,N$, with charges
obeying $\sum_I Q_I^a = -\sum_i Q_i^a$, and superpotential couplings
\begin{equation}
\cL_{\cJ} =  \sum_I  \int d\theta\left\{ \Gamma^I P_I + \sum_j \Gamma^j \Phi^I J_{Ij} \right\} +\text{h.c.},
\end{equation}
where the $P_I$ are polynomials of multi-degree $-Q_I^a$, and $J_{Ij}$ are polynomials of multi-degree $-(Q_I^a+Q_i^a)$.

To completely specify the theory, we also need to describe the $E^I$ in $\cDb_+\Gamma^I = \sqrt{2} E^I$.  For simplicity, we
assume that we can se the $E^I$ to their (2,2) values by field redefinitions.  The generalization is obvious, but requires a more
cumbersome notation.  The theory will be (0,2) supersymmetric provided that
\begin{equation}
P_I Q^a_I \Sigma_a+ J_{Ii} E^i= 0~~\text{for all}~~I.
\end{equation}

Under appropriate combinatorial conditions~\cite{MR1463173,Batyrev:2007cq} and for generic coefficients in the polynomials
$P_I$, the gauge theory in a geometric phase flows to an NLSM with target space a dimension $k=d-1-N$ (quasi-smooth)
Calabi-Yau complete intersection  $\cap_I\{P_I = 0\}$ in $V$.  We assume these conditions are satisfied, and study the linear
model in the geometric phase.

The twisting and counting of zero modes proceeds exactly as in the hypersurface case.  As might be expected, the selection rule
for the $\sigma$ correlators implies that the non-vanishing correlators have precisely $k$ insertions. Furthermore, an
analysis of the combinatorial conditions shows that $d_I \le 0$.  Once we replace $\gamma^0,\psib^0 \to \gamma^I,\psib^I$, the
localization argument goes through verbatim.  Making a similar assumption for the normalization of the zero mode measure, we arrive
at the following result for the $n$-th instanton sector:
\begin{equation}
\label{eq:comp1}
\lad \sigma_{a_1} \cdots \sigma_{a_{k}} \rad_n =  \la \sigma_{a_1} \cdots \sigma_{a_{d-1-N}} \prod_{J=0}^N (-K_J)
\left[\prod_{I=0}^N K_I^{-Q_I^b n_b} \right] \ra_n,
\end{equation}
where we have defined $K_J = \sum_a Q_J^a \sigma_a$.

To write a compact expression for the summed correlators, it is helpful to define
$T =  \prod_J (-K_J)$ and $\delta_b = \prod_I K_I^{-Q^b_I}$.  In terms of these, we may write
\begin{eqnarray}
\lad \sigma_{a_1} \cdots \sigma_{a_{k}} \rad & = & \sum_{n \in \cK^\vee} \la \sigma_{a_1} \cdots \sigma_{a_{d-1-N}} T \prod_b \delta_b^{n_b}  \ra_n\prod_a q_a^{n_a} ,\nonumber\\
~& = & \sum_{m \in \cK^\vee} \prod_a \left\{\oint_{C(0)} \frac{d u_a}{2\pi i u_a}\right\}  \sum_{n \in \cK^\vee} \la  \sigma_{a_1} \cdots \sigma_{a_{k}} T \prod_b \left(\frac{\delta_b}{u_b}\right)^{m_b} \ra_n \prod_c (q_c u_c)^{n_c},\nonumber\\
~&~&~
\end{eqnarray}
where the first line follows by summing the result of eqn.~(\ref{eq:comp1}), and in the second line we have introduced a sum and a
simple contour integral to pick out the relevant summand.  The contours are simply around $u_a = 0$.  Exchanging the sums on
$n$ and $m$, we get a restriction formula for the complete intersections:
\begin{equation}
\lad \sigma_{a_1} \cdots \sigma_{a_{k}} \rad=  \prod_{a}\left\{ \oint_{C(0)} \frac{d u_a}{2\pi i u_a}\right\} \la  \sigma_{a_1} \cdots \sigma_{a_{k}} \Delta \ra (q_c u_c),
\end{equation}
where
\begin{equation}
\Delta(u) = T~\sum_{m \in \cK^\vee} \prod_b \left(\frac{\delta_b}{u_b}\right)^{m_b}.
\end{equation}

This formula is a bit more complicated than the hypersurface case, and we have not found an argument showing that it reproduces
the principal discriminant locus of the theory.  Nevertheless, it seems like a compact and potentially useful way to express the
correlators.  As a simple check of its veracity, we note that it reduces to our previous result in the case of a single hypersurface.

A slightly more involved check is to compare the results obtained here to known examples.  Perhaps the simplest of these is
a (2,2) CICY in $\P^5$ of a quadratic polynomial $P_1$ and a quartic polynomial $P_2$.  The charges for the linear sigma model
are
\begin{equation}
Q = (-4,-2,1,1,1,1,1,1),
\end{equation}
so that $K_1 = -2 \sigma$, and $K_2 = -4 \sigma$, and $T = 8 \sigma^2$.
The singular locus is easily determined by finding the zeroes of the effective potential $\Jt(\sigma)$, which leads to the
locus $1 - 4^5 q  = 0$.
We expect a single non-vanishing correlator, $\lad\sigma^3\rad$, and according to our formula,
\begin{equation}
\lad \sigma^3\rad = \oint_{C(0)} \frac{d u}{2\pi i u} \la \sigma^3 \times \frac{8\sigma^2}{1-4^5 \sigma^6 u^{-1}} \ra (qu).
\end{equation}
Using our result for the correlators of the $\P^n$ V-model (eqn.~(\ref{eq:Pn})), we have $\sigma^6 = q u$, leading to
\begin{equation}
\lad\sigma^3\rad = \frac{8}{1-4^5 q}.
\end{equation}
This correlator has the right singular locus and the right classical ($q\to 0$) limit.

\subsection{Examples of Quantum Restriction}
To illustrate our results, we will now apply the (0,2) restriction formula to two examples.
\subsubsection{Hypersurface in Resolved $\P^4_{1,1,2,2,2}$} \label{ss:HinP411222}
The corresponding V-model was already discussed in section~\ref{sss:P411222}.  To construct the M-model, we
introduce the fields $(\Phi^0,\Gamma^0)$ with charges
\begin{equation}
Q_0^a = \sum_i Q_i^a = \begin{pmatrix} -4 \\ 0 \end{pmatrix}.
\end{equation}
Next, we specify polynomials $P$ and $J_i$ with charges $-Q_0^a$ and $-Q_0^a - Q_i^a$, respectively.  These must
be chosen to satisfy the $\sum_{i=0}^n E^i \cJ_i = 0$ constraint.  With the E deformations given in eqn.~(\ref{eq:Edefex2}),
a choice for the $P$ and $J_i$ is to take
\begin{equation}
P = P_0 + \Delta P, ~~~J_i = \frac{\p P_0}{\p \phi^i},
\end{equation}
with
\begin{eqnarray}
P_0 & = &( \phi_1^8 + \phi_2^8) \phi_6^4 + \phi_3^4+\phi_4^4+\phi_5^4,\nonumber\\
\Delta P &=& 2 (\ep_1 \phi_1^8 + \ep_2 \phi_1^7 \phi_2+\ep_3 \phi_1 \phi_2^7)\phi_6^4.
\end{eqnarray}

Using the quantum restriction formula (eqn.~(\ref{eq:Arestrict}) ) and the Coulomb branch solution to the V-model
(eqn.~(\ref{eq:VP411222})), we obtain
\begin{equation}
\lad \sigma_1^{3-a} \sigma_2^a \rad = -
\left\{ \Res_{z=\frac{1-4^4 q_1}{2}}
+ \Res_{z= \frac{4\ep_2\ep_3 -\ep_1}{2(1+\ep_1)}} + \Res_{z=\infty} \right\} \frac{G(z)P'(z)}{P(z)},
\end{equation}
where
\begin{equation}
G(z) = \frac{16 z^a (1-2z)}{H(z) (1-2 z -4^4 q_1)},
\end{equation}
and $P(z)$ and $G(z)$ are given in eqn.~(\ref{eq:VP411222})).

We find that the three-point functions are given by
\begin{eqnarray}
\lad \sigma_1^3 \rad 			& = & \frac{8}{D}, \cr
\lad \sigma_1^2 \sigma_2 \rad	& = & \frac{4(1-2^8 q_1)}{D}, \cr
\lad \sigma_1 \sigma_2^2 \rad	& = & \frac{4(2^{10} q_1 q_2 -2 q_2 +2^8 \ep_1q_1+2\ep_2\ep_3 -\ep_1)}{(1-4q_2)D}, \cr
\lad  \sigma_2^3 \rad 		& = & 4\left[q_2(1+4q_2 - 2^8 q_1 - 3072 q_1 q_2) + \ep_1^2(1-2^8 q_1)  \right.\cr
&~&~~
+2\ep_1 (-2^{10} q_1 q_2 + 3 q_2 -\ep_2\ep_3) \cr
&~&~~ \left.
 +\ep_2\ep_3(- 2^8 q_1  + 2^{10} q_2 q_1 + 1 - 12 q_2)\right]/(1-4q_2)^2 D,
\end{eqnarray}
where
\begin{equation}
D = (1-2^8 q_1)^2 - 2^{18} q_1^2 q_2 + 2 \ep_1(1-2^8 q_1) - 4 \ep_2 \ep_3
\end{equation}
is the principal component of the singular locus, in agreement with the computation based on the effective
potential for the $\sigma_a$.

\subsubsection{Hypersurface in $\P^2 \times \P^2$} \label{sss:P2P2example}

Here we study (0,2) deformations of the A/2 twisted example of a bi-cubic hypersurface in $\P^2\times \P^2$.
The charges are given by
\begin{equation}
Q = \begin{pmatrix} -3 & 1 & 1 & 1 & 0 & 0 & 0 \\ -3 &0 & 0 & 0 & 1 & 1 & 1 \end{pmatrix},
\end{equation}
with the anticanonical divisor represented by $-K = \sum_i Q_i^a\sigma_a = 3(\sigma_1 + \sigma_2)$.
If we ignore the $E\cdot J = 0$ constraint, then the E-parameters are given by two $3\times 3$
matrices---a simple generalization of the $\P^1\times\P^1$ example analyzed above, as well as a rescaling
of the $(\Phi^0,\Gamma^0)$ multiplets.  The $E\cdot J = 0$ constraint and the field redefinitions we discussed
above eliminate many of these possibilities.  Let us fix the J-parameters so that the following E-deformations are
allowed by the constraint:
\begin{equation}
M_{(1)} = \begin{pmatrix} \sigma_1 & \ep_2 \sigma_2  & 0 \\ \ep_3 \sigma_2 & \sigma_1 + \ep_1\sigma_2 &0\\ 0 & 0 &\sigma_1 \end{pmatrix},~~~
M_{(2)} = \begin{pmatrix} \sigma_2 & \gamma_2 \sigma_1  & 0 \\ \gamma_3 \sigma_1 & \sigma_2 + \gamma_1\sigma_1&0 \\ 0 & 0 &\sigma_2 \end{pmatrix}.
\end{equation}

Applying the quantum restriction formula, we have
\begin{equation}
\lad \sigma_1^a \sigma_2^{3-a} \rad = \la \sigma_1^a \sigma_2^{3-a} \frac{3(\sigma_1+\sigma_2) }{1+3(\sigma_1+\sigma_2)} \ra
= 3 \la \sigma_1^a \sigma_2^{3-a} \frac{(\sigma_1+\sigma_2)}{1 + 3^3 (\sigma_1+\sigma_2)^3} \ra,
\end{equation}
where on the right-hand side the correlators are computed in the (0,2)-deformed linear sigma model for $\P^2\times\P^2$.
The quantum cohomology of this V-model is quite similar to the $\P^1\times\P^1$ theory.  We have
\begin{eqnarray}
\label{eq:P2P2qc}
\det M_{(1)} = \sigma_1^3 + \ep_1 \sigma_1^2 \sigma_2 -\ep_0 \sigma_1 \sigma_2^2 &=& q_1,\nonumber\\
\det M_{(2)} = \sigma_2^3 + \gamma_1 \sigma_2^2 \sigma_1 -\gamma_0 \sigma_2 \sigma_1^2 &=& q_2,
\end{eqnarray}
where $\ep_0 = \ep_2\ep_3$, and $\gamma_0 = \gamma_2 \gamma_3$.
Finally, using the Coulomb branch analysis, we have
\begin{equation}
\lad \sigma_1^a \sigma_2^{3-a} \rad = 3\sum_{\sigma=\sigma_\ast} \sigma_1^a \sigma_2^{3-a} \frac{(\sigma_1+\sigma_2)}{H(1 + 3^3 (\sigma_1+\sigma_2)^3)},
\end{equation}
where
\begin{equation}
H = 3\left[-\ep_0 \sigma_2^4 + 2\ep_1 \sigma_1\sigma_2^3 + (3+\ep_1\gamma_1-\ep_0\gamma_0)\sigma_1^2\sigma_2^2 + 2\gamma_1\sigma_1^3 \sigma_2 -\gamma_0\sigma_1^4 \right],
\end{equation}
and $\sigma_\ast$ are solutions to eqn.~(\ref{eq:P2P2qc}).
From this form we immediately see that the four correlators are not independent but rather
satisfy
\begin{equation}
\label{eq:P2P2rel}
q_2 \lad\sigma_1^3\rad +(\gamma_0 q_1 +\ep_1 q_2) \lad\sigma_1^2\sigma_2\rad
-(\gamma_1 q_1 +\ep_0 q_2) \lad\sigma_1\sigma_2^2\rad - q_1 \lad\sigma_2^3 \rad=0.
\end{equation}
This is a very pretty property, as it has some simple consequences for {\em normalized} Yukawa couplings:  for example,
it shows that if we tune parameters to set three of these couplings to zero, the fourth will also be zero.

To obtain the actual correlators, it is useful to introduce $z = \sigma_2/\sigma_1$, as well as
\begin{equation}
\Ht(z) = \sigma_1^{-4} H,~~~ S(z) = \sigma_1^{-3} \det M_{(1)}.
\end{equation}
The equations satisfied by $\sigma_1$ and $z$ are
\begin{equation}
\sigma_1^3 = S^{-1} q_1,
\end{equation}
\begin{equation}
P(z) = q_2 S - q_1 (z^3 +\gamma_1 z^2 -\gamma_0 z) = 0,
\end{equation}
so that the correlators may be written as
\begin{equation}
\lad \sigma_1^a \sigma_2^{3-a} \rad = 3\sum_{z|P(z)=0} \frac{z^{3-a}(1+z)S}{\Ht(S + 3^3 (1+z)^3)}.
\end{equation}
Repeated application of $P=0$ to eliminate $z^n$ for $n \ge 3$ allows us to recast the correlators into
a simpler form:
\begin{equation}
\lad \sigma_1^a \sigma_2^{3-a} \rad = 3\sum_{w|P(w)=0} \oint_{C(w)} \frac{dz }{2\pi i}\frac{P'(z)}{P(z)} \frac{A_1 z^2+A_2 z + A_3}{ A_4z^2+A_5z + A_6},
\end{equation}
where $A_1,\ldots,A_6$ are easily computable but complicated rational functions of $q,\ep,\gamma$, and $C(w)$ denotes a
small contour in the $z$ plane around $z=w$.  Pulling the contour off the roots of $P(z)$ onto the roots of the denominator and
the point at infinity, we see that the correlators may be computed by completely elementary methods.  The resulting expressions
are, however, quite complicated.  To give a flavor of the results without having to introduce a lot of new notation, we set
$\ep_0=\gamma_0 =\ep_1=0$.  We find it convenient to set $\qt_1 = 3^3 q_1$ and $\qt_2 = 3^3 q_2$.  We obtain the following results:
\begin{eqnarray}
\lad \sigma_1^3 \rad & = & 3\qt_1\left[3\qt_1+3\qt_2 -6+ (4-5\qt_1-2\qt_2)\gamma_1 + (2\qt_1 -1) \gamma_1^2\right] D^{-1}, \nonumber\\
\lad \sigma_1^2 \sigma_2\rad & = & 3 \left[ (1+\qt_2)^2 -\qt_1(1+\qt_2+2\qt_1) +3\qt_1(1+\qt)\gamma_1 - (\qt_1+\qt_1^2)\gamma_1^2\right]D^{-1}, \nonumber\\
\lad \sigma_1\sigma_2^2 \rad & = & 3\left[(1+\qt_1)^2 -\qt_2(1+\qt_1+2\qt_2) -((1+\qt_1)^2+\qt_2(1-2\qt_1) )\gamma_1\right]D^{-1}, \nonumber\\
\lad \sigma_2^3 \rad & = & 3\qt_2\left[3\qt_1+3\qt_2-6+(1+2\qt_1-5\qt_2+4\qt_1\qt_2+\qt_1^2)\gamma_1+(1+\qt_1)^2\gamma_1^2\right]D^{-1},\nonumber\\
~&~&~
\end{eqnarray}
where
\begin{equation}
D= D_0 + D_1\gamma_1^2+D_2\gamma_1^2+D_3\gamma_1^3,
\end{equation}
and
\begin{equation}
\begin{array}{ccl}
D_0 & = & -1+(1+\qt_1)^3+(1+\qt_2)^3-3\qt_1\qt_2(\qt_1+\qt_2-7), \\
D_1 & = & -3\qt_1\left[ 1+2\qt_1 -7\qt_2 +(\qt_1+\qt_2)^2\right],\\
D_2 & = & 3\qt_1\left[(1+\qt_1)^2 -2\qt_2 +\qt_1\qt_2\right],\\
D_3 & = & -\qt_1 (1+\qt_1)^2.
\end{array}
\end{equation}

The correlators satisfy a number of checks:  they match the expected classical (2,2) limit; on the (2,2) locus the
correlators have a symmetry that exchanges $\sigma_1,\sigma_2$ and $q_1,q_2$; they diverge on the correct discriminant
locus; and, they satisfy the relation in eqn.~(\ref{eq:P2P2rel}).  In this example, the complete dependence on the E-parameters
may be determined with current techniques, since all the E-deformations are linear.  While in general the expressions are
quite complicated, there are also simple lessons to be learned.  In particular, the discriminant locus is
easy to compute, and the relation among amplitudes in eqn.~(\ref{eq:P2P2rel}) is easy to generalize to include
all the other E-parameters.

\section{B/2 Twist and Hypersurfaces} \label{s:restrictB}
We now turn to the B/2-twist of the M-Model. The natural guess based on the (2,2) locus results and the simplicity of the A/2-twisted theory is that the B/2-twisted theory should be independent of the $q_a$ and E-parameters.  We have not been able to prove this in full generality, but we have found a large class of models where the result holds.

In what follows, we will first tackle the dependence on K\"ahler parameters, and we will derive sufficient conditions for independence.  A closely related problem was investigated in~\cite{Sharpe:2006qd}, and  we will compare that work and our results in section~\ref{ss:b2kahler}.  Next, we will restrict
to models that satisfy the sufficient conditions and turn to examine E-dependence.  We will argue that the E-deformations should decouple from
K\"ahler-independent B/2-twisted M-models with a Landau-Ginzburg phase.  We will work with an explicit example and find that our expectations are
borne out:  the only dependence on E-parameters is absorbable into the field redefinitions.  These are promising results for a (0,2)-mirror
map between A/2-twisted and B/2-twisted theories, and we hope to prove they hold more generally in the near future.

\subsection{Field Content and Action}
We have already alluded to the B/2-twist in eqn.~(\ref{eq:twist}).  Under the twist, the spins of the fields are shifted as follows:
\begin{equation}
\begin{array}{cc}
 \begin{array}{llll}
   \sigma_a 		&\to& \sigma_{a,z} 			& \left[\in \Gamma(K)\right], \\
   \lambda_{a,+} 	&\to& \lambda_{a} 		& \left[\in \Gamma (\cO)\right],\\
   \lambdab_{a,+} 	&\to& \lambdab_{a,\zb}			& \left[ \in \Gamma (\Kb)\right],
 \end{array}
&
 \begin{array}{llll}
  \sigmab_a		&\to& \sigmab_{a,\zb} 			& \left[\in \Gamma( \overline{K} )\right], \\
   \lambda_{a,-} 	&\to& \lambda_{a,z}	 		& \left[\in \Gamma(K)\right],\\
   \lambdab_{a,-} 	&\to& \lambdab_{a}				& \left[ \in \Gamma(\cOb)\right],
 \end{array}
\end{array}
\end{equation}
\begin{equation}
\begin{array}{cc}
 \begin{array}{llll}
   \phi^i	 		&\to& \phi^i	 			& \left[\in \Gamma(\cO(d_i))\right], \\
   \psi_+^i 			&\to& \psi^i_{\zb}	 		& \left[\in \Gamma (\Kb \otimes \cOb({-d_i}))\right],\\
  \psib_+^i 		&\to&  \psib^i				& \left[ \in \Gamma (\cO({d_i}))\right],
 \end{array}
&
 \begin{array}{llll}
   \phib^i			&\to& \phib^i	 			& \left[\in \Gamma( \cOb({d_i}) )\right], \\
   \gamma_-^i	 	&\to& \gamma^i_z	 		& \left[\in\Gamma(K\otimes\cO(d_i))\right],\\
  \gammab_-^i	 	&\to& \gammab^i			& \left[ \in \Gamma(\cO(-d_i))\right].
 \end{array}
\end{array}
\end{equation}
As in our discussion of the A/2 twist, the holomorphic/anti-holomorphic sections of the bundle in the brackets correspond to the zero modes
of the kinetic operator for the particular field, and the $d_i$ are the degrees in a background with fixed instanton number.  Unlike the
A/2 twist, the B/2 twist of the $\Phi^0,\Gamma^0$ multiplets is identical to the other matter multiplets, and their twisted constituent fields
are obtained by setting $i = 0$ in the expressions.

After dropping gauge multiplet fields without zero modes (e.g. $\sigma_z$), the twisted action takes the form\footnote{Details of the B/2 twist are given
in Appendix \ref{ss:b2twistapp}.}
\begin{equation}
\label{eq:B2action}
\cL =  \cL_{\text{kin}}  + \cL_{\phi} + P \Pb + \phi^0 J_i \Jb_i \phib^0 + \cL_{\text{Yuk}},
\end{equation}
with
\begin{eqnarray}
\cL_{\text{Yuk}}  & = & Q_0^a \lambdab_a \psib^0 \phi^0 + Q_i^a \lambdab_a \psib^i \phi^i
                                   +\gammab^i \sum_{\mu\in S_i} E^{ai}(\phi)  \lambda_a + \gammab^0 Q_0^a \phi^0 \lambda_a \nonumber\\
                       ~ & ~ & -\gamma_z^0 P_{,i} \psi^i_{\zb} - \gamma^i_{z} J_i \psi^0_{\zb} - \gamma^i_{z} \phi^0 J_{i,j} \psi^j_{\zb}\nonumber\\
                       ~ & ~ & -\psib^0 \Jb_i \gammab^i -\psib^i \Pb_{,i} \gammab^0 - \psib^j \phib^0 \Jb_{i,j} \gammab^i.
\end{eqnarray}
Note, we have used field re-definitions to fix $E^0$ to its (2,2) value, but we have allowed for non-linear E-parameters.

The B/2 twist leads to the same $Q_T$ as in the A/2 twisted theory:  namely, $\Qb_+$ becomes the scalar nilpotent operator.
It is then not too surprising that this half-twisted theory localizes onto $\cM_{n;P}$---the same field configurations as its A/2 twisted cousin.
In addition, we expect the massive modes to cancel in determinants, leaving a finite dimensional integral over the zero modes.
The similarities end at this point, since the difference in the twisting leads to a different set of local observables and
different non-vanishing correlators.

Taking our cue from the usual results on the (2,2) locus, we would like to compute correlators of local, gauge-invariant operators
$O_{\alpha} = \phi^0 f_\alpha(\phi)$, where $f_\alpha(\phi)$ is polynomial in the $\phi^i$.  On the (2,2) locus these operators are
just the monomials in the superpotential.  In the (0,2) theory they remain perfectly well-defined operators in the B/2 theory.
The usual selection rule of the $B$-model is unmodified by the (0,2) deformations, and we expect
\begin{equation}
\la O_{1}\cdots O_{s} \ra = 0
\end{equation}
unless $s = d-1$.

\subsection{Vanishing Conditions} \label{ss:vanishing}
To study these correlators in more detail, we work in a geometric phase with K\"ahler cone $\cK$, and fix an instanton number
$n_a \in \cK^\vee$.  We find it convenient to adopt the following splitting of the matter fields: we treat separately
the fields in the $\Phi^0,\Gamma^0$ multiplets and split up the $n$ multiplets according to the degrees $d_i$:
\begin{equation}
I = \{1,\ldots,n\} = I_- \cup I_0 \cup I_+,
\end{equation}
 where
\begin{eqnarray}
I_- &  = & \left\{ i \in I | d_i < 0 \right\}, \nonumber\\
I_0 &  = & \left\{ i \in I | d_i = 0 \right\}, \nonumber\\
I_+ &  = & \left\{ i \in I | d_i > 0 \right\}.
 \end{eqnarray}
We will also have use for the subsets $I_{>1}$ and $I_{<-1}$ defined in the same fashion.

The first simplification comes from working in a geometric phase, where $d_0 \le 0$.   Since the path integral localizes
to $\phi^0 = 0$, as long as $d_0 < 0$, the correlator $\la O_{1}\cdots O_{s} \ra_n$ must vanish due to
a lack of $\phi^0$ zero modes.  Thus, without loss of generality, we may restrict attention to instantons satisfying
\begin{equation}
d_0 = \sum_{i=1}^n Q_i^a n_a = 0.
\end{equation}
This suffices to show the B/2 twisted models for Calabi-Yau hypersurfaces in products of projective spaces localize to constant
maps, since for these examples $d_0 = 0$ implies $n_a = 0$.  Since the correlators have a holomorphic dependence on the
couplings and are not perturbatively renormalized, we conclude that these B/2-twisted theories are independent of the K\"ahler
parameters.

There are plenty of examples where $d_0 = 0$ does not imply $n_a = 0$.  Perhaps the simplest of these is the B/2 twist of the
two-parameter model we discussed in section~\ref{ss:HinP411222}.  The charges for this M-model are
given by
\begin{equation}
\begin{pmatrix} -4 & 0 & 0 & 1 & 1 & 1 & 1 \\
                          0  & 1 & 1 & 0 & 0 & 0 &-2
\end{pmatrix},
\end{equation}
and the dual cone for the smooth geometric phase is just the first quadrant.  Since $d_0 = 4 n_1$, the contributions from
instantons with $n_1 =0, n_2\ge 0$ are allowed.

Another condition on contributing instantons may be obtained by examining the term $\gamma^i_z \phi^0 J_{i,j} \psi^j_{\zb}$ in the
action.  Recall that $\gamma^i_z$ has no zero modes if $i \not\in I_{>1}$ and otherwise has $d_i -1$ zero modes, while
$\psi^i_{\zb}$ has no zero modes if $i \not\in I_{< -1}$ and otherwise has $-d_i-1$ zero modes. Since these zero modes may
only be soaked up in pairs by bringing down the aforementioned term in the action, the instanton contribution will vanish unless
\begin{equation}
\sum_{i \in I_{<-1}} (-d_i -1 ) = \sum_{i\in I>1} (d_i -1).
\end{equation}
Since $d_0 = 0$, it follows that
\begin{equation}
|I_+| = |I_-|.
\end{equation}
This additional condition is sufficient to show that
the B/2-twisted correlators in the two parameter example only receive contributions from constant maps.
In fact, this readily extends to a number of other two-parameter examples, such as those based on hypersurfaces in
$\P^4_{1,1,2,2,6}$, $\P^4_{1,2,2,3,4}$, $\P^4_{1,2,2,2,7}$, and $\P^4_{1,1,1,6,9}$.\footnote{These examples were studied in some
detail in the early days of mirror symmetry.  See, for example,~\cite{Hosono:1993qy}.}

A third condition on the instanton numbers follows by considering the $r$ zero modes of $\lambda_a$.  One of these may be soaked up by
bringing down the term $\gammab^0 Q_0^a \phi^0 \lambda_a$, but to absorb the remaining $r-1$ requires bringing down
powers of $\gammab^i E^{ai} \lambda_a$.  However, not all of these can contribute:  $\gammab^i$
has no zero modes when $d_i > 0$, while the $E^{ai}$ have no zero modes when $d_i <0$; hence, the only contributions
to this coupling can come from fields with $i \in I_0$. It follows then that
\begin{equation}
|I_0| \ge r-1.
\end{equation}
In the examples we have examined this has always turned out to be a weaker condition than the other two, but
for larger gauge groups it may begin to play an important role.

Unfortunately, these elegant conditions are not sufficient to rule out non-trivial instanton contributions in all generality.  Consider the
two-parameter V-model with charges
\begin{equation}
Q = \begin{pmatrix} 1 & 1 & 0 & 0 & 1 & 1 \\ 0 & 0 & 1 & 1 & -1 & -1 \end{pmatrix}.
\end{equation}
The classical cone $\cK_c$ has $ r_1 > 0$ and $r_1+r_2 >0$.  It is divided into two phases, $\cK_1$ with $r_2 >0$ and $\cK_2$ with $r_2 <0$.
The first of these has the exceptional set
\begin{equation}
F = \{\phi_3=\phi_4=0\}\cup\{\phi_1=\phi_2=\phi_5=\phi_6=0\},
\end{equation}
while the second has
\begin{equation}
F = \{\phi_1=\phi_2=\phi_3=\phi_4=0\}\cup\{\phi_5=\phi_6=0\}.
\end{equation}
It is not hard to show that $V_{\cK_1}$ and $V_{\cK_2}$ are isomorphic smooth toric varieties.  We can construct the corresponding M-model
in the usual way, by taking $\phi^0$ with charges $\binom{-4}{0}$, and using the hypersurface
\begin{equation}
P = \phi_1^4 + \phi_2^4 + (\phi_3^4 + \phi_4^4+ \phi_3^2 \phi_4^2) \phi_5^4 + (\phi_3^4 +\phi_4^4) \phi_6^4.
\end{equation}
It is easy to see that the common solutions to $P=dP = 0$ are in the exceptional set, so that $P= 0$ is a smooth hypersurface in $V$.
That will persist for small deformations of $P$, and obviously for small (0,2) deformations away from $J_i = P_{,i}$.

\subsection{B/2-Twisted Theories and K\"ahler Parameters} \label{ss:b2kahler}
We have seen that in a number of models simple restrictions on the instanton numbers rule out contributions to
B/2-twisted amplitudes from non-trivial instantons.  Thus, in these theories the correlators are independent of the K\"ahler parameters.
Unfortunately, as the example in the last section indicates, there are also models where the zero-mode counting arguments are not sufficient
to rule out instanton contributions.  What is one to make of this?

An interesting perspective on this question was found in~\cite{Sharpe:2006qd}.  In that work the following puzzle was pointed out:
the B topological sector of a (2,2) SUSY NLSM with Calabi-Yau target-space may be alternatively described by a standard B-twist or a B/2-twist;
in the former case it is trivial to see that the theory localizes onto constant maps, but in the latter case this is not at all obvious, and it seems
that there is a possibility of non-trivial holomorphic maps contributing to the correlators.

In a number of models simple index theory arguments,
analogous to the fermion zero mode counting discussed in section~\ref{ss:vanishing}, are sufficient to rule out contributions from non-trivial maps.
However, there
are also examples where these arguments are not sufficient.  It was argued in~\cite{Sharpe:2006qd} that on the (2,2) locus
the resolution is as follows:  precisely in the case where the index theory permits a non-trivial contribution, one can show that the corresponding
top-form on the instanton moduli space is {\em exact}.  Thus, the contribution from a non-trivial instanton reduces to terms coming from the
boundary of the instanton moduli space.   If one works with a nice compactification of the instanton moduli space, such as that provided by the GLSM,
the contribution vanishes!

The last assumption of a ``nice'' compactification of the instanton moduli space is natural in the case of the GLSM, and although
the examples considered in~\cite{Sharpe:2006qd} were restricted to non-compact toric Calabi-Yaus, it is natural to expect that the arguments
should be generalizable to the hypersurface case as well.  Furthermore, we believe it should be possible to generalize those results to (0,2)
deformations, but we have not been able to show this is the case.  It would be extremely interesting to show this in full generality for B/2-twisted
M-Models.

Instead of pursuing this general result further, we will now take a more detailed look at some (0,2) examples where the index theory is sufficient to rule
out contributions from non-trivial instantons.  A look at the B/2-twisted action in eqn.~(\ref{eq:B2action}) shows that even in these models there remains an interesting complication:  as expected, the theory depends holomorphically on the parameters in the $J_i$ and $P$; however, there also seems to be a non-trivial dependence on the E-parameters contained in $E^{ai}(\phi)$.  Is this dependence really there?  If so, how do we compute it?  In what follows, we will
argue that there do exist B/2-twisted M-models that are independent of both K\"ahler parameters and E-deformations.

\subsection{Models with a Landau-Ginzburg Phase} \label{ss:lg}
The Calabi-Yau/Landau-Ginzburg (LG) correspondence was one of the first successes of the GLSM approach to compactifications~\cite{Witten:1993yc}.
The basic point is simple:  since the singular locus in the GLSM moduli space is complex co-dimension one, the various phases are
connected by paths in the moduli space consisting of smooth theories.  If the phases of an M-model include a phase where the
low energy theory is described by a Landau-Ginzburg orbifold, then the geometric and LG theories simply correspond to different points in the moduli
space.

Strictly speaking, the Landau-Ginzburg orbifold description only applies in the limit where the K\"ahler parameters are taken to be
arbitrarily deep in the corresponding phase.  For points away from this limit the low-energy theory is  a
finite deformation of the Landau-Ginzburg orbifold by twist-field operators.  The complementary statement from the point of view of the
geometric phase is that the theory receives world-sheet instanton corrections.  In the twisted GLSM, these effects are both
represented by the gauge theory instantons and may be computed by the same techniques in any phase. For each phase, the sum converges
when the $q_a$ are taken to be deep in the corresponding K\"ahler cone, and the resulting rational function may be trivially continued around
the complex co-dimension one singularities to other phases~\cite{Witten:1993yc,Morrison:1994fr}.

The correspondence becomes much simpler if one considers the B-twist of the theory.  The amplitudes
in the B-model are independent of the K\"ahler parameters, so that algebro-geometric computations in the geometric phase are precisely reproduced
by computations at the LG orbifold point.

Let us now consider the B/2-twisted theories that have an LG phase and are independent of the
K\"ahler parameters.\footnote{These include all the models in table~\ref{table:parameter} except for the bi-cubic in $\P^2\times\P^2$,
which does not have an LG phase.}  The CY/LG correspondence exists both on and off the (2,2) locus~\cite{Witten:1993yc},
and as long as the $E^i$ are close to the (2,2) values, by taking the F-I terms deep into the LG phase, the $\Sigma_a$ and
$\Phi^0$ multiplets both acquire arbitrarily large masses and should decouple from the low energy theory, being just set to their VeVs.
Thus, the low energy theory is described by matter multiplets with (0,2) superpotential $ \Gamma^i J_i (\Phi)$ and chiral
constraint $\cDb_+ \Gamma^i = 0$.  Thus, if the B/2-twisted theory is independent of the K\"ahler moduli, we expect it to be
independent of small variations in the E-parameters as well; moreover, it should just reduce to a B/2-twisted LG orbifold.

It is interesting to see how this independence works out in detail, and we will now turn to two examples of B/2 twisted theories that are
independent of K\"ahler moduli and study them in the LG phase. First we will work with the M-model for the quintic, and then turn to
the M-model for a hypersurface in $\P^4_{1,1,2,2,2}$.  The first example is merely a warm-up meant to illustrate how the LG description
emerges in the half-twisted model.  The second case, while not much more complicated than the quintic, will also illustrate how the
E-parameters (both linear and non-linear) decouple from the correlators.

\subsubsection{The Quintic}

The M-model for the quintic has charges
\begin{equation}
Q = (-5,1,1,1,1,1),
\end{equation}
and consequently the D-term
\begin{equation}
-5 |\phi^0|^2 + \sum_{i=1}^5 |\phi^i|^2 = \rho.
\end{equation}
In addition, there is a matter superpotential as in eqn.~(\ref{eq:02Pot}), obeying the (0,2) SUSY constraint
\begin{equation}
\sum_{i} \phi^i J_i = 5 P,
\end{equation}
with $P$ a homogeneous degree $5$ polynomial in the $\phi^i$, and $J_i$ being homogeneous of degree $4$.  The model has
no E-deformations, so we have set the E-parameters to their (2,2) values.

When $\rho>0$, the low energy theory is a NLSM with target space a hypersurface $P=0$ in $\P^4$, with bundle
structure encoded in the $J_i$.  For generic parameters, one finds  $\phi^0=0$.  When $\rho<0$,
the low energy field configurations have $|\phi^0|^2 = -\rho/5$ and $\phi^i=0$, $i=1,\ldots,5$.  Thus, $\phi^0$ acquires
a large VeV and mass, in turn giving a mass to the $\sigma$ field as well as the gauge field.  The resulting low energy
theory is a LG orbifold, since  since the vacuum $\phi^0\neq 0$, $\phi^i=0$ preserves a $\Z_5$ gauge symmetry, whose
generator acts on the $\phi^i$ via
\begin{equation}
\label{eq:quinticorb}
 \phi^i \mapsto e^{2\pi i/5} \phi^i.
\end{equation}
To study the correlators in more detail, we must construct the B/2 twisted theory.  We observe that in the LG phase, the $\GU(1)_L\times\GU(1)_R$
charges given in table~\ref{table:SimpleR_2} are slightly awkward, as they assign charges to $\phi^0$ --- a field with a vacuum
expectation value.  The resolution is simple:  since the $\GU(1)_L\times\GU(1)_R$ symmetries are only defined up to global gauge
transformations, we may use the latter to make a judicious choice for the charges of the former.  A convenient choice is given in
table~\ref{table:RLGquintic}.  Since the B/2 twist involves the difference of these charges, the twisting of the various fields
remains unmodified from the geometric phase.
\begin{table}[t]
\begin{center}
\begin{tabular}{|c|c|c|c|c|c|c|c|}
\hline
$~			$&$\theta^+ 	$&$\Phi^i		$&$\Gamma^i	$&$\Phi^0		$&$\Gamma^0	$&$\Sigma_a	 $&$\Upsilon_a 	 $\\ \hline
$\GU(1)_R	$&$1			$&$\ff{1}{5}	$&$\ff{1}{5}	$&$0			$&$0			$&$1			 $&$1			 $\\ \hline
$\GU(1)_L		$&$0			$&$\ff{1}{5}	$&$-\ff{4}{5}	$&$0			$&$-1		$&$-1		 $&$0			 $\\ \hline
\end{tabular}
\end{center}
\caption{The $\GU(1)_R$ and $\GU(1)_L$ symmetry charges for the quintic in the LG phase.}
\label{table:RLGquintic}
\end{table}

%\begin{table}[t]
%\begin{center}
%\TABLE[t]{
%\begin{tabular}{|c|c|c|c|c|c|c|c|}
%\hline
%$~			$&$\theta^+ 	$&$\Phi^i		$&$\Gamma^i	$&$\Phi^0		$&$\Gamma^0	$&$\Sigma_a	 $&$\Upsilon_a 	 $\\ \hline
%$\GU(1)_R	$&$1			$&$\ff{1}{5}	$&$\ff{1}{5}	$&$0			$&$0			$&$1			 $&$1			 $\\ \hline
%$\GU(1)_L		$&$0			$&$\ff{1}{5}	$&$-\ff{4}{5}	$&$0			$&$-1		$&$-1		 $&$0			 $\\ \hline
%\end{tabular}
%%\end{center}
%\caption{The $\GU(1)_R$ and $\GU(1)_L$ symmetry charges for the quintic in the LG phase.}
%\label{table:RLGquintic}
%}
%\end{table}
Since this theory is independent of the K\"ahler moduli, we may restrict to constant maps, in which case the zero
mode action takes a simple form:
\begin{eqnarray}
\cL &=& -5 \lambdab \psib^0 \phi^0 + \lambdab \psib^i \phi^i + \gammab^i \phi^i \lambda - 5 \gammab^0\phi^0 \lambda \nonumber\\
~   &~&~+P\Pb + \gammab^i \Jb_i \psib^0 + \gammab^0 \Pb_{,i} \psib^i \nonumber\\
~   &~&+|\phi^0|^2 J_i \Jb_i + \gammab^i \phib^0 \Jb_{i,j} \psib^j.
\end{eqnarray}
Here $\phi^0$ is fixed to its vacuum value, i.e. $|\phi^0|^2 = -\rho/5$.

Examining this action we see that, aside from the factors of $\phi^0$, the last line is precisely the zero mode action for a (0,2) Landau-Ginzburg
model with (0,2) potential given by $\Gamma^i J_i$.  However, the remaining terms may look puzzling for a moment.  The resolution is simple:
since the correlators must be independent of $\rho$, we may take $\rho \to -\infty$ without affecting the results.  In this limit the terms in the first line should
simply soak up the $\lambdab,\lambda,\gammab^0,\psib^0$ zero modes, the terms in the second line should not contribute, and the $\phi^0$ dependence
in the last line should cancel the $\phi^0$ factors in the operator insertions.

To see this explicitly, we perform the following change of variables:
\begin{equation}
\begin{array}{ll}
\phi^i = (\phi^0)^{-1/4} \phi'^i, & \lambdab = (\phi^0)^{-1} \lambdab', \\
\gammab^i = (\phi^0)^{-1/4} \gammab'^i, & \lambda = (\phi^0)^{-1} \lambda'.
\end{array}
\end{equation}
This leads to a change of the measure:
\begin{equation}
D[\text{fields}] = (\phi^0)^{-3/4} D[\text{fields}'].
\end{equation}
Gauge invariance and the consequent quasi-homogeneity properties of the couplings imply that under
this rescaling the action becomes
\begin{eqnarray}
\cL &=& -5 \lambdab' \psib^0 - 5 \gammab^0\lambda' + (\phi^0)^{-5/4} \lambdab' \psib^i \phi'^i + (\phi^0)^{-3/2} \gammab^i \phi^i \lambda  \nonumber\\
~   &~&~+|\phi^0\phib^0|^{-5} P\Pb + (\phib^0)^{-1/2} \gammab'^i \Jb_i \psib^0 + (\phib^0)^{-1} \gammab^0 \Pb_{,i} \psib^i \nonumber\\
~   &~&+J_i \Jb_i + \gammab'^i \Jb_{i,j} \psib^j,
\end{eqnarray}
where $P, J_i,$ and $J_{i,j}$ are all functions of $\phi'^i$.
The correlators we wish to compute are
\begin{equation}
\la \phi^0 f_1 (\phi^i) \phi^0 f_2 (\phi^i)\phi^0 f_3 (\phi^i) \ra_{\text{GLSM}},
\end{equation}
where the $f_\alpha$ are degree $5$ polynomials in the $\phi^i$.  Applying the change of coordinates, we see that the powers of $\phi^0$ from the measure
cancel those from the insertions, and thus, up to terms that vanish as $\rho \to -\infty$,
\begin{equation}
\la \phi^0 f_1 (\phi^i) \phi^0 f_2 (\phi^i)\phi^0 f_3 (\phi^i) \ra_{\text{GLSM}} \propto \la f_1(\phi') f_2(\phi') f_3(\phi') \ra_{\text{LG-Orb}},
\end{equation}
where the $(0,2)$ potential is indeed given by $ J_i (\Phi')$, and the orbifold action is given in eqn.~(\ref{eq:quinticorb}).  At this point, the correlator may be evaluated
by simple LG techniques~\cite{Vafa:1990mu,Vafa:1989xc,Melnikov:2007xi}.

\subsubsection{Hypersurface in Resolved $\P^4_{1,1,2,2,2}$}
Now we return to the example already studied in section~\ref{ss:HinP411222}.  Here the LG phase is the cone defined by
$\rho_2 < 0$ and $2 \rho_1 + \rho_2 < 0$.  The classical gauge theory moduli space in this phase has $\phi^i = 0$ for $i = 1,\ldots,5$, and
\begin{equation}
-8|\phi^0|^2 = 2 \rho_1 + \rho_2,~~~-2 |\phi^6|^2 = \rho_2.
\end{equation}
A finite $\Z_8$ subgroup of the gauge group is left unfixed by the VeVs of $\phi^0$ and $\phi^6$.  The action of its generator on the fields
is given by
\begin{equation}
(\phi^0,\phi^1,\phi^2,\phi^3,\phi^4,\phi^5,\phi^6) \to (\phi^0,\zeta_8 \phi^1,\zeta_8 \phi^2,\zeta_8^2\phi^3,\zeta_8^2\phi^4,\zeta_8^2\phi^5,\phi^6),
\end{equation}
where $\zeta_8$ is an eighth root of unity.  This finite gauge group is a subgroup of the R-symmetry action with charges
\begin{equation}
R_i = (0,\ff{1}{8},\ff{1}{8},\ff{1}{4},\ff{1}{4},\ff{1}{4},0).
\end{equation}

In this two-parameter example it is not immediately clear how to define the LG limit point.  A limit that suits our purposes and lies
deep in the corresponding K\"ahler cone is to take
\begin{equation}
\begin{array}{ccc}
\rho_2 &\sim& -M^2 \\
2\rho_1+\rho_2 &\sim& -M^2
\end{array}
,~~~M\to \infty.
\end{equation}

The only contributions to the B/2-twisted correlators in this theory come from the zero instanton sector, and the zero mode action
is given by
\begin{eqnarray}
\label{eq:b2ex2}
\cL & = & -4 \lambdab_1 \psib^0 \phi^0 + \lambdab^1 \psib^6 \phi^6 - 2\lambdab_2 \psib^6 \phi^6 + Q_i^a \lambdab_a \psib^i \phi^i \nonumber\\
~   & ~ & -4 \gammab^0 \lambda_1 \phi^0 + \gammab^6 \lambda_1 \phi^6 -2 \gammab^6 \lambda_2 \phi^6 + \gammab^i E^{ai} \lambda_a
\nonumber\\
~   & ~ & + P\Pb + |\phi^0|^2 J_6 \Jb_6 + \gammab^0 \Pb_{,i} \psib^i + \gammab^0 \Pb_{,6} \psib^6 \nonumber\\
~   & ~ & + \phib^0 \gammab^6 \Jb_{6,j} \psib^j + \phib^0 \gammab^i \Jb_{i,6} \psib^6 + \phib^0 \gammab^6 \Jb_{6,6} \psib^6
               + \gammab^i \Jb_i \psib^0 + \gammab^6 \Jb_6 \psib^0 \nonumber\\
~  & ~  & + |\phi^0|^2 J_i \Jb_i + \phib^0 \gammab^i \Jb_{i,j} \psib^j.
\end{eqnarray}
Note that we have allowed for non-linear E-deformations, and we have also used the fact that parameters in $E^0$ and $E^6$ may be fixed
to their (2,2) values by field redefinitions of $\sigma_a$, $\phi^0$ and $\phi^6$.

The next step is to scale out $\phi^0$ and $\phi^6$ out of the action to the extent possible.  To do this, we start with some simple field redefinitions:
\begin{equation}
\phi^i = t_1^{Q^1_i} t_2^{Q_i^2} s^{R_i} \phi'^i,~~~\gammab^i = \tb_1^{Q^1_i} \tb_2^{Q_i^2} \sbr^{R_i} \gammab'^i,
\end{equation}
and we choose
\begin{equation}
t_1 = (\phi^0)^{-1/4}, ~~~ t_2 = (\phi^0)^{-1/8} (\phi^6)^{-1/2}.
\end{equation}
Then, gauge invariance and the R-symmetry imply
\begin{eqnarray}
P(\phi,\phi^6) & = & t_1^4 s P(\phi', 1), \nonumber\\
J_i(\phi,\phi^6) & = & t_1^4 t_2^{-1} s^{7/8} J_i (\phi', 1) ~~~\text{for}~i=1,2,\nonumber\\
J_i(\phi,\phi^6) & = & t_1^3 s^{3/4} J_i (\phi', 1) ~~~~~~~\text{for}~i=3,4,5,\nonumber\\
J_6(\phi,\phi^6) & = & t_1^3 t_2^2 s J_6(\phi', 1).
\end{eqnarray}
Next, we choose $s$ to scale the $J_i$ uniformly for all $i$.  This is achieved by setting $s = t_1^{-8}t_2^8$.
In this case, the field redefinition simplifies to
\begin{equation}
\phi^i = (\phi^6)^{-1} \phi'^i,~~~\gammab^i = (\phib^6)^{-1} \gammab'^i,
\end{equation}
and we have
\begin{eqnarray}
P(\phi,\phi^6) & = &(\phi^6)^{-4} P(\phi', 1), \nonumber\\
J_i(\phi,\phi^6) & = & (\phi^6)^{-3} J_i (\phi', 1),\nonumber\\
J_6(\phi,\phi^6) & = & (\phi^6)^{-5} J_6(\phi', 1).
\end{eqnarray}

Recalling that $\phi^6$ scales as $M$, it is easy to see from the action that the terms involving $E^{ai}$ are suppressed
by $M^{-3}$ relative to the other Yukawa couplings involving the $\lambda_a$ fermions.  Applying  this field redefinition to
a three-point function of
$O_\alpha = \phi^0 f_{\alpha}(\phi,\phi^6)$, we find that as $M\to\infty$,
\begin{equation}
\la O_1 O_2 O_3 \ra  \propto  (\phi^0 (\phi^6)^{-3})^5 \int d^2\phi'^i d\gammab'^i d\psib^i ~f_1(\phi', 1) f_2 (\phi', 1) f_3 (\phi', 1)  e^{-\cL'} +O(M^{-1}),
\end{equation}
with
\begin{equation}
\cL' = |\phi^0 (\phi^6)^{-3}|^2 J_i (\phi', 1) \Jb_i (\phib',1) + \phib^0 (\phib^6)^{-3} \gammab'^i \Jb_{i,j} \psib^j.
\end{equation}
Aside from the factors of $\phi^0(\phi^6)^{-3}$, we recognize the zero mode integral of a (0,2) Landau-Ginzburg theory with
potential given by the $J_i(\phi',1)$.  Solving this by the usual saddle-point techniques, we find that the bosonic and fermionic determinants
produce just the right factor of $(\phi^0 (\phi^6)^{-3})^{-5}$ to cancel the contribution from the $\lambdab,\lambda$ integration, change in the
measure, and the insertions.  Thus, we conclude that
\begin{equation}
\la O_1 O_2 O_3\ra_{\text{GLSM}} \propto \la f_1 f_2 f_3 \ra_{\text{LG-Orb}}.
\end{equation}
As expected, these B/2-twisted amplitudes are independent of the E-deformations.

\section{Conclusions} \label{s:conclusion}
We have obtained a number of results in half-twisted (0,2) linear sigma models for Calabi-Yau hypersurfaces in toric varieties.  First,
we obtained a count of linear model parameters and described field redefinitions that render some of these redundant.  Second, we showed that
a quantum restriction formula relates the genus zero A/2-twisted amplitudes to the (0,2) quantum cohomology of the ambient toric variety. Finally,
we derived a set of sufficient conditions for the B/2-twisted theories to be independent of the K\"ahler parameters, and we argued that for
models that satisfy the conditions and have a Landau-Ginzburg phase, the B/2-twisted correlators are also independent of the bundle deformations
associated to the ambient toric variety.

There are two important loose ends that require attention.  First, the solution of the A/2-twisted model must be extended to non-linear E-deformations.
Second, it is important to look for a general proof that B/2-twisted M-models are independent of K\"ahler parameters and E-deformations.  These
results will be useful for a general formulation of (0,2) mirror symmetry in the types of theories we have considered.

Even without these general results in hand, it seems worthwhile to examine the theories we have already identified as having the requisite
properties in the B/2 sector.  A natural interpretation of our results is that the A/2-twisted M-model depends on the $N(V)$ ``toric'' deformations
associated to the ambient variety, while the B/2-twisted theory depends on the $N(M)-N(V)$  ``polynomial'' deformations.  Since the models we
examined all have well-known (2,2) mirrors, it is natural to ask whether the (0,2) deformations of the mirror theories respect the splitting we
advocate.

Supposing that those matters are settled in favor of (0,2) mirror symmetry for linear sigma models, to make contact with physical observables,
we will still have match the linear model deformations to moduli of the SCFT, and determine the K\"ahler potential.  These are not easy tasks,
but our success gives us hope that perhaps even in questions regarding the K\"ahler potential progress may be made by considering
additional structure beyond (0,2) supersymmetry in these vacua.  Perhaps these additional structures (such as the $\GU(1)_L$ current
algebra) may enable us to extend some of the results of~\cite{Dixon:1989fj} off the (2,2) locus.

Finally, one should attempt to extend our techniques to theories without a (2,2) locus.  The techniques we have developed should
apply to a number of phenomenologically interesting models simply at the price of additional book-keeping.  How to proceed
to search for mirror pairs, non-renormalization theorems and make contact with the SCFT coordinates is much less clear, but
the exactly soluble (0,2) models studied in~\cite{Blumenhagen:1996vu} may provide some clues.  Still, it seems that applying our
techniques will be a valuable step in unraveling the quantum effects in these theories.

\acknowledgments
We would like to thank R.~Donagi, J.~Guffin, B.~Nill, R.~Plesser, S.~Sethi, and E.~Sharpe for
useful discussions.  I.M. would like to especially thank R.~Plesser for discussions of
quantum restriction.  J.McO. would like to thank the Simons Workshop in Mathematics and Physics 2008 for their hospitality whilst this work was being completed. We would also like to thank A.~Eaton and E.~Honisch for discussions
on modern English usage.  J.McO. is supported by the Ledley Fellowship.  This article is based upon work supported in part by the National
Science Foundation under Grants PHY-0094328 and PHY-0401814.

\appendix
\section{Linear Model Conventions} \label{app:conv}
In this section we describe some of the details in analyzing the half-twisted $(0,2)$-GLSM models in this paper. We first describe the $(0,2)$-GLSM field content, supersymmetry transformations and action. We then describe the A/2- and B/2-twisted theories. Unless otherwise specified, we follow the conventions in \cite{Witten:1993yc}. We parameterize $(0,2)$ superspace  by coordinates $x^\pm$, $\theta^+$, $\thetab^+$ and work in Wess-Zumino
gauge.  Denoting the gauge-covariant derivative by $\nabla$, we write the superspace derivatives as
\begin{equation}
\cD_+  = \p_{\theta^+} - i \thetab^+ \nabla_+,~~~ \cDb_+ = -\p_{\thetab^+} + i \theta^+ \nabla_+.
\end{equation}

The field content of the (0,2) GLSMs considered in this paper splits into gauge field multiplets, chiral matter multiplets, and fermioninc
matter multiplets.  The superspace expansions of these in Wess-Zumino gauge is given by
\begin{eqnarray}
V_{a,-} & = & v_{a,-} - 2i \theta^+ \lambdab_{a,-} - 2i \thetab^+ \lambda_{a,-} + 2 \theta^+ \thetab^+ D_a, \nonumber\\
\Upsilon_a & = & i \cDb_+ V_{a,-} + \theta^+ \p_- v_{a,+} \nonumber\\
~&=& -2 (\lambda_{a,-} - i \theta^+(D_a -i f_{a,01}) -i \theta^+\thetab^+ \p_+ \lambda_{-,a} ), \nonumber\\
\Phi^i &=& \phi^i + \sqrt{2} \theta^+ \psi_+^i -i \theta^+\thetab^+ \nabla_+ \phi^i, \nonumber\\
\Sigma_a & = & \sigma_a +\sqrt{2} \theta^+\lambda_{a,+} - i \theta^+\thetab^+ \p_+\sigma_a,\nonumber\\
\Gamma^i &=& \gamma_-^i - \sqrt{2} \theta^+ G^i - i \theta^+\thetab^+ \nabla_+\gamma_-^i - \sqrt{2}\thetab^+ E^i(\Phi,\Sigma) \nonumber\\
~&= & \gamma_-^i -\sqrt{2} \theta^+ G^i -\sqrt{2}\thetab^+ E^i(\phi,\sigma)\nonumber\\
~&~&~ - i\theta^+\thetab^+\left[\nabla_+\gamma_-^i + 2i E^i_{~,j} \psi_+^j + 2 i E^i_{~,a} \lambda_{a,+} \right].
\end{eqnarray}

The corresponding supercharges are
\begin{equation}
\cQ_+ = \p_{\theta^+} + i \thetab^+ \nabla_+,~~~\cQb_+ = -\p_{\thetab^+} -i \theta^+\nabla_+.
\end{equation}
\subsection{$(0,2)$-GLSM}
In order to study twisting and localization, it will behoove us to determine the supersymmetry transformations of the component fields.
These are generated by
\begin{equation}
\delta_{\ep} = \ep^+ \cQ_+ - \epb^+ \cQb_+,
\end{equation}
and in Wess-Zumino gauge the action is as follows:
\begin{enumerate}
\item {\bf Vector Multiplets:}
\begin{equation}
\begin{array}{llllll}
\delta v_-  &=& -2i(\epb^+\lambda_- + \ep^+\lambdab_-),~~ & \delta v_+ &=& 0 ,\\
\delta\lambda_- & = &-i\ep^+(D-if_{01}) ,~~ & \delta\lambdab_- & = & +i\epb^+(D+if_{01}),\\
\delta f_{01} &=& i \ep^+ \p_+\lambdab_- + i\epb^+\p_+\lambda_- ,~~ & \delta D  &= &\epb^+\p_+\lambda_- -\ep^+\p_+\lambdab_-.
\end{array}
\end{equation}

\item {\bf Bosonic Chiral Multiplets (including $\Sigma$):}
\begin{equation}
\begin{array}{ccccccc}
\delta \phi			&=&+\sqrt{2}\ep^+\psi_+,		&~~~	&\delta\psi_+		&=&- i \sqrt{2}\epb^+ \nabla_+ \phi ,\\
\delta \phib		&=&-\sqrt{2}\epb^+\psib_+,		&~~~	&\delta\psib_+		&=&+ i \sqrt{2}\ep^+ \nabla_+ \phib.
\end{array}
\end{equation}

\item {\bf Fermionic Matter Multiplets:}
\begin{equation}
\begin{array}{lclclcl}
\delta \gamma_-  &=& -\sqrt{2}\ep^+ G-\sqrt{2}\epb^+ E, &~~~ & \delta G & = & i\sqrt{2}\epb^+( \nabla_+\gamma_- +i\ff{\p E}{\p\phi}\,\psi_++i\ff{\p E}{\p\sigma}\,\lambda_+),\\
\delta\gammab_- &=& -\sqrt{2}\epb^+\Gb-\sqrt{2}\ep^+ \Eb, &~~~ & \delta \Gb & = & i \sqrt{2}\ep^+(\nabla_+\gammab_--i\ff{\p\overline{E}}{\p\phib}\,\psib_+ -i\ff{\p\overline{E}}{\p\sigmab}\,\lambdab_+ ).
\end{array}
\end{equation}

\end{enumerate}
%The reader is invited (in fact, begged) to check that $Q_+^2 = \Qb_+^2 = 0$, while $\{Q_+,\Qb_+\}$ gives
%\begin{equation}
%\left[\delta,\delta'\right] = 2 i (\epb^+ \ep'^+ - \epb'^+ \ep^+) \nabla_+
%\end{equation}
%on everything (replacing $\nabla_+ \to \p_+$ for neutral fields), except for the gauge field.  On it we find
%\begin{eqnarray}
%\left[\delta,\delta'\right]v_+ 		&=&  0\\
%\left[\delta,\delta'\right]v_- 		&=& 2 i (\epb^+ \ep'^+ - \epb'^+ \ep^+) (-2 f_{01}).
%\end{eqnarray}

\subsubsection{The $(0,2)$-GLSM Action}

The Lagrangian is of the form
\be
\cL = \cL_{\Upsilon,\, {\rm KE}} + \cL_{\Sigma,\, {\rm KE}} + \cL_{\Phi,\, {\rm KE}} + \cL_{\cJ}  + \cL_{\rm FI},\label{eq:action_1}
\ee
where the first four terms are the kinetic terms for the gauge multiplet, $\Sigma$-multiplet, the $\Phi$-multiplet and the $\Gamma$-multiplet respectively. The last two terms are the matter potential and the Fayet-Ilioupoulos and theta-angle term. Working first in Minkowski space, with signature $(-,+)$ and $\p_\pm = \p_0\pm\p_1$, the terms are:
\begin{enumerate}
\item {\bf Gauge Kinetic Term:}
\begin{equation}
\cL_{\Upsilon,\, {\rm KE}}  =   \ff{1}{8e_0^2}\int d\theta^+ d \thetab^+ ~\Upsilonb_a \Upsilon_a\nonumber\\
=\frac{1}{2 e_0^2} \left[ 2\lambdab_{a,-} i \p_+ \lambda_{a,-} + D_a^2 + f_{a,01}^2 \right].
\end{equation}

\item {\bf $\Sigma$ Kinetic Term:}
\begin{equation}
\cL_{\Sigma,\, {\rm KE}} = \ff{i}{2e_0^2}\int d\theta^+ d\thetab^+  ~\Sigmab_A \p_- \Sigma_A = \ff{1}{e_0^2}[\p_+\sigmab_A \p_-\sigma_A + \lambdab_{A,+}i\p_- \lambda_{A,+}].
\end{equation}

\item {\bf $\Phi$ Kinetic Term:}
\begin{eqnarray}
\cL_{\Phi,\, {\rm KE}} &=& \ff{i}{2}\int d\theta^+ d\thetab^+  ~\Phib^i (\p_- + i Q_i^a V_{a,-} ) \Phi^i \nonumber,\\
~   &=& \ff{1}{2} \nabla_+ \phib^i \nabla_- \phi^i + \ff{1}{2} \nabla_- \phib^i \nabla_+ \phi^i + \psib_+^i i\nabla_- \psi_+^i\nonumber,\\
~   &~&~+Q_i^a D_a \phib^i\phi^i -i \sqrt{2} Q_i^a \lambdab_{a,-} \psib_+^i \phi^i - i\sqrt{2} Q_i^a \lambda_{a,-} \psi_+^i \phib^i.
\end{eqnarray}

\item {\bf $\Gamma$ Kinetic Term:}
\begin{eqnarray}
\cL_{\Gamma,\, {\rm KE}} & = &\ff{1}{2} \int d\theta^+ d\thetab^+ ~ \Gammab^I \Gamma^I =
\gammab_-^I i\nabla_+\gamma_-^I + G^I \Gb^I - E^I \Eb^I \nonumber,\\
~&~&~-\gammab_-^I (E^I_{~,j} \psi_+^j + E^I_{~,A} \lambda_{A,+})
          -(\Eb^I_{~,j} \psib_+^j + \Eb^I_{~,A} \lambdab_{A,+} ) \gamma_-^I.
\end{eqnarray}

\item {\bf The F-I Term:}
\begin{equation}
\cL_{\text{F-I}} =  \ff{1}{4}\int d\theta^+~ \Upsilon_a ( i r ^a + \theta^a /2\pi)|_{\thetab^+=0} + ~\text{h.c.} = -D_a \rho^a + \ff{\theta^a}{2\pi} f_{a,01}.
\end{equation}

\item {\bf The Matter Potential:}
\begin{eqnarray}
\cL_{\cJ} &=& -\ff{1}{\sqrt{2}} \int d\theta^+~ \Gamma^I \cJ_I(\Phi)|_{\thetab^+=0} + ~\text{h.c.} \nonumber\\
~ &=&~G^I \cJ_I(\phi) + \Gb^I \cJb_I(\phib) +\gamma_{-}^I \cJ_{I,j} \psi_+^j + \psib_{+}^j \cJb_{I,j} \gammab_{-}^I \nonumber\\
~ &=&~G^0 P + \Gb^0 \Pb + G^i \phi^0 J_i + \Gb^i \phib^0 \Jb_i \nonumber\\
~ &~&~~+ \gamma_-^0 P_{,i} \psi_+^i + \gamma_-^i J_i \psi_+^0 + \gamma_-^i \phi^0 J_{i,j} \psi_+^j \nonumber\\
~ &~&~~+ \psib_+^i \Pb_{,i} \gammab_-^0 + \psib_+^0 \Jb_i \gammab_-^i + \psib_+^j \Jb_{i,j} \phib^0 \gammab_-^i.
\end{eqnarray}
In the last few lines we have assumed that there are as many $\Gamma^I$ as there are $\Phi^i$, and we used the
form of $\cJ$ relevant to the hypersurface example:
\begin{equation}
\cJ_0 = P,~~~ \cJ_i = \phi^0 J_i.
\end{equation}
%The SUSY variation of this term is (up to total derivatives)
%\begin{equation}
%\delta L  =  \sqrt{2}\ep \psib^j_+ \p_j (E^I \cJ_I) +\text{h.c.}.
%\end{equation}
\end{enumerate}
The first five terms are individually supersymmetric. The matter potential is supersymmetric provided the constraint $E^I \cJ_I =0 $ is satisfied.

For our purposes it is more useful to have the action in Euclidean space.   This is easily achieved by substituting
\begin{equation}
\p_+ \to 2 i \p_{\zb},~~~ \p_- \to 2i \p_z,~~~f_{01} \to -i f_{12}.
\end{equation}
into the Minkowski expressions and flipping the sign of the action.

\subsection{The Half-Twist} \label{ss:halftwist}
Motivated by the GLSMs with $(2,2)$ supersymmetry, we demand that our model has the symmetries given in Table~\ref{table:R_2}.
\TABLE[t]{
\begin{tabular}{|c|c|c|c|c|c|c|c|}
\hline
$~			    $&$\theta^+ $&$\Phi^i	$&$\Gamma^I		    $&$\cJ_I		$&$E^I			 $&$\Sigma_a	    $&$\Upsilon_A 	 $\\ \hline
$\GU(1)_R		$&$1		$&$q_i		$&$q_I				$&$1-q_I		$&$1+q_I		$&$1			 $&$1			 $\\ \hline
$\GU(1)_L		$&$0		$&$q_i		$&$q_I-1			$&$1-q_I		$&$q_I-1		$&$-1			 $&$0			 $\\ \hline
\end{tabular}
\caption{The $\GU(1)_R$ and $\GU(1)_L$ symmetry charges.}
\label{table:R_2}
}
It is easy to verify explicitly that the classical action respects these symmetries, provided that the $\cJ_I$ and $E^I$ can be assigned
the requisite charges.  These chiral symmetries are, in general, anomalous in the presence of non-trivial gauge fields, with anomaly
functions proportional to $(\sum_{\text{left}} q Q^a  - \sum_{\text{right}} q Q^a)n_a$, where $q$ is the global symmetry charge, $Q^a$
the gauge charge, and $n_a$ the instanton number.  In the case at hand we have
\begin{eqnarray}
\GU(1)_R &:& \sum_I q_I Q_I^a - \sum_i (q_i-1) Q_i^a,\nonumber\\
\GU(1)_L  &:& \sum_I (q_I-1) Q_I^a - \sum_i q_i Q_i^a.
\end{eqnarray}
On the (2,2) locus the familiar result holds:  the vectorial combination is always non-anomalous, while the axial combination has an anomaly
proportional to $\sum_i Q_i^a$.

To work out the twist, we need charges corresponding to the generators
\begin{equation}
J_A  = \ff{1}{2} ( J_R + J_L), ~~~ J_B = \ff{1}{2} (J_R - J_L).
\end{equation}
The two symmetries are important in each twist:  one is used for the twist, and the other becomes the ghost number of the
twisted theory.  We list both symmetries in Table \ref{table:R_3}.
\TABLE[t]{
\begin{tabular}{|c|c|c|c|c|c|c|c|}\hline
$~	     		$&$\theta^+	    $&$\Phi^i	$&$\Gamma^I		        $&$\cJ_I		$&$E^I			 $&$\Sigma_a	     $&$\Upsilon_A 	 $\\ \hline
$\GU(1)_A		$&$\ff{1}{2}	$&$q_i		$&$q_I-\ff{1}{2}		$&$1-q_I		$&$q_I			 $&$0			 $&$\frac{1}{2}	$\\ \hline
$\GU(1)_B		$&$\ff{1}{2}	$&$0		$&$\ff{1}{2}			$&$0			$&$1			 $&$1			 $&$\frac{1}{2}	$\\ \hline
\end{tabular}
\caption{The $\GU(1)_A$ and $\GU(1)_B$ charges relevant for the half-twists}
\label{table:R_3}
}
To twist, we redefine the Lorentz charges as in eqn.~(\ref{eq:twist}).
Note that our sign convention in (\ref{eq:twist}) differs to that in \cite{Witten:1993yc,Witten:1991zz}. Both $A/2$- and $B/2$-twists result in $\Qb_+$ becoming a world-sheet scalar.  Thus, in the half-twisted models, the supercharge $\Qb_+$ becomes the BRST-charge $Q_T$.
%The field content and Lorentz charge after twisting is given by:
%\begin{equation}
%\begin{array}{c|c|c|c|c|c|c|c|c|c|c|c|c|c|c|}
%~	&\phi^i		&\phib^i		&\psi_+^i		&\psib_+^i		&\gamma_-^I	&\gammab_-^I	 &G^I		 &\Gb^I		
%	&\sigma		&\sigmab		&\lambda_+	&\lambdab_+	&\lambda_-	&\lambdab_- 		\\	 \hline
%T	&0			&0			&\half		&\half		&-\half		&-\half		&0		&0		
%	&0			&0			&\half		&\half		&-\half		&-\half			\\	\hline
%A	&q_i			&-q_i			&q_i-\half 		&\half-q_i		&q_I-\half		&\half-q_I		 &q_I-1	&1-q_I	
%	&0			&0			&-\half		&\half		&\half		&-\half			\\	\hline
%T'	&-q_i			&q_i			&1-q_i		&q_i			&-q_I		&q_I-1		&1-q_I	 &q_I-1		
%	&0			&0			&1			&0			&-1			&0				\\	\hline
%B	&0			&0			&-\half 		&\half		&\half		&-\half		 &0		&0	
%	&1			&-1			&\half		&-\half		&\half		&-\half			\\	\hline
%T''	&0			&0			&1			&0			&-1			&0			&0		 &0		
%	&-1			&+1			&0			&1			&-1			&0				\\	\hline
%\end{array}\nonumber
%\end{equation}
We will restrict attention to the case $I=i$ and $A=a$---necessary conditions for the theory to have a (2,2) locus.  In these
theories we may take the charges to be  $q_i=0$ for $i>0$ and $q_0=1$.  These charges are a bit ambiguous in the presence of the
gauge symmetry.  Our choice makes the symmetries transparent in the geometric phase of a CY hypersurface GLSM, where $\phi^0 = 0$ and
the $\phi^i$ are constrained to lie on the hypersurface.  In other phases a different assignment is more suitable.

\subsection{The $A/2$-Twist}
It is useful to relabel the twisted fields in accordance with the modified Lorentz charges.  This is carried out in tables~\ref{table:atwist_1}~,~\ref{table:atwist_2}.
\TABLE[t]{
\begin{tabular}{|l|c|c|c|c|c|c|}\hline
${\rm Old~Name}     $&$ \sigma_a	$&$\sigmab_a	$&$\lambda_{a,+}		$&$\lambdab_{a,+}	 $&$\lambda_{a,-}	 $&$\lambdab_{a,-}	$\\\hline
${\rm Lorentz~Charge}$&$0		    $&$0			$&$1				    $&$0				 $&$-1			 $&$0				 $    \\\hline
${\rm New~Name}      $&$\sigma_a	$&$\sigmab_a	$&$\lambda_{a,\zb}	$&$\lambdab_a		 $&$\lambda_{a,z}	 $&$\chib_a		$	 \\\hline
\end{tabular}
\caption{The $A/2$-twisted Lorentz charges and labels for gauge field multiplets}
\label{table:atwist_1}
}
\TABLE[t]{
\begin{tabular}{|l|c|c|c|c|c|c|c|c|}\hline
${\rm Old~Name}       $&$ \phi^0	$&$\phib^0		 $&$\psi_+^0	 $&$\psib_+^0	  $&$\gamma_-^0	 $&$\gammab_-^0	 $&$G^0	         $&$\Gb^0	$\\ \hline
${\rm Lorentz~Charge} $&$-1		    $&$+1			 $&$0		     $&$+1			  $&$-1			 $&$0			 $&$0		     $&$0		$\\ \hline
${\rm New~Name}       $&$\phi_z^0	$&$\phib_{\zb}^0 $&$\psi^0	     $&$\psib^0_{\zb} $&$\gamma^0_z	 $&$\gammab_0	 $&$G^0	         $&$\Gb^0	$\\ \hline\hline
${\rm Old~Name}       $&$\phi^i		$&$\phib^i		 $&$\psi_+^i	 $&$\psib_+^i	  $&$\gamma_-^i	 $&$\gammab_-^i	 $&$G^i		     $&$\Gb^i	$\\	\hline
${\rm Lorentz~Charge} $&$0	    	$&$0			 $&$1			 $&$0			  $&$0			 $&$-1			 $&$1		     $&$-1		$\\	\hline
${\rm New~Name}       $&$\phi^i		$&$\phib^i		 $&$\psi^i_{\zb} $&$\psib^i		  $&$\gamma^i	 $&$\gammab^i_{z} $&$G^i_{\zb}	 $&$\Gb^i_z	$\\	\hline
\end{tabular}
\caption{The $A/2$-twisted Lorentz charges and labels for chiral matter multiplets}
\label{table:atwist_2}
}
The functions $E$ and $\cJ$ become:
\begin{eqnarray}
E^0_z = i\sqrt{2}B^a\sigma_a \phi^0_z,     &~~~& \cJ_0 = P(\phi^1,\ldots,\phi^n),\nonumber\\
E^i  = i\sqrt{2}E^{ai}(\phi) \sigma_a, &~~~& \cJ_{i,z} = \phi^0_z J_i(\phi^1,\ldots,\phi^n).
\end{eqnarray}
Note that on the $(2,2)$-locus we have:
\begin{equation}
E^0_z = i\sqrt{2} Q^a_0 \sigma_a \phi^0_z,~~~E^i = i\sqrt{2} Q_i^a \sigma_a \phi^i,~~~J_i = \frac{\p P}{\p\phi^i}.
\end{equation}

\subsubsection{$Q_T$-Transformations}\label{ss:Aloc}

We give the action of BRST charge $Q_T  =\Qb_+$ in Euclidean space in terms of the new fields.
\begin{enumerate}
\item {\bf Gauge Field Multiplet:}
\begin{equation}
\begin{array}{lcl}
(v_{a,z},v_{a,\zb}) \to (\lambda_{a,z},0),		&~~~& \sigma_a \to 0, \\
\lambda_{a,z}\to 0,						&~~~& \sigmab_a \to \sqrt{2}\lambdab_{a},\\
\chib_{a} \to -i (D_a+f_{a}),				&~~~& \lambda_{a,\zb} \to -2 \sqrt{2} \p_{\zb}\sigma_a, \\
(D_a,f_a) \to -2 i \p_{\zb} \lambda_{a,z}(1,-1),			&~~~& \lambdab_a \to 0.
\end{array}
\end{equation}			
We have written $f_{a,12}$ as  $f_a$.
\item {\bf Bosonic Chiral Multiplets:}
\begin{equation}
\begin{array}{lcl}
\phi^0_{z} \to 0,						&~~~&	\phi^i \to 0, \\
\phib^0_{\zb} \to \sqrt{2}~\psib^0_{\zb},	&~~~&	\phib^i \to \sqrt{2}~\psib^i, \\
\psi^0 \to -2\sqrt{2} \nabla_{\zb}\phi^0_{z},	&~~~&	\psi^i_{\zb} \to -2\sqrt{2} \nabla_{\zb} \phi^i, \\
\psib^0_{\zb} \to 0,					&~~~&	\psib^i \to 0.
\end{array}
\end{equation}
\item {\bf Fermionic Matter Multiplets:}
\begin{equation}
\begin{array}{lcl}
\gamma^0_{z} \to \sqrt{2} E^0_z,						&~~~&	\gamma^i \to \sqrt{2} E^i, \\
\gammab^0 \to \sqrt{2}~\Gb^0,							&~~~&	\gammab^i_{z} \to \sqrt{2}~\Gb^i_{z}, \\
G^0 \to \sqrt{2}(2\nabla_{\zb}\gamma^0_z + \frac{\p E^0_z}{\p\phi^0_z} \psi^0+ \frac{\p E^0_z}{\p\sigma_a} \lambda_{a,\zb}),		
&~~~&	
G^i_{\zb} \to \sqrt{2} (2\nabla_{\zb}\gamma^i+ \frac{\p E^i}{\p\phi^j} \psi^j_{\zb}+ \frac{\p E^i}{\p\sigma_a} \lambda_{a,\zb}) , \\
\Gb^0 \to 0,					&~~~&	\Gb^i_{z} \to 0.
\end{array}
\end{equation}
\end{enumerate}

\subsubsection{The A/2-Twisted Action} \label{ss:Aaction}

We end the section by giving the A/2-twisted action using eqn.~(\ref{eq:action_1}) and Tables~\ref{table:atwist_1},\ref{table:atwist_2}.
For convenience, we separate out the $Q_T$-exact terms.
\begin{enumerate}
\item {\bf Gauge Kinetic Term:}
\begin{equation}
\cL_{\Upsilon,\, {\rm KE}} = \ff{1}{e_0^2} \left[ 2 \lambda_{a,z} \p_{\zb} \chib_a + \ff{1}{2} f_a^2 -\ff{1}{2} D_a^2 \right].
\end{equation}
This is $Q_T$ exact with $\cL = \{ Q_T, V\}$ and
\be
V = \frac{i}{2e_0^2} \bar\chi_a(f_a-D_a).
\ee

\item {\bf $\Sigma$ Kinetic Term:}
\begin{equation}
 \cL_{\Sigma,\, {\rm KE}} = \ff{1}{e_0^2} \left[4 \p_z \sigma_a \p_{\zb}\sigmab_a + 2 \lambda_{a,\zb} \p_z \lambdab_a \right].
\end{equation}
This is $Q_T$ exact with $\cL = \{ Q_T, V\}$ and
\be
V = -\frac{\sqrt{2}}{e_0^2}\lambda_{a,\bar z} \del_z \bar \sigma_a
\ee

\item {\bf $\Phi^0$ Kinetic Term:}
\begin{eqnarray}
 \cL_{\Phi^0,\, {\rm KE}} & = & 2 \nabla_{\zb} \phib^0_{\zb} \nabla_z \phi^0_z + 2 \nabla_z \phib^0_{\zb} \nabla_{\zb} \phi^0_z + 2 \psib^0_{\zb} \nabla_z \psi^0\nonumber\\
~&~&~ - Q_0^a D_a \phib^0_{\zb} \phi^0_{z} + i\sqrt{2} Q_0^a \chib_a \psib^0_{\zb} \phi^0_z + i\sqrt{2} Q_0^a \lambda_{a,z} \psi^0\phib^0_{\zb}.
\end{eqnarray}
This is $Q_T$ exact with $\cL = \{ Q_T, V\}$ and
\be
V = -\sqrt{2} \psi^0 \nabla_z \bar\phi^0_{\bar z} - iQ_0^a \bar\chi_a \bar\phi^0_{\bar z}\phi^0_z.
\ee

\item {\bf $\Phi^i$ Kinetic Term:}
\begin{eqnarray}
 \cL_{\Phi^i,\, {\rm KE}} & = & 2 \nabla_{\zb}\phib^i \nabla_z \phi^i + 2 \nabla_z \phib^i \nabla_{\zb} \phi^i + 2 \psi^i_{\zb} \nabla_z \psib^i \nonumber\\
~&~&~-Q_i^a D_a \phib^i \phi^i + i\sqrt{2} Q_i^a \chib_a \psib^i \phi^i + i\sqrt{2} Q_i^a \lambda_{a,z} \psi^i_{\zb} \phib^i.
\end{eqnarray}
This is $Q_T$ exact with $\cL = \{ Q_T, V\}$ and
\be
V = -\sqrt{2} \psi_{\bar z}^i \nabla_z \bar\phi^i - iQ_i^a \bar\chi_a \bar\phi^i\phi^i.
\ee

\item {\bf $\Gamma^0$ Kinetic Term:}
\begin{eqnarray}
 \cL_{\Gamma^0,\, {\rm KE}} & = & 2 \gamma^0_z \nabla_{\zb} \gammab^0 - G^0 \Gb^0 + E^0_z \Eb^0_{\zb} \nonumber\\
~&~&~+\gammab^0 (\frac{\p E^0_z}{\p \phi^0_z} \psi^0 + \frac{\p E^0_z}{\p\sigma_a} \lambda_{a,\zb} )
           -\gamma^0_z (\frac{\p \Eb^0_{\zb}}{\p\phib^0_{\zb}} \psib^0_{\zb}+ \frac{\p\Eb^0_{\zb}}{\p\sigmab_a} \lambdab_a).
\end{eqnarray}
This is $Q_T$ exact with $\cL = \{ Q_T, V\}$ and
\be
V = \frac{1}{\sqrt{2}} \left(-\bar\gamma^0 G^0 + \gamma^0_z \bar E^0_{\bar z} \right) .
\ee

\item {\bf $\Gamma^i$ Kinetic Term:}
\begin{eqnarray}
 \cL_{\Gamma^i,\, {\rm KE}} & = & 2 \gammab^i_z \nabla_{\zb} \gamma^i + E^i \Eb^i - G^i_{\zb} \Gb^i_{z} \nonumber\\
~&~&~+\gammab^i_z (\frac{\p E^i}{\p\phi^j} \psi^j_{\zb} + \frac{\p E^i}{\p\sigma_a} \lambda_{a,\zb} )
           - \gamma^i (\frac{\p\Eb^i}{\p\phib^j} \psib^j + \frac{\p\Eb^i}{\p\sigmab_a} \lambdab_a ).
\end{eqnarray}
This is $Q_T$ exact with $\cL = \{ Q_T, V\}$ and
\be
V = \frac{1}{\sqrt{2}} \left(-\bar\gamma^i_z G^i_{\bar z} + \gamma^i \bar E^i\right) .
\ee

\item {\bf F-I Term:}
\begin{equation}
\cL = D_a \rho^a + i \frac{\theta^a}{2\pi} f_a.
\end{equation}
This can be written as the sum of a $Q_T$-closed and $Q_T$-exact term:
\be
\cL = \half (D_a - f_a) (\rho^a - \frac{i\theta^a}{2\pi}) + \left\{Q_T , V\right\},
\ee
with
\be
V = \frac{i}{2}\bar \chi_a (\rho^a + \frac{i\theta^a}{2\pi}).
\ee

\item {\bf Matter Potential:}
\begin{eqnarray}
\cL_{\cJ} & = & - G^0 P -\Gb^0 \Pb - G^i_{\zb} \phi^0_z J_i -\Gb^i_z \phib^0_{\zb} \Jb_i \nonumber\\
~&~&~-\gamma^0_z P_{,i} \psi^i_{\zb} - \psib^i \Pb_{,i} \gammab^0 - \gamma^i J_i \psi^0 -\psib^0_{\zb} \Jb_i \gammab^i_z\nonumber\\
~&~&~-\gamma^i J_{i,j} \phi^0_z \psi^j_{\zb} - \psib^j \Jb_{i,j} \phib^0_{\zb} \gammab^i_z,
\end{eqnarray}
where we used our explicit form of the $\cJ_i$.

All anti-holomorphic pieces are $Q_T$-exact, while the remainder are $Q_T$-closed:
\bea
\cL = -G^0P - G^i_{\bar z}\phi^0_z J_i - \gamma^0_z P_{,i}\psi^i_{\bar z}  -\gamma^i J_i\psi^0 - \gamma^i J_{i,j} \phi^0_z \psi^j_{\bar z} + \{ Q_T, V\},
\eea
where
\be
V = \frac{-1}{\sqrt{2}} \left(\bar P \bar\gamma^0 + \bar\gamma_z^i \Jb_i \bar\phi^0_{\bar z}\right).
\ee
\end{enumerate}

\subsection{The $B/2$-Twist} \label{ss:b2twistapp}
We proceed analogously for the $B/2$-twisted model.
The $B/2$-twisted field content is given in Tables~\ref{table:btwist_1},~\ref{table:btwist_2}.
\TABLE[t]{
\begin{tabular}{|l|c|c|c|c|c|c|}\hline
${\rm Old~Name}       $&$\sigma_a		$&$\sigmab_a		$&$\lambda_{a,+}		 $&$\lambdab_{a,+}	 $&$\lambda_{a,-}	 $&$\lambdab_{a,-}	$ \\	 \hline
${\rm Lorentz~Charge} $&$-1			$&$1				$&$0				    $&$1				 $&$-1			 $&$0				 $\\	 \hline
${\rm New~Name}       $&$\sigma_{a,z}	$&$\sigmab_{a,\zb}$&$\lambda_{a}		 $&$\lambdab_{a,\zb}$&$\lambda_{a,z}	 $&$\lambdab_{a}		$\\	\hline
\end{tabular}
\caption{The $B/2$-twisted Lorentz charges and labels for gauge field multiplets}
\label{table:btwist_1}
}
\TABLE[t]{
\begin{tabular}{|l|c|c|c|c|c|c|c|c|}\hline
${\rm Old~Name}       $&$\phi^i		$&$\phib^i		$&$\psi_+^i		$&$\psib_+^i		 $&$\gamma_-^i	 $&$\gammab_-^i		 $&$G^i	 $&$\Gb^i	$\\\hline
${\rm Lorentz~Charge} $&$0		    $&$0		    $&$1		    $&$0			    $&$-1			 $&$0			 	    $&$0		 $&$0$		\\\hline
${\rm New~Name}       $&$ \phi^i	$&$\phib^i		$&$\psi^i_{\zb}	$&$\psib^i		    $&$\gamma^i_z	 $&$\gammab^i		    $&$G^i	 $&$\Gb^i$	\\\hline
\end{tabular}
\label{table:btwist_2}
\caption{The $B/2$-twisted Lorentz charges and labels for chiral matter multiplets. As opposed to the $A/2$-twist here $i=0,\ldots,n$. }
}
The $E$ and $\cJ$ get re-written as
\begin{eqnarray}
E^0_z = i\sqrt{2}B^a\sigma_{a,z} \phi^0,     &~~~& \cJ_0 = P(\phi^1,\ldots,\phi^n),\nonumber\\
E^i_z  = i\sqrt{2}E^{ai}(\phi) \sigma_{a,z} ,  &~~~& \cJ_{i} = \phi^0 J_i(\phi^1,\ldots,\phi^n).
\end{eqnarray}
The $(2,2)$ locus amounts to setting
\begin{equation}
E^0_z = i\sqrt{2} Q^a_0 \sigma_{a,z} \phi^0,~~~E^i_z = i\sqrt{2} Q_i^a \sigma_{a,z} \phi^i,~~~J_{i} = \frac{\p P}{\p\phi^i}.
\end{equation}

\subsubsection{$Q_T$-Transformations}

The action of BRST charge $Q_T=\Qb_+$ is again easy to write down.
\begin{enumerate}
\item {\bf Gauge Field Multiplets:}
\begin{equation}
\begin{array}{lcl}
\sigma_{a,z} \to 0,						&~~~&	(v_{a,z}, v_{a,\zb} ) \to (\lambda_{a,z}, 0) , \\
\sigmab_{a,\zb} \to \sqrt{2}~\lambdab_{a,\zb},	&~~~&	(D_a,f_a) \to -2i \p_{\zb} \lambda_{a,z} (1,-1) \\
\lambda_a \to -2\sqrt{2} \p_{\zb}\sigma_{a,z},	&~~~&	\lambda_{a,z} \to 0 \\
\lambda_{a,\zb} \to 0,					&~~~&	\lambdab_{a} \to -i (D_a+f_a).
\end{array}
\end{equation}
\item {\bf Matter Multiplets:}
\begin{equation}
\begin{array}{lcl}
\phi^i \to 0,						&~~~&	\gamma^i_z \to \sqrt{2}~E^i_z, \\
\phib^i \to \sqrt{2}~\psib^i,				&~~~&	\gammab^i \to \sqrt{2}\Gb^i \\
\psi^i_{\zb} \to -2\sqrt{2} \nabla_{\zb}\phi^i,&~~~&	G^i \to \sqrt{2} (2 \nabla_{\zb} \gamma^i_z +  \frac{\p E^i_z}{\p \phi^j} \psi^i_{\zb} +  \frac{\p E^i_z}{\p \sigma_{a,z}} \lambda_a), \\
\psib^i \to 0,						&~~~&	\Gb^i \to 0.
\end{array}
\end{equation}
\end{enumerate}

\subsubsection{The B/2-Twisted Action}

We give the twisted action as well as the $Q_T$-closed and $Q_T$-exact pieces.
\begin{enumerate}
\item {\bf Gauge Kinetic Term}
\begin{equation}
\cL = \ff{1}{e_0^2} \left[ 2 \lambda_{a,z} \p_{\zb} \bar\lambda_a + \ff{1}{2} f_a^2 -\ff{1}{2} D_a^2 \right].
\end{equation}
This is $Q_T$ exact with $\cL = \{ Q_T, V\}$ and
\be
V = \frac{i}{2e_0^2} \bar\lambda_a(f_a-D_a).
\ee
\item {\bf $\Sigma$ Kinetic Term}
\begin{equation}
\cL = \ff{1}{e_0^2} \left[2 \p_z \sigmab_{a,\zb} \p_{\zb}\sigma_{a,z} + 2 \p_{\zb} \sigmab_{a,\zb}\p_{z} \sigma_{a,z} + 2 \lambdab_{a,\zb} \p_z \lambda_a \right].
\end{equation}
This is $Q_T$ exact with $\cL = \{ Q_T, V\}$ and
\be
V = -\frac{\sqrt{2}}{e_0^2}\lambda_{a} \del_z \bar \sigma_{a\,\bar z}.
\ee

\item {\bf $\Phi^i$ Kinetic Term}
\begin{eqnarray}
\cL & = & 2 \nabla_{\zb}\phib^i \nabla_z \phi^i + 2 \nabla_z \phib^i \nabla_{\zb} \phi^i + 2 \psi^i_{\zb} \nabla_z \psib^i \nonumber\\
~&~&~-Q_i^a D_a \phib^i \phi^i + i\sqrt{2} Q_i^a \lambdab_a \psib^i \phi^i + i\sqrt{2} Q_i^a \lambda_{a,z} \psi^i_{\zb} \phib^i.
\end{eqnarray}
This is $Q_T$ exact with $\cL = \{ Q_T, V\}$ and
\be
V = -\sqrt{2} \psi_{\bar z}^i \nabla_z \bar\phi^i - iQ_i^a \bar\lambda_a \phi^i\bar\phi^i.
\ee

\item {\bf $\Gamma^i$ Kinetic Term}
\begin{eqnarray}
\cL & = & 2 \gamma^i_z \nabla_{\zb} \gammab^i + E^i_z \Eb^i_{\zb} - G^i \Gb^i \nonumber\\
~&~&~+\gammab^i (\frac{\p E^i_z}{\p\phi^j} \psi^j_{\zb} + \frac{\p E^i_z}{\p\sigma_{a,z}} \lambda_{a} )
           - \gamma^i_z (\frac{\p\Eb^i_{\zb}}{\p\phib^j} \psib^j + \frac{\p\Eb^i_{\zb}}{\p\sigmab_{a,\zb}} \lambdab_{a,\zb} ).
\end{eqnarray}
This is $Q_T$ exact with $\cL = \{ Q_T, V\}$ and
\be
V = \frac{1}{\sqrt{2}} \left(-\bar\gamma^i G^i + \gamma^i_z \bar E^i_{\bar z}\right) .
\ee

\item {\bf F-I Term}
\begin{equation}
\cL = D_a \rho^a + i \frac{\theta^a}{2\pi} f_a.
\end{equation}
This can be written as the sum of a $Q_T$-closed and $Q_T$-exact term:
\be
\cL = \half (D_a - f_a) (\rho^a - \frac{i\theta^a}{2\pi}) + \left\{Q_T , V\right\},
\ee
with
\be
V = \frac{i}{2}\bar \lambda_a (\rho^a + \frac{i\theta^a}{2\pi}).
\ee

\item {\bf Matter Potential}
\begin{eqnarray}
\cL & = & - G^0 P -\Gb^0 \Pb - G^i \phi^0 J_i -\Gb^i \phib^0 \Jb_i \nonumber\\
~&~&~-\gamma^0_z P_{,i} \psi^i_{\zb} - \gamma^i_z J_i \psi^0_{\zb} - \gamma^i_z \phi^0 J_{i,j} \psi^j_{\zb}  \nonumber\\
~&~&~-\psib^0 \Jb_{i} \gammab^i - \psib^i \Pb_{,i} \gammab^0 - \psib^j \phib^0 \Jb_{i,j} \gammab^i.
\end{eqnarray}
Again, the explicit form of the $\cJ$ was used. All anti-holomorphic pieces are $Q_T$-exact, while the remainder are $Q_T$-closed:
\bea
\cL = - G^0 P  - G^i \phi^0 J_i -\gamma^0_z P_{,i} \psi^i_{\zb} - \gamma^i_z J_i \psi^0_{\zb} - \gamma^i_z \phi^0 J_{i,j} \psi^j_{\zb}  \nonumber\\+ \{ Q_T, V\},
\eea
where
\be
V = \frac{-1}{\sqrt{2}} \left(\bar P \bar\gamma^0 + \bar\gamma^i \bar J_i \bar\phi^0\right).
\ee
\end{enumerate}

%\bibliographystyle{utphys}
%\bibliography{bigref}
%\bibliographystyle{/Users/lmel/BIB/utphys}
%\bibliography{/Users/lmel/BIB/bigref}
\providecommand{\href}[2]{#2}\begingroup\raggedright\endgroup

\end{document}